\documentclass{aa}
\usepackage{graphicx}
\usepackage{hyperref}
\usepackage{txfonts}
\usepackage{placeins}
\usepackage{natbib}
\begin{document}
   \title{Water, hydrogen cyanide, carbon monoxide, and dust production from distant comet 29P/Schwassmann-Wachmann 1 \thanks{{\it Herschel} is an ESA space observatory with science instruments
  provided by European-led Principal Investigator consortia and with important contribution from NASA. }\fnmsep
\thanks{Based on observations carried out under project numbers 243-07, 151-09, D22-09, 144-10 and 001-21  with the IRAM 30 m telescope. IRAM  is  supported  by  INSU/CNRS  (France),  MPG  (Germany)  and IGN (Spain).}}

\author{D.~Bockel\'ee-Morvan\inst{1} \and N.~Biver\inst{1} \and C.A. Schambeau\inst{2,3} \and J. Crovisier\inst{1} \and C. Opitom\inst{4} \and M.~de Val Borro\inst{5} \and E.~Lellouch\inst{1} \and P.~Hartogh\inst{6} \and B. Vandenbussche\inst{7}  \and E. Jehin\inst{8} \and M.~Kidger\inst{9} \and M.~K\"{u}ppers\inst{9} \and D.C.~Lis\inst{10} \and R.~Moreno\inst{1}   \and S.~Szutowicz\inst{11} \and V.~Zakharov\inst{12,1}}

\institute{LESIA, Observatoire de Paris, Universit\'e PSL, Sorbonne Universit\'e, Universit\'e  Paris Cit\'e, CNRS, 5 place Jules Janssen, 92195 Meudon, France\\ 
\email{dominique.bockelee@obspm.fr} 
\and Florida Space Institute, University of Central Florida, 12354 Research Parkway, Partnership 1, Orlando, FL 32826, USA
\and Department of Physics, University of Central Florida, Orlando, FL 32816, USA
\and Institute for Astronomy, University of Edinburgh, Royal Observatory, Edinburgh, EH9 3HJ, UK
\and Astrochemistry Laboratory, Goddard Space Flight Center, NASA, 8800 Greenbelt Rd., Greenbelt, MD 20771, USA
\and Max-Planck-Institut f\"ur Sonnensystemforschung, Justus-von-Liebig-Weg~3, 37077 G\"{o}ttingen, Germany
\and Instituut voor Sterrenkunde, Katholieke Universiteit Leuven, Celestijnenlaan 200D, Bus-2410, 3000, Belgium
\and Space sciences, Technologies \& Astrophysics Research (STAR) Institute, University of Li\`ege, Li\`ege, All\'ee du 6 Ao\^ut 17, 4000, Belgium
\and 
European Space Agency European Space Astronomy Centre, Camino Bajo el Castillo, s/n Urbanizaci\'on Villafranca del Castillo 28692 Villanueva de la Ca\~nada, Madrid, Spain 
\and Jet Propulsion Laboratory, California Institute of Technology, 4800 Oak Grove Drive, Pasadena, CA, 91109, USA
\and Centrum Bada\'n Kosmicznych Polskiej Akademii Nauk (CBK PAN), Bartycka 18A, Warszawa 00-716, Poland
\and INAF - Istituto di Astrofisica e Planetologia Spaziali, Area Ricerca Tor Vergata, Via Fosso del Cavaliere 100, 00133 Rome, Italy}

   \date{Received}

 \abstract{29P/Schwassmann-Wachmann 1 is a distant Centaur/comet, showing persistent CO-driven activity and frequent outbursts. } 
 {We aim to better characterize its gas and dust activity from multiwavelength observations performed during outbursting and quiescent states.}
{We used the HIFI, PACS and SPIRE instruments of the {{\it Herschel}} space observatory on several dates in 2010, 2011, and 2013 to observe the H$_2$O 557 GHz and NH$_3$ 573 GHz lines and to image the dust coma in the far-infrared. Observations with the IRAM 30 m telescope
 were undertaken in 2007, 2010, 2011, and 2021 to monitor the CO production rate through the 230 GHz line, and to search for HCN at 89 GHz. The 70 and 160 $\mu$m PACS images were used to measure the thermal flux from the nucleus and the dust coma. Modeling was performed to constrain the size of the sublimating icy grains and to derive the dust production rate.} 
 {HCN is detected for the first time in comet 29P (at 5$\sigma$ in the line area). H$_2$O is detected as well, but not NH$_3$. H$_2$O and HCN line shapes differ strongly from the CO line shape, indicating that these two species are released from icy grains. CO production rates are in the range (2.9--5.6) $\times$ 10$^{28}$ s$^{-1}$ (1400--2600 kg s$^{-1}$). A correlation between the CO production rate and coma brightness is observed, as is a correlation between CO and H$_2$O production. The correlation obtained between the excess of CO production and excess of dust brightness with respect to the quiescent state is similar to that established for the continuous activity of comet Hale-Bopp.  The measured $Q$(H$_2$O)/$Q$(CO) and $Q$(HCN)/$Q$(CO) production rate ratios  are 10.0 $\pm$ 1.5 \% and 0.12 $\pm$ 0.03 \%, respectively, averaging the April-May 2010 measurements ($Q$(H$_2$O) = (4.1 $\pm$ 0.6) $\times$ 10$^{27}$ s$^{-1}$, $Q$(HCN) = (4.8 $\pm$ 1.1) $\times$ 10$^{25}$ s$^{-1}$).  We derive three independent and similar values of the effective radius of the nucleus, $\sim$ 31 $\pm$ 3 km, suggesting an approximately spherical shape.  The inferred dust mass-loss rates during quiescent phases are in the range 30--120 kg s$^{-1}$, indicating a dust-to-gas mass ratio $<$ 0.1 during quiescent activity. We conclude that strong local heterogeneities exist on the surface of 29P, with quenched dust activity from most of the surface, but not in outbursting regions.}
{The volatile composition of the atmosphere of 29P strongly differs from that of comets observed within 3 au from the Sun. The observed correlation between CO, H$_2$O and dust activity may provide important constraints for the outburst-triggering mechanism.}
{}


\authorrunning{Bockel\'ee-Morvan et al.}
\titlerunning{Comet 29P/Schwassmann-Wachmann 1}
 
   \maketitle
%

\section{Introduction}

Comet 29P/Schwassmann-Wachmann 1 is a periodic comet orbiting on a nearly circular orbit with
a small inclination ($i$ = 9.4$^{\circ}$) at 6 au from the Sun. It
is also classified as a Centaur, which is a transition object
between the trans-neptunian and Jupiter-family dynamical
populations. Comet 29P is the most notable occupant of the short-lived dynamical Gateway, a temporary low-eccentricity region exterior to Jupiter through which the majority of Jupiter-family comets pass \citep{2019ApJ...883L..25S}.
The properties of its nucleus are poorly constrained. Its size is estimated to be $\sim$ 30 km in radius \citep{Stansberry2004,2013ApJ...773...22B,2015Icar..260...60S,2021PSJ.....2..126S}.


Comet 29P is well known for its permanent activity and its
episodic outbursts, which can change its visual brightness from typically
$m_{\rm v}$ = 16 to 11 during major outbursts
\citep[e.g.][]{Trigo2008,Trigo2010,Miles2016}. The outbursts are observed with some periodicity (about every 57 d), which is thought to correspond to the rotation period of the nucleus, and which suggests that the triggering mechanism involves the insolation of specific regions \citep{Trigo2010,Miles2016}. Carbon monoxide is permanently detectable
 in the coma with a production
rate of typically 3--5 $\times$ 10$^{28}$ s$^{-1}$, and is thought
to be the main driver of the activity \citep{Senay1994,
Crovisier1995,Festou2001,Gunnarsson2002,Gunnarsson2008,Paganini2013}.
Dust outbursts seem not always to be associated with an increase in the CO production \citep{2020AJ....159..136W}.
In addition to CO, H$_2$O  \citep[in the infrared,][]{Ootsubo2012} and daughter species  CO$^+$, CN, and
possibly N$_2^+$ 
\citep[in the visible,][]{Cochran1991,Korsun2008,2016P&SS..121...10I} have been detected in comet 29P. At 6 au from the Sun,
water sublimation from the nucleus is expected to be very
inefficient. The amorphous-to-crystalline water transition phase
that may proceed inside the nucleus is thought to be responsible for
the outbursts  \citep{1987ApJ...313..893P,1990ApJ...363..274P,Enzian1997, Kossacki2013}.

We present in this paper observations of 29P obtained in 2010-2013 with
the {\it Herschel} space observatory \citep{Pilbratt2010} in the
framework of the guaranteed-time key programme ``Water and related
chemistry in the Solar System'' \citep{hart09}, which targeted several comets \citep[e.g.][]{2010A&A...518L.149B,2010A&A...521L..50D,Biver2012,2012A&A...544L..15B,2014A&A...562A...5B,2014A&A...564A.124D}. Searches for
H$_2$O (557 GHz) and NH$_3$ (573 GHz) lines were performed with the Heterodyne Instrument
for the Far-Infrared  \citep[HIFI,][]{2010HIFI}, which led to the first far-infrared detection of water. A previous attempt to detect the 557 GHz H$_2$O line in comet 29P using the Odin space telescope was unsuccessful \citep{2007P&SS...55.1058B}. Continuum images at 70 and 160 $\mu$m were obtained using the Photodetector
Array Camera and Spectrometer \citep[PACS,][]{Pacs2010}, and at 250, 350 and 500 $\mu$m with the Spectral and Photometric Imaging REceiver \citep[SPIRE,][]{2010A&A...518L...3G}. Unlike the PACS observations, those with SPIRE did not lead to a conspicuous detection. We also
gather in this paper observations of CO and HCN carried out
in 2007, 2010, 2011 and 2021 with the 30 m antenna of the Institut de
radioastronomie millim\'etrique (IRAM), as well as optical
photometry observations that place the {\it Herschel} and IRAM data
in context.

The observations are described in Sect.~\ref{sec:obs}. The gas
production rates are derived in Sect.~\ref{sec:gas}. Section~\ref{sec:CO-mv-correlation} studies the correlations between production rates and dust activity. In
Sect.~\ref{sec:gas-ori} we present observational evidence for the predominant release of H$_2$O and HCN molecules by icy grains in the atmosphere of 29P. The H$_2$O observations are analyzed with a model simulating the sublimation of icy grains released during an outburst.  Section~\ref{sec:PACS-analysis} presents an analysis of the nucleus and dust thermal emissions observed with PACS.  In Sect.~\ref{sec:SPIRE-analysis} the SPIRE data are discussed. A summary follows in Sect.~\ref{sec:summary}. The models that are used to describe the dynamics, thermal properties, and sublimation of icy grains are presented in the appendix. A preliminary summary of these observations was given by \citet{2010DPS....42.0304B}.

\section{Observations}
\label{sec:obs}

\begin{table*}[t]
\caption{Log of the {\it Herschel} observations of 
29P.}\label{tab:1}
\begin{tabular}{lcccccccc}
\hline\hline\noalign{\smallskip}
Date (UT) & $r_h$ & $\Delta$ & Instrument &   ObsId &  Measurement &  Int.$^a$  & $m_{\rm R}^b$ & $\Delta T_{\rm outburst}^c$  \\
dd.dd/mm/yyyy &  (au)     &  (au) &  &    &   & (min) & & (day) \\
\hline\noalign{\smallskip}
19.05/04/2010 & 6.206 & 5.814 & HIFI &1342195094 &  H$_2$O 1$_{10}$--1$_{01}$, NH$_3$ 1$_{0}$--0$_{0}$ & 59 & 13.2 & 3.0(D)\\
11.02/05/2010 & 6.210 & 6.165 & HIFI & 1342196411 &  H$_2$O 1$_{10}$--1$_{01}$, NH$_3$ 1$_{0}$--0$_{0}$ & 48 & 15.3 & 25(D), 5.5(E)\\
30.24/12/2010  & 6.244 & 5.875 & HIFI & 1342212132& H$_2$O 1$_{10}$--1$_{01}$, NH$_3$ 1$_{0}$--0$_{0}$ & 51 & 16.2 & $>$ 81 \\
10.49/06/2010 & 6.215 & 6.634 & PACS & 1342198444/45 & Photo 70 \& 160 $\mu$m & 24 & 16.1 & 36(E), 17(F)  \\
02.70/01/2011 & 6.244 & 5.822 & PACS & 1342212281/82 & Photo 70 \& 160 $\mu$m & 95 & 16.6 & $>$ 84 \\
17.75/02/2013 & 6.231 & 5.820 & PACS & 1342263832-35 & Photo 70 \& 160 $\mu$m &  169 & 16.4 & $>$ 42\\
10.57/06/2010 & 6.215 & 6.635 & SPIRE & 1342198449 & Photo 250, 350, \& 500 $\mu$m & 55 & 16.1 & 36(E), 17(F)  \\
 \hline\noalign{\smallskip}
\end{tabular}

$^a$ Integration time. $^b$ Nuclear $R$-magnitude of comet 29P in a 10\arcsec-diameter aperture. $^c$  Time after the outbursts listed in Table~\ref{tab:outburst} with the label given within the brackets.
\end{table*}

\begin{table*}
\caption[]{H$_2$O 1$_{10}$--1$_{01}$, NH$_3$ 1$_{0}$--0$_{0}$, and HCN $J$(1--0) line areas and Doppler shifts, together with the gas production rates. } \label{tab:2other}
\begin{tabular}{lcccccccc}
\hline\hline\noalign{\smallskip}
&&&&&\multicolumn{3}{c}{Production rate (s$^{-1}$)}\\
\cline{6-8}\\
UT date & $r_{h}$  & Molec. & Line area$^a$     &Velocity shift &  Nucleus$^b$  & Icy grains$^c$  & Icy grains$^c$  \\[0.cm]
(dd.dd/mm/yyyy)  & (au)       &      &  (mK km~s$^{-1}$)         & (km~s$^{-1}$)   &     & $L_p$=10$^4$ km &  $L_p$= 5$\times$10$^4$ km\\
\cline{1-8}\\[-0.2cm]
19.05/04/2010 & 6.206 &  H$_2$O  & 19.0 $\pm$ 2.9$^d$ & +0.04 $\pm$ 0.04 & (4.6$\pm$0.8)$\times$10$^{27}$ & (1.6$\pm$0.3)$\times$10$^{27}$ & (4.4$\pm$0.8)$\times$10$^{27}$ \\
11.02/05/2010 & 6.210 &  H$_2$O  & 13.9 $\pm$ 3.4$^d$  & $-$0.16 $\pm$ 0.08  & (3.5$\pm$0.9)$\times$10$^{27}$ & (1.3$\pm$0.3)$\times$10$^{27}$& (3.5$\pm$0.9)$\times$10$^{27}$ \\
30.24/12/2010 & 6.244 &   H$_2$O  & $<$ 9.8$^d$ & $-$ & $<$2.4$\times$10$^{27}$& $<$0.8$\times$10$^{27}$ & $<$2.3$\times$10$^{27}$\\
\cline{1-8}\\[-0.2cm]
 19.05/04/2010 & 6.206  & NH$_3$  & $<$ 13 & $-$ & $<$5.6$\times$10$^{27}$& & \\
11.02/05/2010 & 6.210 & NH$_3$  & $<$ 15 & $-$ & $<$7.1$\times$10$^{27}$& & \\
30.24/12/2010 & 6.244  & NH$_3$  & $<$ 14 & $-$ & $<$6.0$\times$10$^{27}$& & \\
\cline{1-8}\\[-0.2cm]
30.80/12/2007$^e$ & 5.981  & HCN  & 19 $\pm$ 8 & $-$ & (3.2$\pm$1.3)$\times$10$^{25}$ & (2.7$\pm$1.1)$\times$10$^{25}$& (4.8$\pm$2.0)$\times$10$^{25}$\\
12.06/02/2010 & 6.194 &  HCN  & 10 $\pm$ 10 & $-$ & (1.9$\pm$1.9)$\times$10$^{25}$& (1.6$\pm$1.6)$\times$10$^{25}$& (2.9$\pm$2.9)$\times$10$^{25}$\\
05.00/05/2010$^f$ & 6.210  & HCN  & 21 $\pm$ 5 & $-$ & (4.8$\pm$1.1)$\times$10$^{25}$& (4.1$\pm$0.9)$\times$10$^{25}$& (7.2$\pm$1.7)$\times$10$^{25}$ \\
11.18/01/2011 & 6.245 & HCN  & 37 $\pm$ 9 & $-$ & (8.2$\pm$2.0)$\times$10$^{25}$ & (7.0$\pm$1.7)$\times$10$^{25}$ & (1.2$\pm$0.3)$\times$10$^{26}$ \\
14.95/11/2021$^g$ & 5.931 & HCN  & $<$ 18  & $-$ & $<$ 3$\times$10$^{25}$ & $<$ 2.6 $\times$10$^{25}$& $<$ 4.5 $\times$10$^{25}$ \\
Average 2007--2011 & 6.2\phantom{00}  & HCN & 21 $\pm$ 4  & $-$0.04 $\pm$ 0.07 & (4.4$\pm$0.8)$\times$10$^{25}$& (3.7$\pm$0.7)$\times$10$^{25}$ & (6.6$\pm$1.2)$\times$10$^{25}$ \\
 \hline
\end{tabular}

{\bf Notes.} 3-$\sigma$ upper limits are given in case of non
detection. $^{(a)}$ Line area in main-beam brightness
temperature scale. For HCN $J$(1--0), sum of the three hyperfine
components.  $^{(b)}$ In the assumption of nucleus production, and assuming
a coma temperature of 6 K (see Sect.~\ref{sec:exci}). $^{(c)}$ In the assumption of production from icy grains at the cometocentric distance $L_{\rm p}$,  with release at a temperature of 100 K (see Sect.~\ref{sec:exci}).
 $^{(d)}$ Line area measured on
HRS spectra. $^{(e)}$ Mean date for the average of measurements
performed on 29.8 and 31.8 Dec 2010. 
$^{(f)}$ Mean date for the average of measurements
performed on  17.87 April, 30.79 April, 22.6 May, and 28.78 May 2010. $^{(g)}$ Mean date for the average of measurements performed in November 2021.
\end{table*}

\subsection{HIFI observations}
\label{obs:hifi}

\begin{figure}[h]
\includegraphics[width=10.cm, angle = 0]{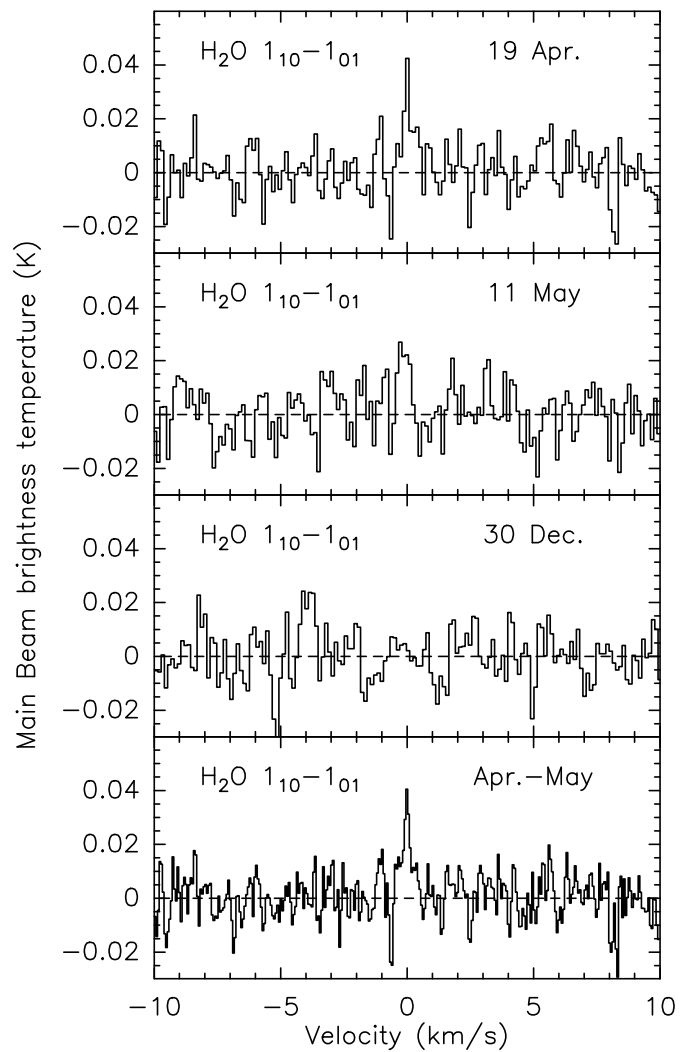}
 \caption{H$_2$O 1$_{10}$--1$_{01}$ line observed in comet 29P in 2010 with the HIFI instrument of {\it Herschel}.
 The UT date of the observation is indicated in the upper right corner.
 The velocity scale is in the comet rest frame. The spectra, acquired with the HRS, have been smoothed to a spectral resolution of 129 m s$^{-1}$, except for the bottom spectrum
 which shows the average of the spectra obtained on 19 April and 11 May at a spectral resolution of 67 m s$^{-1}$. } \label{fig:1}
\end{figure}

 Observations with the {\it Herschel}/HIFI instrument were
performed on 19 April, 11 May, and 30 December 2010, when the
comet was at $r_h$ = 6.2 au from the Sun. A log of the
observations, with the geometrical parameters (heliocentric distance $r_h$ and the comet-observer
distance $\Delta$), is presented in Table~\ref{tab:1}. The H$_2$O
1$_{10}$--1$_{01}$ and NH$_3$ $1_0-0_0$ lines, at 556.9360 and
572.5498 GHz, respectively, were observed simultaneously in the
lower and upper sidebands of band 1b of the HIFI receiver. They were
observed in the two orthogonal horizontal (H) and vertical (V)
polarizations. The observing mode was frequency-switching (FSW)
with a frequency throw of 94.5 MHz. Spectra were acquired with
both the Wide Band Spectrometer (WBS) and High Resolution
Spectrometer (HRS). The spectral resolution of the WBS is 1.1 MHz.
The HRS was used in the high-resolution mode (125 kHz spectral
resolution corresponding to $\sim$0.07 km s$^{-1}$). The
integration time was typically about 1~h for each
measurement (Table~\ref{tab:1}). The half-power beam width (HPBW)
is 38.1\arcsec~at 557 GHz \citep{Teyssier2017}. The comet was tracked using the ephemeris from
JPL Horizons.

The pointing for {\it Herschel} observations taken between
30 March 2010 and 14 June 2011 was
offset due to a warm star-tracker.  As a consequence, the HIFI
observations of comet 29P experienced small pointing offsets.  We used HIPE v12.0\footnote{The last version of HIPE was 15.0, but the different versions do not affect the data reduction.} to calculate the
 improved pointing corrections using the most accurate representation of
 the star tracker focal length.  We also  took the pointing offset between H and V
 polarisation beams  of 6.6\arcsec~in band 1b into account \citep[about 20\% of the full width
 at half-maximum of the beam,][]{Teyssier2017}. The largest
 offset of the comet nucleus corresponds to about 5\arcsec~from the center of
 the synthetic beam; it occurred for the April 2010 H+V average observation. The average
 pointing offsets for the May 2010 and December 2010 are 3.8 and
 3.3\arcsec, respectively.

Figure~\ref{fig:1} shows the H$_2$O spectra obtained with the HRS
spectrometer, averaging the two polarizations. Intensities are
given in units of main-beam brightness temperature, assuming
a main-beam efficiency of 0.62, and a forward efficiency of 0.96  \citep{Shipman2017,Teyssier2017}.
H$_2$O is detected in April and May 2010, with a
signal-to-noise ratio of  6.6 and 4.1, respectively. The  signal-to-noise ratio in the line area is 7.4, averaging the two periods. When these April and May 2010 data are averaged, the H$_2$O line is approximately
centered at the zero Doppler velocity in the comet rest frame
($\Delta v$ = --0.08 $\pm$ 0.05 km s$^{-1}$), and the line width of the
April-May averaged spectrum is 0.48 $\pm$ 0.07 km s$^{-1}$.
However, the spectrum obtained on 30 December, 2010 shows no indication of 
a line. The NH$_3$ $1_0-0_0$ line is not detected in any of the observed periods.

Measured line areas, or their upper limits, are given in
Table~\ref{tab:2other}. We also provide the mean
velocity shift of the line with respect to the comet frame in this table.

\begin{figure*}[h!]  
\centering
\begin{minipage}[t]{18cm}
\includegraphics[width=8.5cm]{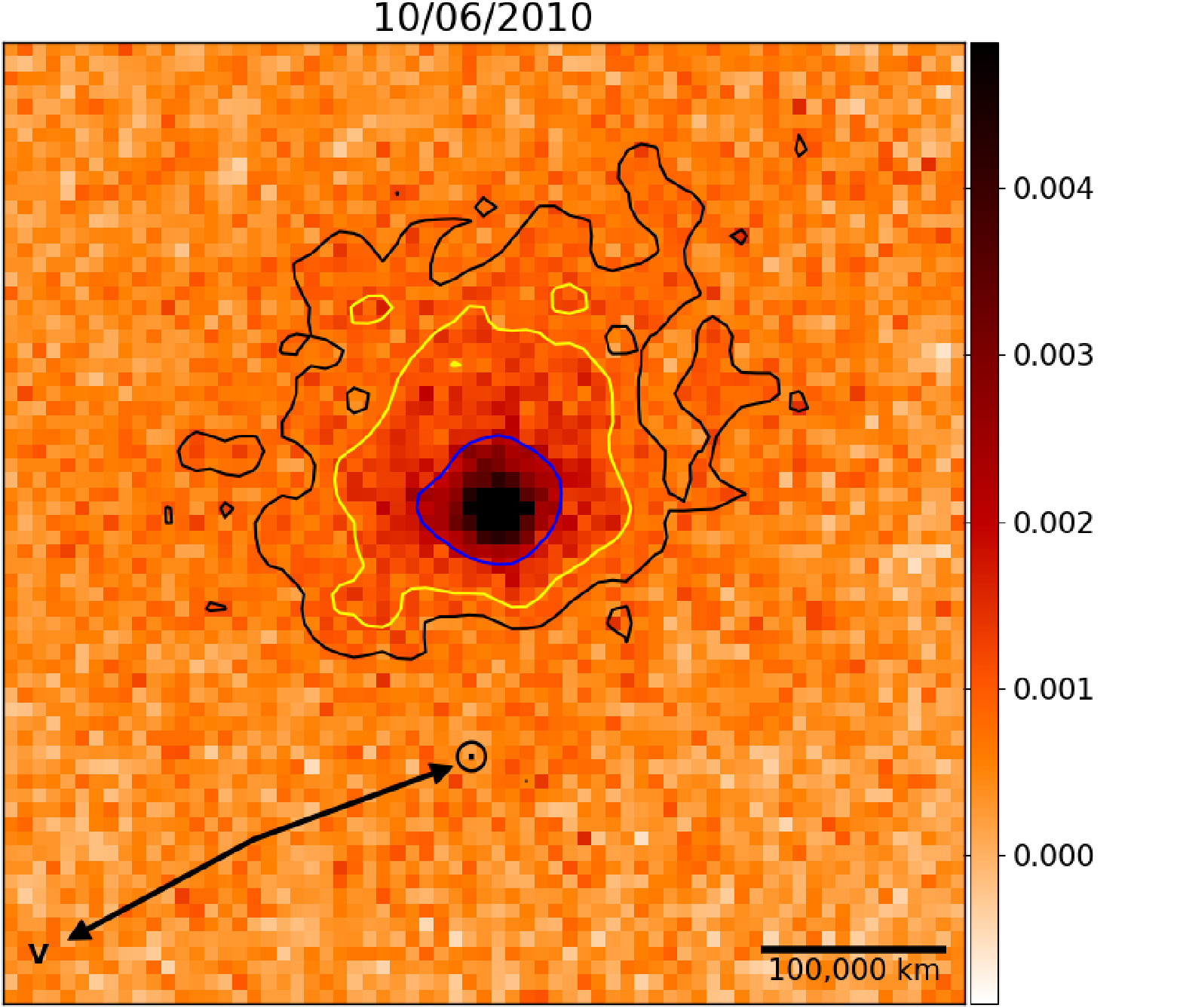}\hfill
\includegraphics[width=8.5cm]{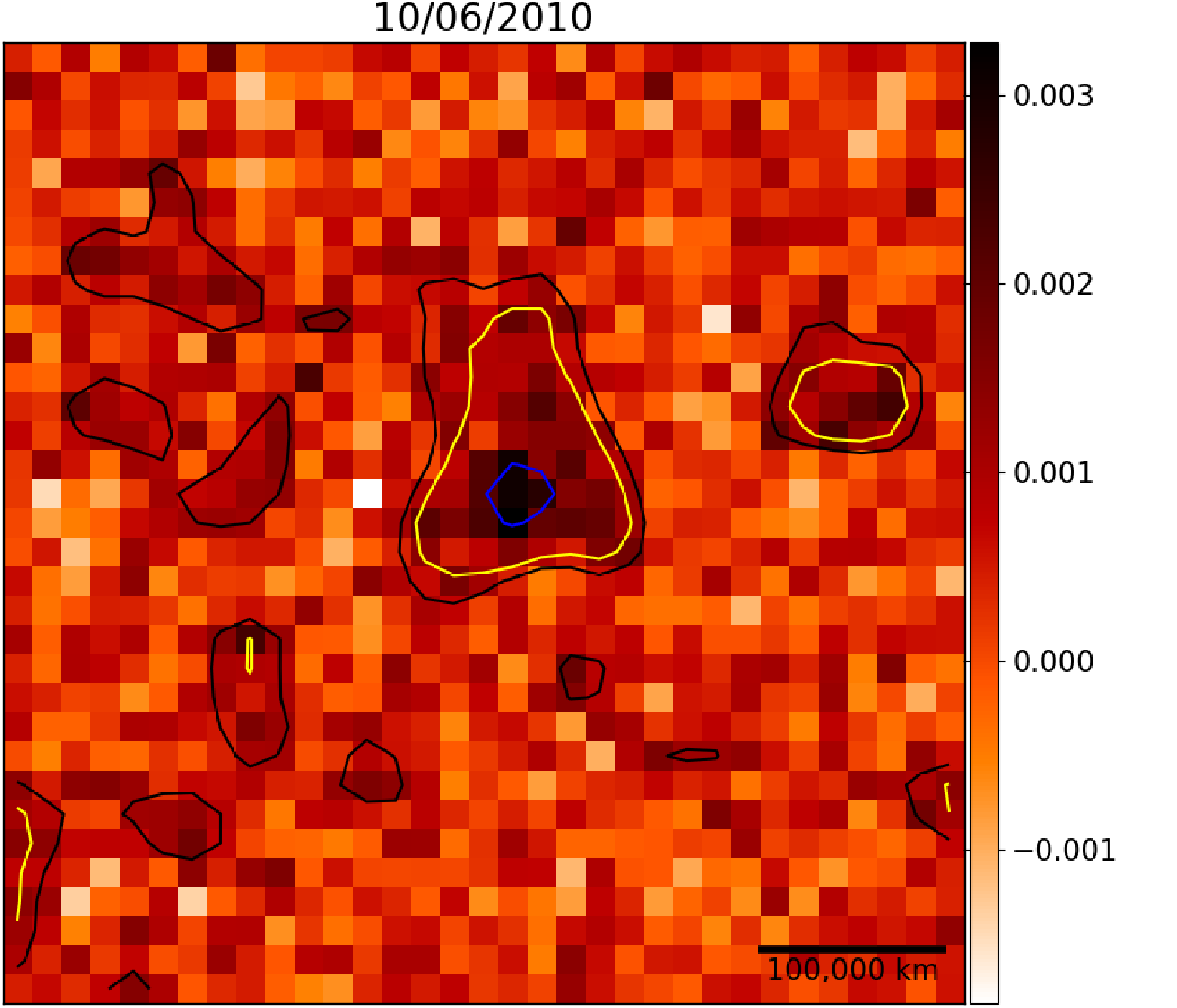}
\vspace{+0.2cm}
\end{minipage}
\begin{minipage}[t]{18cm}
\includegraphics[width=8.5cm]{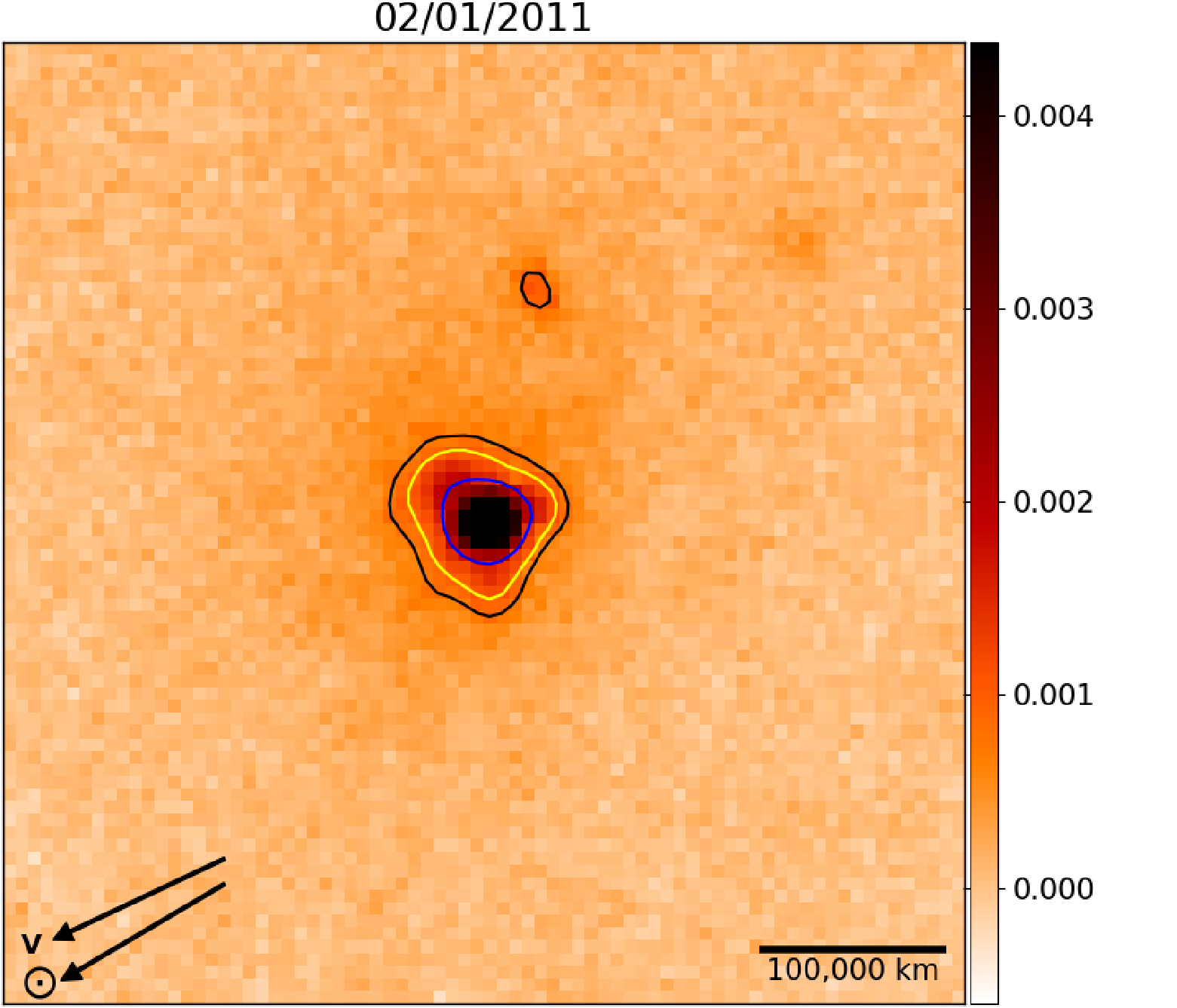}\hfill
\includegraphics[width=8.5cm]{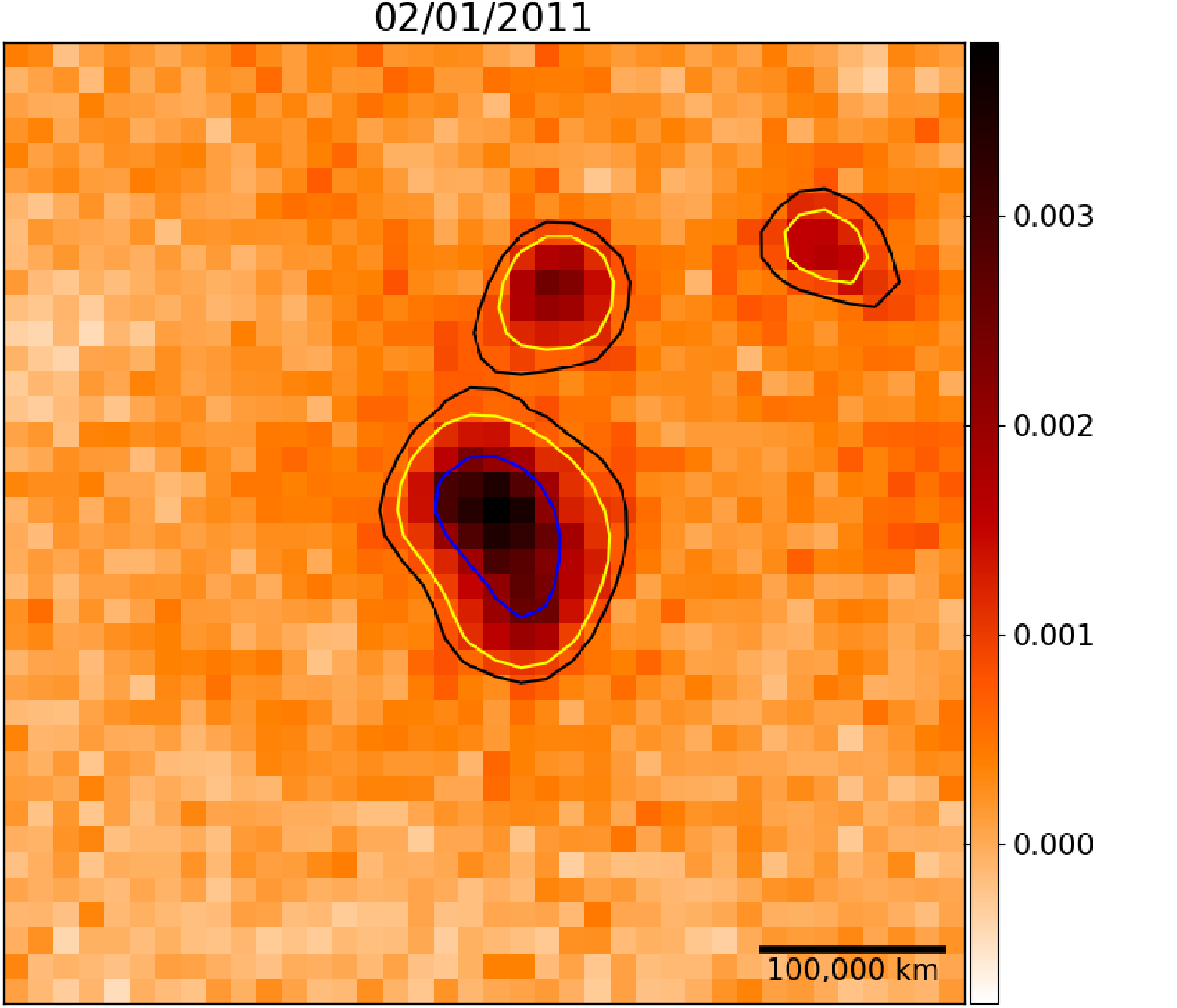}
\vspace{+0.2cm}
\end{minipage}
\begin{minipage}[t]{18cm}
\includegraphics[width=8.5cm]{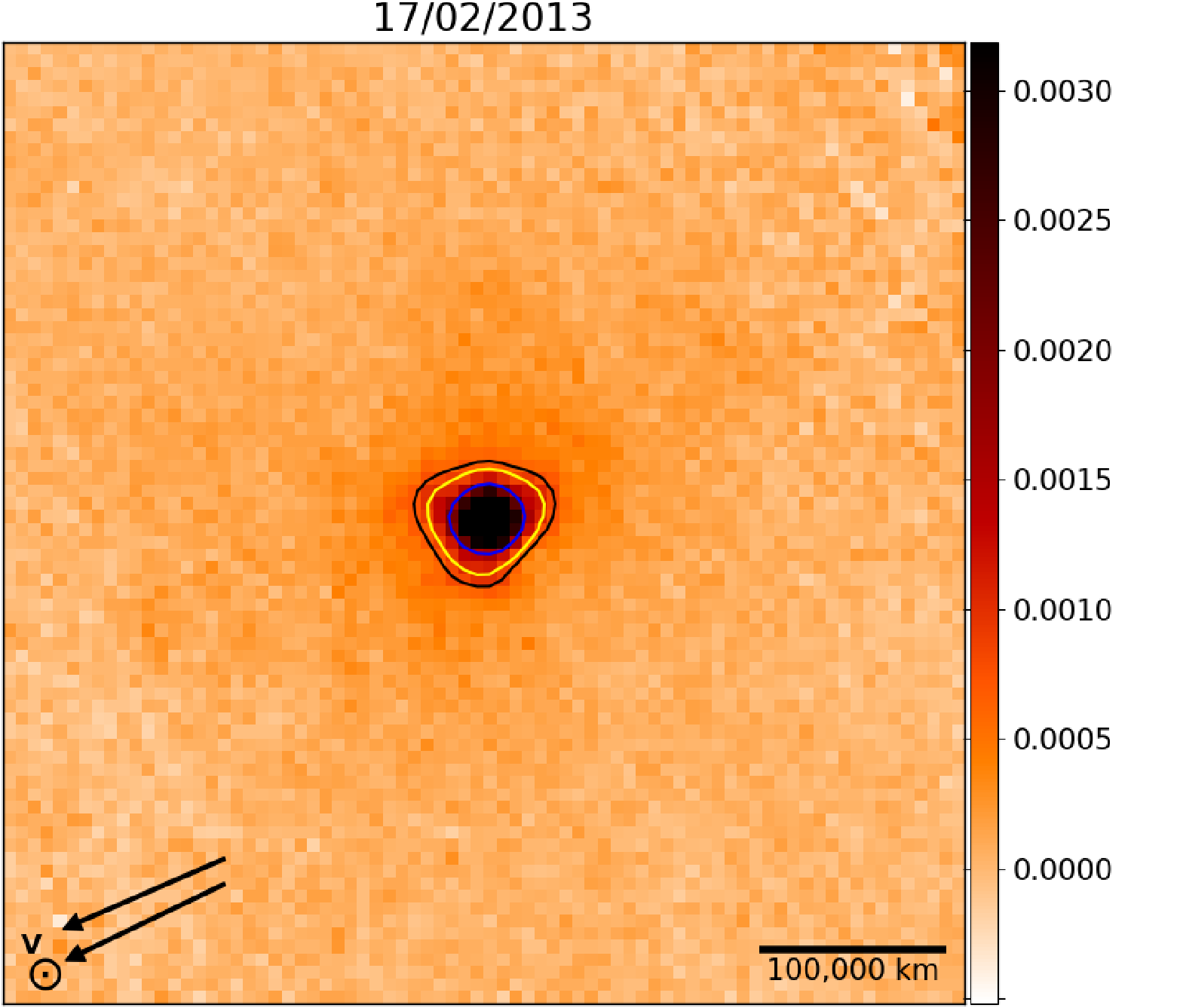}\hfill
\includegraphics[width=8.5cm]{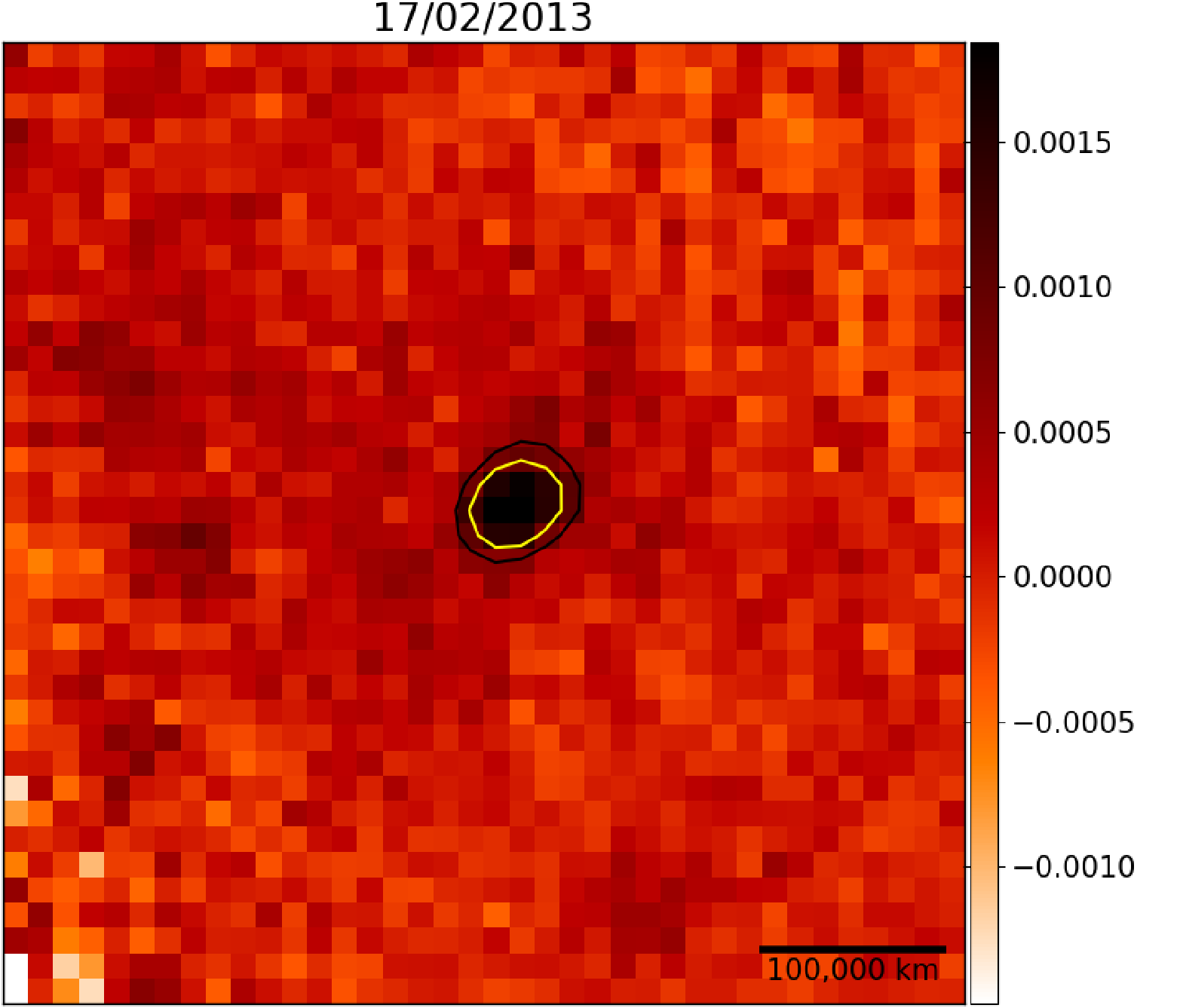}
\end{minipage}
\caption{Cropped PACS images of 29P in the 60-85
$\mu$m band (left) and in the 125--210 $\mu$m band (right). Dates from top to bottom are 10 June 2010, 2 January
2011, and 17 February 2013. Flux per pixel (1.6 and 3.2\arcsec~for the 70 and 160 $\mu$m images, respectively) is given in Jy (color bar).  The projected skyplane field of view is the same for each image (5.13 $\times$ 10$^5$ km $\times$ 5.13 $\times$ 10$^5$ km). Arrows indicate the skyplane-projected Sun direction and comet-projected trajectory. Negative pixel values are the result of the local background subtraction. {Comet 29P was in quiescent state at the three dates (Table~\ref{tab:1})}.}
\label{fig:PACS-IMAGE}
\end{figure*}

\begin{figure}[!h]
\begin{center}
\includegraphics[width=9.5cm,bb = 120 30 444 732]{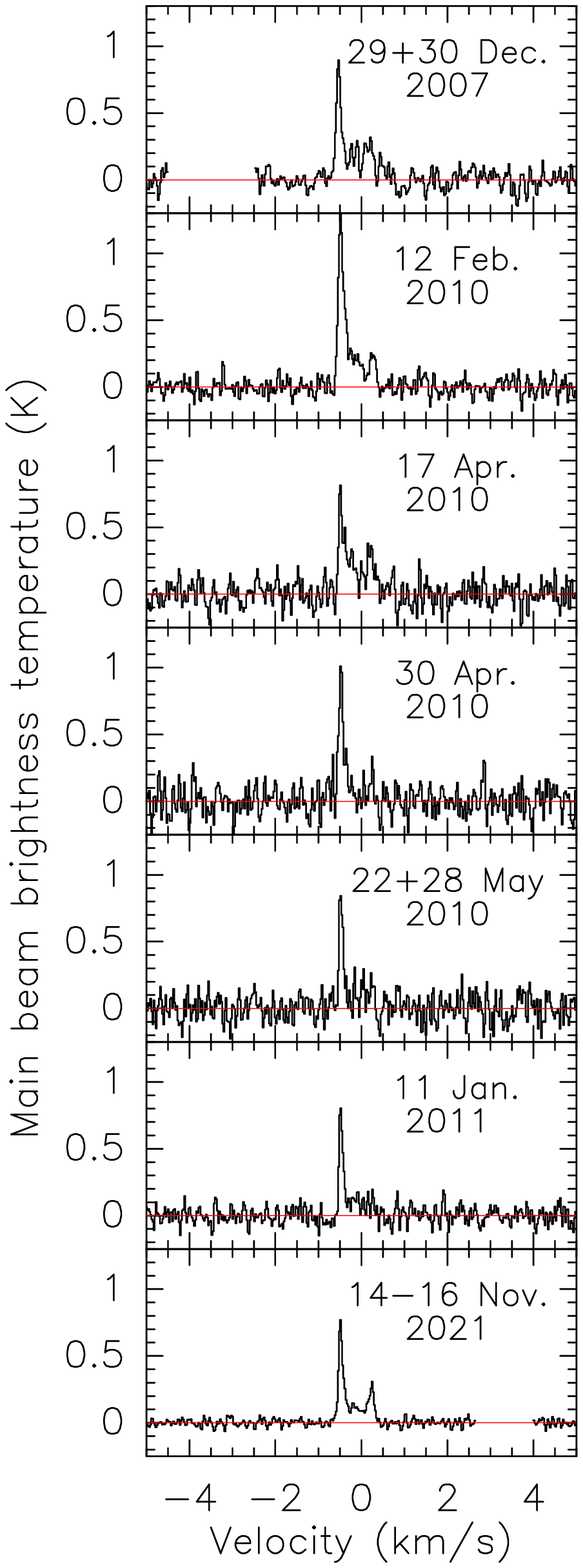}
\end{center}
 \caption{CO $J$(2--1) line observed in comet 29P with the IRAM 30 m telescope, from 2007 to 2021. Channels corresponding to the CO line from Earth's mesosphere (2007 spectrum obtained in FSW mode) and CO galactic lines (the 2021 spectrum includes data in PSW mode) are blanked.
 The velocity scale is in the comet rest frame. The spectral resolution is 51 m s$^{-1}$.} \label{fig:COspectra}
\end{figure}

\subsection{PACS observations}
The {\it Herschel}/PACS imaging observations were obtained on 10 June 2010,
that is one to two months after the HIFI measurements of April-May
2010,  on 2 January 2011, that is three days after the H$_2$O
observations of December 2010, and on 17 February 2013 (Table~\ref{tab:1}).
In photometer mode, the PACS instrument takes images
simultaneously in two of its three filters at 70 $\mu$m,
100 $\mu$m and 160 $\mu$m (red, green, and blue) that cover the
60-85 $\mu$m, 85-125 $\mu$m, and 125-210 $\mu$m ranges,
respectively.  The maps presented here were taken in the red and
blue bands with orthogonal scanning directions with respect to the
detector array using the medium-scan slewing speed of 20\arcsec/s.
For the May 2010 observations, we used three scan legs with a
9.9\arcmin~length and a 2.5\arcmin~leg separation, while the
January 2011 observation have eight scan legs with a
5\arcmin~length and 0.3\arcmin~leg separation. The mini-scan map mode was used in February 2013 (eight legs with 3\arcmin~length and 0.03\arcmin~leg separation). The pixel sizes are 6.4\arcsec~$\times$~6.4\arcsec~and
3.2\arcsec~$\times$~3.2\arcsec~for the red and blue channels,
respectively. On 17.75 February 2013, one of the two PACS red arrays was not operational \citep[][]{Exter2018}. This issue did not affect the data quality, but the size of the 160 $\mu$m image is smaller and the comet is offset from the center of the image.

We downloaded and used Level 2.5 Unimap maps produced by the PACS scan-map pipeline from the Herschel Science Archive\footnote{\url{http://archives.esac.esa.int/hsa/whsa/}} \citep{Exter2018}. For the Level 2.5 maps, the blue images were resampled to a pixel scale of 1.6\arcsec/pixel and the red images to 3.2\arcsec/pixel. The Level 2.5 maps were calibrated to Jy/pixel values and include a local background removal. Additionally, inspection of the Level 2.5 maps beyond the region of coma contributions revealed a low-level residual background from each image that was removed before their analysis. 

The PACS 70 $\mu$m and 160 $\mu$m images are shown in Fig.~\ref{fig:PACS-IMAGE}
for the three different epochs.  The 70 $\mu$m image obtained on 10 June 2010 is more extended than others. This is further discussed in Sect.~\ref{sec:dust-thermal}.

\subsection{SPIRE observations}\label{sec:spire}

The {\it Herschel}/SPIRE imaging observations were undertaken on 10 June 2010, approximately 2 hours after the PACS data acquisition (Table~\ref{tab:1}).
In photometry mode, the SPIRE instrument takes images with fields of view (FOV) of 4$'$ $\times$ 8$'$ simultaneously in three filters centered on 250 $\mu$m, 350 $\mu$m, and 500 $\mu$m.
29P was imaged using the small-map mode which involved scanning the telescope across the sky at 30\arcsec/s in two nearly orthogonal scan paths.
Level 2 scan maps were acquired from the {\it Herschel} Science Archive.
For 29P, the small-scan maps used for analysis were those generated for Solar System objects, consisting of calibrated maps in Jy/beam, corrected for the proper motion of 29P  \citep{spire_handbook}. The Level 2 scan maps have a circular FOV with a radius of $\sim$ 5$'$ that includes observational coverage from each of the individual detector scans.The HPBW of SPIRE photometer is 17.9$\arcsec$, 24.2$\arcsec$, and 35.4$\arcsec$~ at 250 $\mu$m, 350 $\mu$m, and 500$\mu$m, respectively. 

 The SPIRE images are shown in Fig. ~\ref{fig:SPIRE-IMAGE}. A marginal signal is observed at the position of comet 29P, especially in the 250 $\mu$m image. However, the images are crowded by signals from astronomic sources with similar or higher intensity. 



\subsection{IRAM 30 m observations}
\label{obs:iram}

\begin{table*}[t!]
\caption{Log of the IRAM-30m observations of
29P.}\label{tab:IRAM}
\begin{tabular}{lccccccclc}
\hline\hline\noalign{\smallskip}
Date (UT) & $r_h$ & $\Delta$ & $\tau^a$ &  Int. & Mode & Lines & $m_{\rm R}^b$ & $\Delta T_{\rm outburst}^c$\\
dd.dd/mm/yyyy &  (au)     &  (au) &  &  (min) & & & &  (day)\\
\hline\noalign{\smallskip}
%
%
 29.80--29.83/12/2007 & 5.981 & 5.009  & 0.1 &  42 &  FSW & CO $J$(2--1), HCN $J$(1--0) & 15$\pm$1$^d$ & 0.2(A)\\
 30.80--30.83/12/2007 & 5.981 & 5.012  & 0.1 &   36 & FSW & CO $J$(2--1) & 13.0 & 1.2(A) \\
 31.81--31.82/12/2007 & 5.981 & 5.014  & 0.07&  12 & FSW  & CO $J$(2--1), HCN $J$(1--0) & 13.1 & 2.2(A) \\
 12.04--12.07/02/2010 & 6.194 & 5.207  & 0.08  & 32 & WSW   &  CO $J$(2--1), HCN $J$(1--0) &13.0 & 9.6(C) \\
 17.84--17.90/04/2010 & 6.206 & 5.795  & 0.4   &  56 & WSW   &  CO $J$(2--1), HCN $J$(1--0) & 12.9 & 1.8(D) \\
 30.74--30.84/04/2010 & 6.208 & 5.999  & 0.57  &  70 & WSW   &  CO $J$(2--1), HCN $J$(1--0) & 15.2 & 15(D) \\
 22.61--22.67/05/2010 & 6.212 & 6.347  & 0.48  & 66  & WSW   &  CO $J$(2--1), HCN $J$(1--0) & 15.5 & 37(D), 17(E)\\
 28.75--28.80/05/2010 & 6.213 & 6.442 & 0.4--1.1 & 42 & WSW   &  CO $J$(2--1), HCN $J$(1--0) & 14.7 & 43(D), 23(E)  \\
 &&&&&&&&4.4(F)\\
 11.16--11.21/01/2011 & 6.245 & 5.697 &  0.22  & 50  & WSW   &  CO $J$(2--1), HCN $J$(1--0) & 16.6 & $>$ 93\\
 13.92--13.96/11/2021 & 5.930 & 5.017 &  0.24  & 45  & WSW+FSW &  CO $J$(2--1), HCN $J$(1--0) & 16.1 &  47(G), 21(H)  \\
 &&&&&&&&10(I)\\
 14.92--14.97/11/2021 & 5.931 & 5.011 &  0.10  & 46  & PSW+FSW &  CO $J$(2--1), HCN $J$(1--0) & 16.0 &  48(G),  22(H) \\
 &&&&&&&&11(I)\\ 
 15.98--15.99/11/2021 & 5.931 & 5.006 &  0.08  & 12  & FSW    &  CO $J$(2--1), CH$_3$OH $J$(5--4) & 16.1 &   49(G), 23(H)\\
  &&&&&&&&12(I)\\
 \hline\noalign{\smallskip}
\end{tabular}

$^a$ Atmospheric opacity at 225 GHz. $^b$ Nuclear red magnitude in a 10\arcsec-diameter aperture. $^c$ Time after outbursts listed in Table~\ref{tab:outburst}, with the label given within the brackets. $^d$ Interpolated from the reported nuclear magnitudes of 15.9 on 28.97 December 2007, and 14.0 on 29.91 December 2007. 
\end{table*}

\begin{table*}
\caption[]{CO $J$(2--1) line areas, Doppler shifts, and production rates. } \label{tab:2CO}
\begin{tabular}{lcccccc}
\hline\hline\noalign{\smallskip}
&&&&&&\multicolumn{1}{c}{Jet component}\\
UT date & $r_{h}$  & Molecule & Line area$^a$ &Velocity shift &  Prod. rate$^b$  & Prod. rate$^{b,~c}$ \\[0.cm]
(dd.dd/mm/yyyy) & (au)        &      &  (mK km~s$^{-1}$)         & (km~s$^{-1}$)   &  (s$^{-1}$) & (s$^{-1}$) \\
\cline{1-7}\\[-0.2cm]
29.82/12/2007 & 5.981 & CO & 271 $\pm$ 18 & $-$0.25 $\pm$ 0.03 & (4.8 $\pm$ 0.3) $\times$ 10$^{28}$& (2.6 $\pm$ 0.2) $\times$10$^{28}$\\
30.82/12/2007 & 5.981 & CO & 301 $\pm$ 17 & $-$0.19 $\pm$ 0.02 & (5.1 $\pm$ 0.3) $\times$ 10$^{28}$& (2.2 $\pm$ 0.2) $\times$10$^{28}$\\
31.81/12/2007 & 5.981 & CO & 292 $\pm$ 30 & $-$0.26 $\pm$ 0.05 & (4.9 $\pm$ 0.5) $\times$ 10$^{28}$& (2.7 $\pm$ 0.3) $\times$10$^{28}$\\
12.05/02/2010 & 6.194 & CO & 332 $\pm$  9 & $-$0.28 $\pm$ 0.01 & (5.6 $\pm$ 0.2) $\times$ 10$^{28}$& (3.4 $\pm$ 0.1) $\times$10$^{28}$\\
17.87/04/2010 & 6.206 & CO & 265 $\pm$ 14 & $-$0.18 $\pm$ 0.02 & (4.8 $\pm$ 0.3) $\times$ 10$^{28}$& (2.1 $\pm$ 0.1) $\times$ 10$^{28}$\\
30.79/04/2010 & 6.208 & CO & 188 $\pm$ 17 & $-$0.36 $\pm$ 0.05 & (3.6 $\pm$ 0.3) $\times$ 10$^{28}$& (2.7 $\pm$ 0.2) $\times$ 10$^{28}$\\
22.64/05/2010 & 6.212 & CO & 201 $\pm$ 18 & $-$0.24 $\pm$ 0.04 & (4.0 $\pm$ 0.4) $\times$ 10$^{28}$& (2.2 $\pm$ 0.2) $\times$ 10$^{28}$\\
28.78/05/2010 & 6.213 & CO & 189 $\pm$ 26 & $-$0.21 $\pm$ 0.05 & (3.9 $\pm$ 0.5) $\times$ 10$^{28}$& (1.9 $\pm$ 0.3) $\times$ 10$^{28}$\\
11.18/01/2011 & 6.245 & CO & 159 $\pm$ 10 & $-$0.27 $\pm$ 0.03 & (2.9 $\pm$ 0.2) $\times$ 10$^{28}$& (1.7 $\pm$ 0.1) $\times$ 10$^{28}$\\
13.94/11/2021 & 5.930 & CO & 185 $\pm$ 10 & $-$0.30 $\pm$ 0.03 & (3.0 $\pm$ 0.2) $\times$ 10$^{28}$& (1.9 $\pm$ 0.1) $\times$ 10$^{28}$\\
14.95/11/2021 & 5.931 & CO & 201 $\pm$  5 & $-$0.21 $\pm$ 0.01 & (3.3 $\pm$ 0.1) $\times$ 10$^{28}$& (1.7 $\pm$ 0.1) $\times$ 10$^{28}$\\
15.99/11/2021 & 5.931 & CO & 190 $\pm$  9 & $-$0.20 $\pm$ 0.02 & (3.1 $\pm$ 0.2) $\times$ 10$^{28}$& (1.5 $\pm$ 0.1) $\times$ 10$^{28}$\\
\hline
\end{tabular}

{\bf Notes.} $^{(a)}$ Line area on the main-beam brightness
temperature scale.  $^{(b)}$ Assuming nucleus production, and 
a coma temperature of 6 K  (see Sect.~\ref{sec:exci}). $^{(c)}$ Derived from the line area measured between $-$0.7
and $-$0.3 km s$^{-1}$.
\end{table*}

In support of the {\it Herschel} observations, comet 29P was
observed from the ground at millimeter wavelengths with the IRAM
30 m telescope. We also include in this paper observations
undertaken in 2007 and 2021. The log of the observations is presented in Table~\ref{tab:IRAM}.

Observations in 2007 were performed in frequency-switching
mode (FSW; throw of 7.2 MHz) with the A100/B100 and A230/B230 receivers
used in parallel. This combination of receivers allowed us to
simultaneously observe  the HCN $J$(1--0) and CO $J$(2--1) lines at
88.632 GHz and 230.538 GHz, respectively, in horizontal and
vertical polarizations. Spectra were acquired with the VESPA
autocorrelator at a spectral resolution of 20 kHz (66 and 25 m
s$^{-1}$, at 89 and 230 GHz, respectively). This high spectral resolution is needed to resolve the narrow blueshifted peak of the CO line (Fig.~\ref{fig:COspectra}).

For the observations undertaken in 2010, 2011, and 2021, we used the EMIR
front-end, installed at the telescope in 2009. EMIR 230 GHz and 90
GHz receivers were used simultaneously, to observe the CO
$J$(2--1) and HCN $J$(1--0) lines. Observations in 2010--2011 were undertaken in
beam-switching mode (WSW), using the wobbling secondary mirror, with the
sky reference position at 3\arcmin~from the comet.  Those of 2021 were obtained either in WSW,
  in FSW, or in position-switching mode (PSW) with a reference at 5\arcmin. The 2007
data contain spectra observed with VESPA at a spectral
resolution of 20 kHz.

The daily integration time was between 12 and 70 min (Table~\ref{tab:IRAM}). The IRAM
HPBW is 10.7\arcsec~and 27.8\arcsec~at 230 GHz and 89 GHz,
respectively. The main-beam efficiency was estimated by observing
planets to $\sim$ 0.73 at 89 GHz and in the range 0.48--0.57 at
230 GHz (depending on the date). The forward efficiency is 0.95
and 0.91 at 89 and 230 GHz, respectively.

The CO $J$(2--1) line is readily detected on individual days 
(Fig.~\ref{fig:COspectra}).  This line was first detected in 29P at the James Clerk Maxwell Telescope (JCMT)
\citep{Senay1994}. It was then observed numerous times at IRAM, 
at the Swedish ESO Submillimetre Telescope (SEST), or with the Arizona Radio Observatory 10 m Submillimeter Telescope (SMT)
\citep{Crovisier1995,Festou2001,Gunnarsson2002,Gunnarsson2008,2020AJ....159..136W}. The CO spectra
present the characteristic CO line shape
observed in this comet, namely, a blueshifted line (velocity shift $\Delta
v$ = $-$0.2 to $-$0.3 km s$^{-1}$, Table~\ref{tab:2CO}), with a strong and narrow (full width at half maximum of 0.123 $\pm$ 0.005 km s$^{-1}$) peak
at $v$ = $-$0.5 km s$^{-1}$. The high S/N November 2021 spectrum also distinctly shows  a peak at +0.25 km s$^{-1}$. 

\begin{figure}[h!]
\begin{center}
\includegraphics[width=5.5cm, angle = 270]{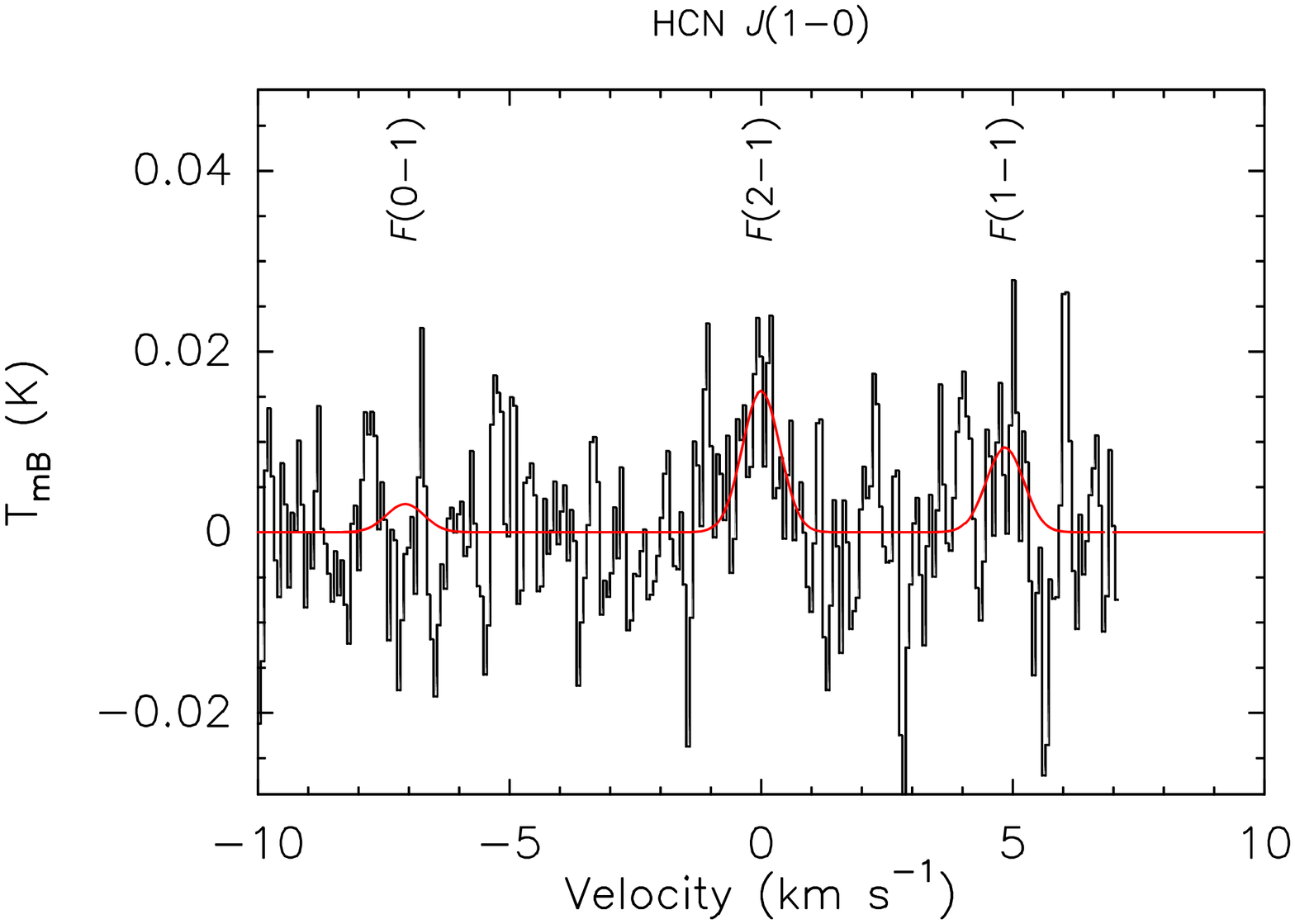}
\end{center}
 \caption{HCN $J$(1--0) line observed in comet 29P with the IRAM 30 m telescope, averaging 2007 to 2011 data. A Gaussian fit
 to the $F$(2--1) main hyperfine component is shown by the red
 line. The Gaussian curves centered at the velocity of the
 $F$(0--1) and $F$(1--1) were rescaled assuming statistical
 weight ratios. The vertical scale is the main-beam brightness
 temperature.
 The velocity scale is in the comet rest frame. The spectral resolution is 66 m s$^{-1}$.} \label{fig:HCN}
\end{figure}

The HCN $J$(1--0) line is detected marginally in December 2007,
April--May 2010, and January 2011, but not in November 2021. The upper limit for 2021 is consistent with most other measurements (Table~\ref{tab:2other}).  When the 2007--2011 data are averaged, the signal
to noise ratio is 5.2 in the line area (Table~\ref{tab:2other}, Fig.~\ref{fig:HCN}). This is the first detection of HCN in comet 29P. From a
Gaussian fit to the main $F$(2--1) hyperfine component, the width
of the line is 0.88 $\pm$ 0.41 km s$^{-1}$. As for water, the HCN
line does not present a significant velocity offset ($\Delta v$ =
--0.04 $\pm$ 0.07 km s$^{-1}$, Table~\ref{tab:2other}), in contrast to
the CO line.

\subsection{Context from optical observations}
\label{sec:optical}

Comet 29P is the target of several photometric monitoring campaigns with the aim 
to understand the origin of its outbursts. \citet{Trigo2008}
established an outburst frequency of 7.3 outbursts/year. We list
in Table \ref{tab:outburst} relevant outbursts (labeled by letters) that 
occurred before one of our observations, and their amplitude $\Delta m_{\rm R}$. The elapsed times $\Delta T_{\rm outburst}$ between the outburst time and the HIFI and IRAM observations are given in Tables~\ref{tab:1}and \ref{tab:IRAM}, respectively. We also provide for each observing date the $R$ magnitude (referred to as the nuclear magnitude) $m_{\rm R}$ measured within a 10\arcsec~diameter aperture (or the visual magnitude in a 13\arcsec~diameter aperture which is comparable to $m_{\rm R}$), taken from the LESIA data base\footnote{\url{https://lesia.obspm.fr/comets}}, Minor Planet Center\footnote{\url{https://minorplanetcenter.net/db_search}}, M. Kidger homepage\footnote{\url{http://www.observadores-cometas.com/}}, R. Miles page on British Astronomical Association website\footnote{ \url{https://britastro.org/node/25120}},  and  \citet{Miles2016}. $m_{\rm R}$ values at the date of {\it Herschel} and IRAM observations are given in Tables~\ref{tab:1} and~\ref{tab:IRAM}, respectively.

The PACS continuum observations were obtained during quiescent activity ($m_{\rm R}$ $\sim$ 16.4, Table~\ref{tab:1}). The first two H$_2$O observations took place 3.0 and 25.0 days
after the major outburst of 16.8 April 2010 ($\Delta m_{\rm R}$ = 3.9, outburst D). Two other outbursts (E \& F) of small amplitude occurred in May 2010, with outburst E ($\Delta m_{\rm R}$ = 1.0) only 5.5 days before the second observation.  As for the third H$_2$O observation on 30 December 2010, the comet was in a quiescent phase since mid-October 2010. In Table \ref{tab:outburst}, we list the outburst (B) of 9.71 November 2009 because H$_2$O was detected with the {\it Akari}
telescope nine days after this relatively faint outburst
\citep{Ootsubo2012}.

The CO and HCN observations in December 2007 and February 2010 were obtained close in time to major outbursts A and C, respectively. This is the case especially for the 29.80--29.83 December 2007 data. R. Miles (personal communication) estimates
the time of outburst A to 29.42$\pm$0.37 December 2007 (updating the value given in \citet{Miles2016}). Using three 29P images from R. Ligustri\footnote{Available on S. Yoshida home page \url{http://www.aerith.net/}} obtained on  31.778 December 2007, 1.833 January 2008, and 8.842 January 2008, we have estimated the outburst time from the expanding shell to 29.61$^{+0.3}_{-0.5}$ December 2007 (with an expansion rate of 0.154 km s$^{-1}$). The resulting elapsed time $\Delta T_{\rm outburst}$ between outburst A and the first CO December 2007 observation is in the range [--0.1 d, 0.7 d] with a central value at +0.2 d.

The comet was quiescent at the time of the January 2011 CO and HCN observations. The November 2021 observations were conducted about one month and a half after its major outburst of 27.8 September 2021 ($\Delta m_{\rm R}$ = 4.5, outburst G). Outbursts are also reported for 16.88 October ($\Delta m_{\rm R}$ = 0.35), 23.75 October ($\Delta m_{\rm R}$ = 2.5, outburst H) and 3.4 November 2021 ($\Delta m_{\rm R}$ = 0.6, outburst I). However, 29P was back to a quiescent state when observed at IRAM on 14 to 16 November 2021 ($m_{\rm R}$ $\sim$ 16, Table~\ref{tab:IRAM}).

To study how the gas production rates correlate with dust activity (see Sect.~\ref{sec:CO-mv-correlation}), we corrected the apparent magnitude  $m_{\rm R}$ (= $m_{\rm R}(\Delta,r_{\rm h},\theta)$) for the geocentric distance and phase angle $\theta$ according to
\begin{equation}
m_{\rm R}(1,r_{\rm h},0)= m_{\rm R}(\Delta,r_{\rm h},\theta) - 5{\rm log}_{\rm 10}(\Delta) + 2.5{\rm log}_{\rm 10}(\phi[\theta]),
\label{eq:mr-correction}
\end{equation}
\noindent
where $\phi(\theta)$ is the phase function normalized to $\phi$ = 0$^\circ$ from \citet{2011AJ....141..177S}.
Admittedly, this is not the most appropriate geocentric correction as the magnitude is measured in a fixed angular aperture. In addition, a heliocentric correction should be considered to take into account the $r_{\rm h}^{-2}$ dependence of the solar light scattering on the dust particles. Since the spanned ranges of $r_{\rm h}$ and $\Delta$ are small along the orbit of 29P, we nonetheless used the commonly used correction given in Eq.~\ref{eq:mr-correction}.

\begin{table}[h]
\caption[]{Relevant 29P outbursts.} \label{tab:outburst}
\begin{tabular}{rcccc}
\hline\hline\noalign{\smallskip}
Outburst date & Peak $m_{\rm R}$ & $\Delta m_{\rm R}$ & Ref. & Label\\
 \hline\noalign{\smallskip}
 29.61$^{+0.3}_{-0.5}$\phantom{00}Dec 2007 & 12.7 & 3.4 & (1, 2) & A \\
 9.71$\pm$0.4\phantom{00}Nov 2009 & 13.5 & 2.5 & (1,3) & B \\
 2.48$\pm$0.15\phantom{00}Feb 2010 &  11.6 & 4.6 & (1,3) & C\\
 16.05$\pm$0.11\phantom{00}Apr 2010 &  12.8 & 3.9 & (1,3)& D\\
 5.5\phantom{00}May 2010 & 15.2 & 1.0 & (4) & E \\
 24.40$\pm$0.4\phantom{00}May 2010  & 14.7 & 1.2 & (1) & F\\
 27.8\phantom{00}Sep 2021 & 11.5 & 4.5 & (5) & G \\
 23.75\phantom{00}Oct 2021 & 13.1 & 2.5 & (5) & H \\
 3.41\phantom{00}Nov 2021 & 15.2 & 0.6 & (5) & I \\
 \hline\noalign{\smallskip}
\end{tabular}

{\bf Notes.} References: (1) \citet{Miles2016}; (2) this work; (3)
\citet{Trigo2010}; (4) from Spanish amateur data (Kidger homepage); (5) R. Miles/J.-F Soulier.
\end{table}

\section{Gas production rates}
\label{sec:gas}

\subsection{Modeling}
\label{sec:exci}
To compute gas production rates, we modeled the excitation
processes and radiative transfer in the coma following previous
works \citep{Biver1997,Biver1999,zakharov2007}.  Processes include
collisions, excitation of the vibrational bands by the solar
radiation, radiation trapping, and spontaneous decay. The excitation model computes the evolution of the populations of the rotational levels as the molecules expand radially in the coma.

Only collisions with CO molecules were considered because CO is the
dominant molecule in the coma of 29P. Indeed, CO$_2$, found to be
relatively abundant in many comets, has an abundance relative to
CO lower than 1\% in 29P \citep{Ootsubo2012}. As derived from
this work  (Tables~\ref{tab:2other}, \ref{tab:abun}), water is also a minor constituent
of the atmosphere of this distant comet. We assumed collisional
cross-sections $\sigma_c$(CO--CO) = 2 $\times$ 10$^{-14}$ cm$^2$, $\sigma_c$(H$_2$O--CO) =
5 $\times$ 10$^{-14}$ cm$^2$, $\sigma_c$(NH$_3$--CO) = 2 $\times$ 10$^{-14}$ cm$^2$, and $\sigma_c$(HCN--CO)= 10$^{-14}$ cm$^2$ \citep{Biver1999}.
Collision rates were computed taking  the relative
masses of the colliding molecules into account. An important parameter for
modeling collisional excitation is the gas temperature, 
which we assumed to be 6 K. This value is a compromise between the
upper limit of 8 K derived from the line width of the blueshifted
component of the $J$(2--1) line (see Fig.~\ref{fig:COspectra}),
the value of 4 K estimated from CO $J$(2--1) maps
\citep{Gunnarsson2008}, and the CO rotational temperature of 4.9
$\pm$ 1.2 K, determined from infrared spectroscopy
\citep{Paganini2013}.  This low gas temperature is consistent with values expected at a few hundred kilometers from the nucleus of 29P on the basis of gas-dynamics calculations \citep{Crifo1999}. For molecules released by the nucleus, the level populations evolve from local thermal equilibrium (LTE) in the collisional region to fluorescence equilibrium in the outer coma. The size of the LTE region is a function of the molecule. Molecules close to the nucleus, where the gas is warmer, do not contribute significantly to the measured signals because the large FOVs exceed 10$^4$ km in radius.

As discussed in Sect.~\ref{sec:gas-ori}, the characteristics of the H$_2$O and HCN lines suggest that these molecules are predominantly produced from icy grains at cometocentric distances $ L_{\rm p} >$ 10$^4$ km where collisions with CO molecules are rare. Therefore, we also investigated the evolution of the level populations of H$_2$O and HCN molecules released at $L_p$ = 10$^4$ and 5$\times$10$^4$ km. We assumed that their initial rotational temperature is equal to 100 K, which corresponds to the expected equilibrium temperature of grains with radii $>$ 20 $\mu$m (Sect.~\ref{sec:gas-ori}).  Calculations were also made with an initial rotational temperature of 170 K to investigate the release from 2-$\mu$m organic grains. For this icy-grain production model, the molecules expand radially (a simplification that admittedly is not physically realistic) from $L_p$ to outward. This truncated density distribution was used to infer production rates in the icy-grain model cases (Table~\ref{tab:2other}).  

\begin{figure}[b]
\begin{center}
\includegraphics[width=9.0cm]{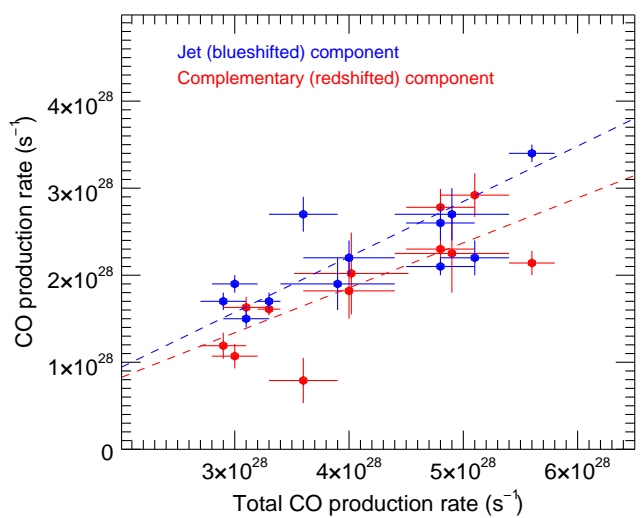}
\end{center}
 \caption{CO production rates in the jet component (blue symbols) and complementary component (red symbols) as a function of the total CO production rate. They are inferred from the line areas measured between --0.7 and --0.3 km s$^{-1}$, and between --0.3 and +0.4 km s$^{-1}$, respectively. Values for the jet component and total production rates are given in Table~\ref{tab:2CO}. The dashed blue and red lines show linear fits to the data points that correspond to the jet and complementary components, respectively.}
 \label{fig:CO-blue-red}
\end{figure}

\subsection{CO production rate}

Table~\ref{tab:2CO} displays production rates derived for CO. The calculations take the peculiar shape of the CO line into account that has  already been discussed in
several papers \citep[e.g.][]{Gunnarsson2002,Gunnarsson2008}. This shape is interpreted and modeled here as due to the combination of a CO jet with a
45$^{\circ}$ half-opening angle, expanding toward the Sun at a
velocity of 0.5 km s$^{-1}$, and a complementary outgassing
out of the jet cone expanding at 0.3 km s$^{-1}$. The total production rates
given in Table~\ref{tab:2CO} assume that the production rate in
the jet component is 60\% of the total production.  We also provide in
Table~\ref{tab:2CO} the production rate in the jet
component, derived from the line areas measured between $-$0.7
and $-$0.3 km s$^{-1}$ and using the same jet parameters as given above. The CO production rate 
in the jet component is between 43 and 75\% of the total CO production rate, with a mean value of 54\%, which is consistent with the previous assumption about the relative contribution in the two components. 
We do not observe any significant trend between the relative contributions of the two components and the total CO production rate (Fig.~\ref{fig:CO-blue-red}).  The CO production rate on the various days is between 3 and 6 $\times$ 10$^{28}$ s$^{-1}$, which is consistent with previous measurements
\citep{Senay1994,Crovisier1995,Festou2001,Gunnarsson2002,Gunnarsson2008,Ootsubo2012,Paganini2013,2020AJ....159..136W}.

\begin{figure}[h!]
\begin{center}
\includegraphics[width=5.5cm, angle = 270]{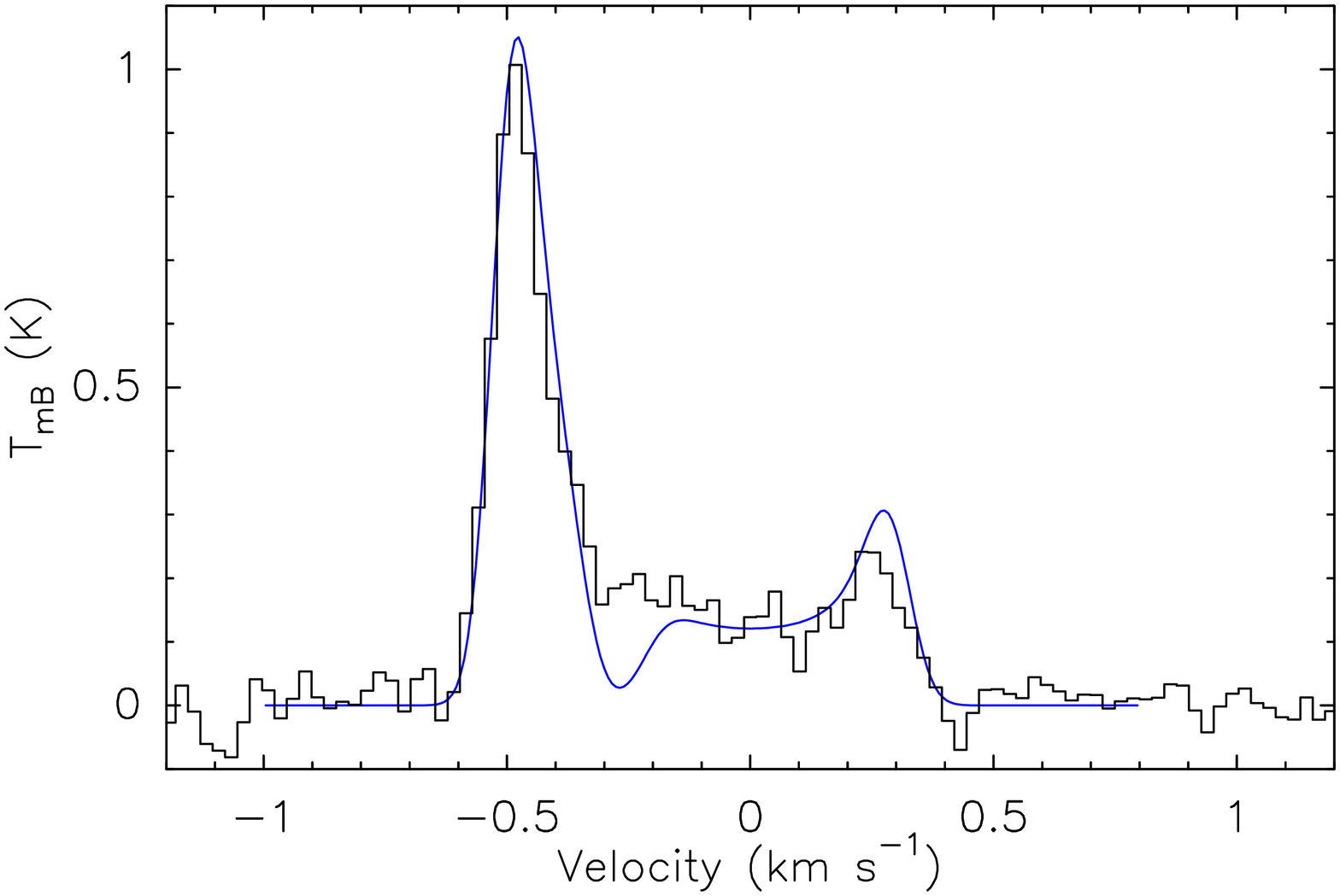}
\includegraphics[width=5.5cm, angle = 270]{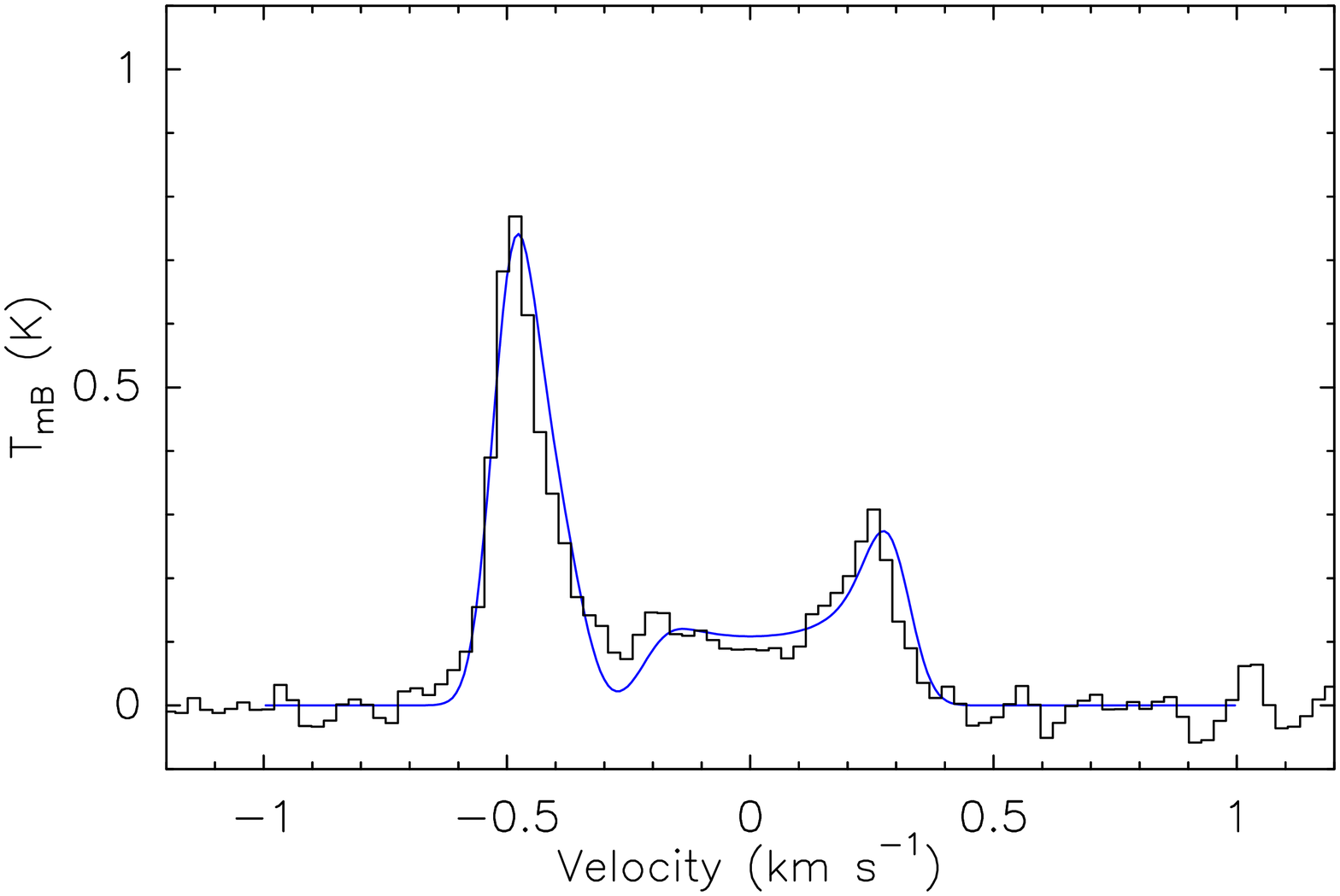}
\end{center}
 \caption{Synthetic CO $J$(2-1) spectra (blue) superimposed on observed IRAM spectra (black). Top: Average data from 29 December 2007 to 11 January 2011 (mean $r_h$ = 6.1 au and mean $\Delta$ = 5.5 au). The production rate in the sunward jet of 45$^{\circ}$ semi-aperture is $2.7\times10^{28}$ s$^{-1}$, and the total CO production rate is $4.5\times10^{28}$ s$^{-1}$. Bottom: Average November 2021 spectrum.  The production rate in the sunward jet is $1.7\times10^{28}$ s$^{-1}$, and the total production rate is $3.2\times10^{28}$ s$^{-1}$.
The outflow velocities within and outside the jet are assumed to be $v_{\rm exp}$ = 0.5 and 0.3 km s$^{-1}$, respectively. } \label{fig:COmod}
\end{figure}

Figure~\ref{fig:COmod} shows synthetic CO spectra that reproduce to first approximation the average IRAM 2007--2011  and November 2021  spectra.   A more realistic model providing a better fit would be the model used by \citet{Festou2001} \citep[see
also][]{Gunnarsson2008}, where the outgassing rate and
expansion velocity both vary continuously with solar zenith angle.

\subsection{H$_2$O production rate}
\label{sec:H2O-prod}
In contrast to the CO line, the HCN and H$_2$O lines have
approximately symmetric shapes (Sects~\ref{obs:hifi} \&~\ref{obs:iram}). Therefore,
we assumed isotropic outgassing and adopted a velocity of 0.3 km
s$^{-1}$, consistent with the half-width of these lines
(0.23 $\pm$ 0.04, and 0.4 $\pm$ 0.2 km s$^{-1}$ for H$_2$O and
HCN, respectively, Sect.~\ref{obs:hifi} and \ref{obs:iram}). The
same assumptions were made to derive the upper limits on the
NH$_3$ production rate.
 
A low level of water production is measured, with a  mean value of
$Q$(H$_2$O) = (4.1 $\pm$ 0.6) $\times$ 10$^{27}$ s$^{-1}$ for April--May
2010, for the nucleus model which assumes water release from the nucleus (Table~\ref{tab:2other}). Using CO production rates measured during this period, we
derive a $Q$(H$_2$O)/$Q$(CO) ratio of 10.0 $\pm$ 1.5 \%. A
3$\sigma$ upper limit $Q$(H$_2$O)/$Q$(CO) $<$ 8\% is measured for
the period 30 December 2010 to 11 January 2011. 

Both H$_2$O and CO
were detected on 19 November 2009 ($r_h$ = 6.18 au) with the {\it
Akari} telescope, through their vibrational bands at 2.7 and 4.3
$\mu$m, respectively \citep{Ootsubo2012}. The water production
rate derived from these measurements is (6.3 $\pm$0.5) $\times$
10$^{27}$ s$^{-1}$ (i.e. 1.5 times higher than the {\it Herschel}
value) for a CO production rate of (2.9$\pm$0.2) $\times$ 10$^{28}$
s$^{-1}$. Therefore, the $Q$(H$_2$O)/$Q$(CO) ratio derived from the {\it
Akari} data is 22$\pm$ 2\%. However, \citet{Ootsubo2012} assumed
 CO and water outflow velocities of 0.31 km s$^{-1}$. Using our velocity assumptions instead, we derive $Q$(H$_2$O) =  (5.9 $\pm$ 0.5) $\times$ 10$^{27}$ s$^{-1}$, $Q$(CO) = (3.8$\pm$0.3) $\times$ 10$^{28}$
s$^{-1}$, and $Q$(H$_2$O)/$Q$(CO) = 15 $\pm$ 2 \%, which is marginally higher than the 
 {\it Herschel} value. We note that the FOVs 
for the two data sets are similar.

The water production rates derived for the icy-grain model  with the nominal grain temperature assumption of 100 K are almost identical to those of the nucleus production model  for $L_p$ =  5$\times$10$^4$ km. They are about three times lower for $L_p$ = 10$^4$ km (Table~\ref{tab:2other}). For $L_p$ = 10$^4$ km, the average population within the HIFI field of view ($\sim$ 8.$\times$ 10$^4$ km radius) of the H$_2$O 1$_{\rm 10}$ rotational level is indeed higher for the icy-grain model than for the nucleus-production model.  For a grain temperature of 170 K, the derived  production rates are 5\% lower.

\subsection{HCN production rate}
The derived HCN production rate  
determined for the 2007--2011 period  is
4.4 $\times$ 10$^{25}$ s$^{-1}$ when we assume direct release from the nucleus (Table~\ref{tab:2other}). The value is almost the same (within 20--50\%) when  production from icy grains is considered.

The HCN production rate typically is a factor of 100 and 1000 lower than the H$_2$O and CO
production rates, respectively. Using the April-May 2010
data alone and considering the nucleus-production model, we find $Q$(HCN)/$Q$(CO) = (1.2 $\pm$ 0.3)
$\times$ 10$^{-3}$, and $Q$(HCN)/$Q$(H$_2$O) = (1.2 $\pm$ 0.3)
$\times$ 10$^{-2}$. From the detection of CN in optical
spectra of comet 29P obtained in December 1989,
\citet{Cochran1991} measured a CN production rate $Q$(CN) = 8
$\times$ 10$^{24}$ s$^{-1}$. This is a factor of 5 lower on
average than the HCN production rate. This discrepancy might
be related to the extended nature of the HCN production, as
discussed in Sect.~\ref{sec:gas-ori}, or to comet variability.

 The HCN abundance
relative to water is a factor of 10 higher than values found in
comets at 1 au from the Sun, which are typically 0.1--0.2 $\times$
10$^{-2}$ \citep{dbm2004}. However, compared with C/1995 O1
(Hale-Bopp) at 6 au \citep{Biver2002} (we extrapolated the water
production rate measured outbound at 5 au from the Sun to 6 au and used the
$Q$(HCN) measured at 6 au outbound), the $Q$(HCN)/$Q$(H$_2$O) and
$Q$(HCN)/$Q$(CO) ratios in 29P are consistent within a factor of about three with the values measured in Hale-Bopp at 6 au
post-perihelion (Table~\ref{tab:abun}).

\subsection{NH$_3$ production rate}
For NH$_3$, the derived 3$\sigma$ upper limit for the average of
April and May 2010 data is 4.5 $\times$ 10$^{27}$ s$^{-1}$
(nucleus-production model, Table~\ref{tab:2other}).  This upper limit is a factor of two lower than the previous best limit from \citet{Paganini2013}. The abundance of NH$_3$ relative to water
($<$ 1.1) is not constraining compared to values measured in
comets near 1 au from the Sun \citep[0.005, e.g.][]{Biver2012}.

Table~\ref{tab:abun}~summarizes the molecular abundances relative to CO measured in 29P and Hale-Bopp at 6 au from the Sun, and in other comets. This table illustrates the strong differences in coma composition between distant comets and comets at $r_{\rm h}$ $\sim$ 1 au.

\begin{table}
\caption[]{Abundances relative to CO. } \label{tab:abun}
\begin{tabular}{llll}
\hline\hline\noalign{\smallskip}
Quantity & 29P & Comets & Hale-Bopp$^a$  \\
& $r_h$ $\sim$ 6 au & $r_h$ $\sim$ 1 au & $r_h$ $\sim$ 6 au\\
 \hline\noalign{\smallskip}
$Q$(H$_2$O)/$Q$(CO) & \phantom{$<$~}0.10 & 5$-$200$^{a,b,c}$& $\leq$ 0.08$^g$ \\
$Q$(HCN)/$Q$(CO) & \phantom{$<$~}0.001 & 0.01$-$0.5$^{a,b,c}$ &  0.003  \\
$Q$(NH$_3$)/$Q$(CO) & $<$~0.1  & 0.03$-$1.0$^{b,c,d}$ & $-$ \\
$Q$(CO$_2$)/$Q$(CO) & $< 0.01^e$ & 0.5--5$^f$& $-$ \\
\hline\noalign{\smallskip}
\end{tabular}
{\bf Notes.} Abundances derived assuming molecule release from the
nucleus, a condition that is not verified at $r_h$ = 6 au from the
Sun.  $^{(a)}$ \citet{Biver2002}. $^{(b)}$ \citet{Dello16}. $^{(c)}$\citet{Lippi21}, excluding values  from the hyperactive comet 103P/Hartley 2. $^{(d)}$ \citet{Disanti2017}. $^{(e)}$ \citet{Ootsubo2012}. $^{(f)}$  \citet{Ahearn2012}, excluding the atypical value of 100 measured in 103P/Hartley 2.$^{(g)}$ Extrapolating the $Q$(H$_2$O) trend observed post-perihelion.
\end{table}

\section{Correlation between gas production and dust outbursts}
\label{sec:CO-mv-correlation}

Several HIFI and IRAM observations were obtained soon after outbursts (Sect.~\ref{sec:optical}). Therefore, it is possible to investigate whether outgassing is correlated to the dust activity for either the quiescent or the outbursting stages. 

\subsection{Correlation of CO to dust }

\begin{figure}[h!]
\begin{center}
\includegraphics[width=9cm]{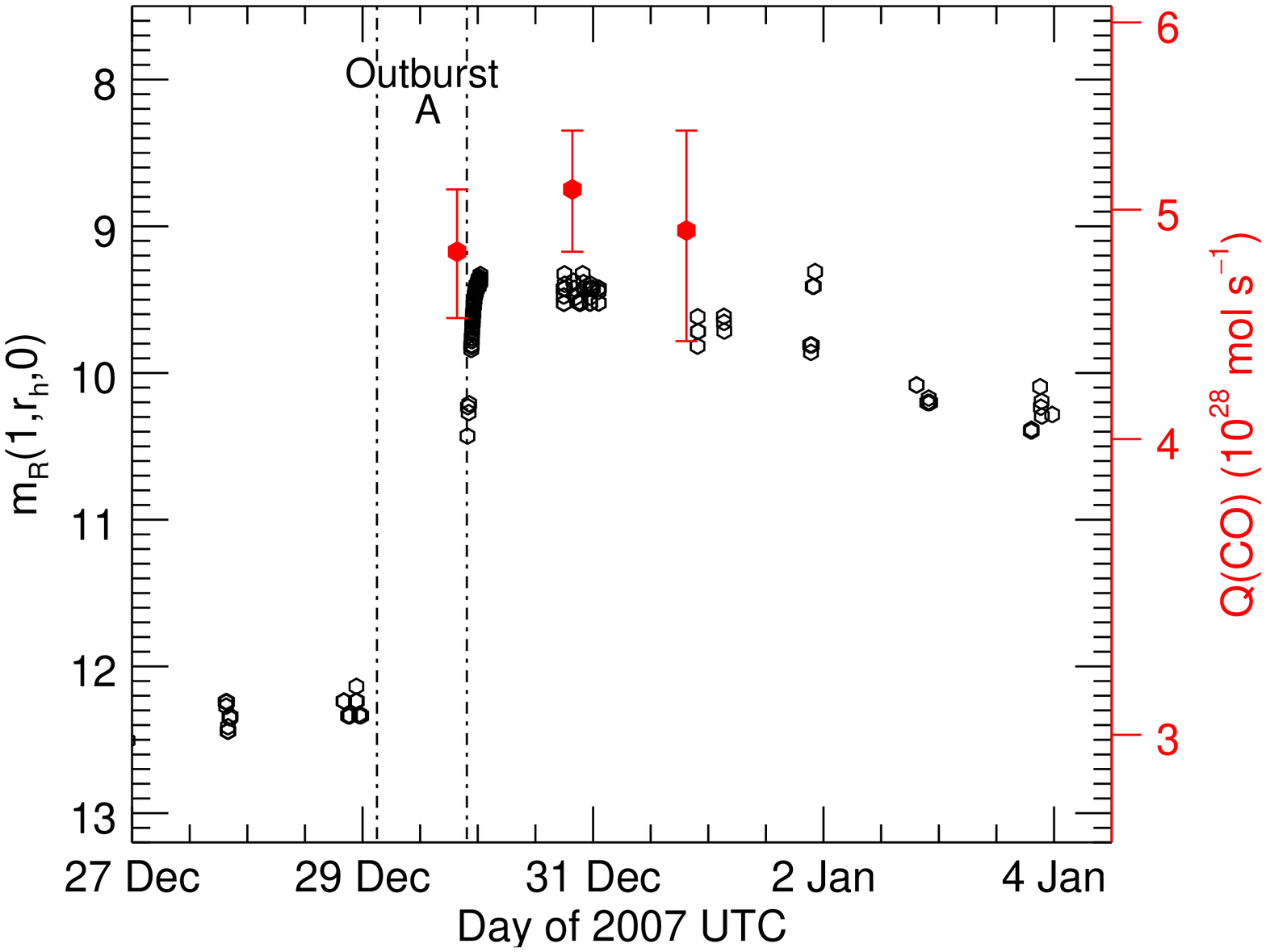}
\includegraphics[width=9cm]{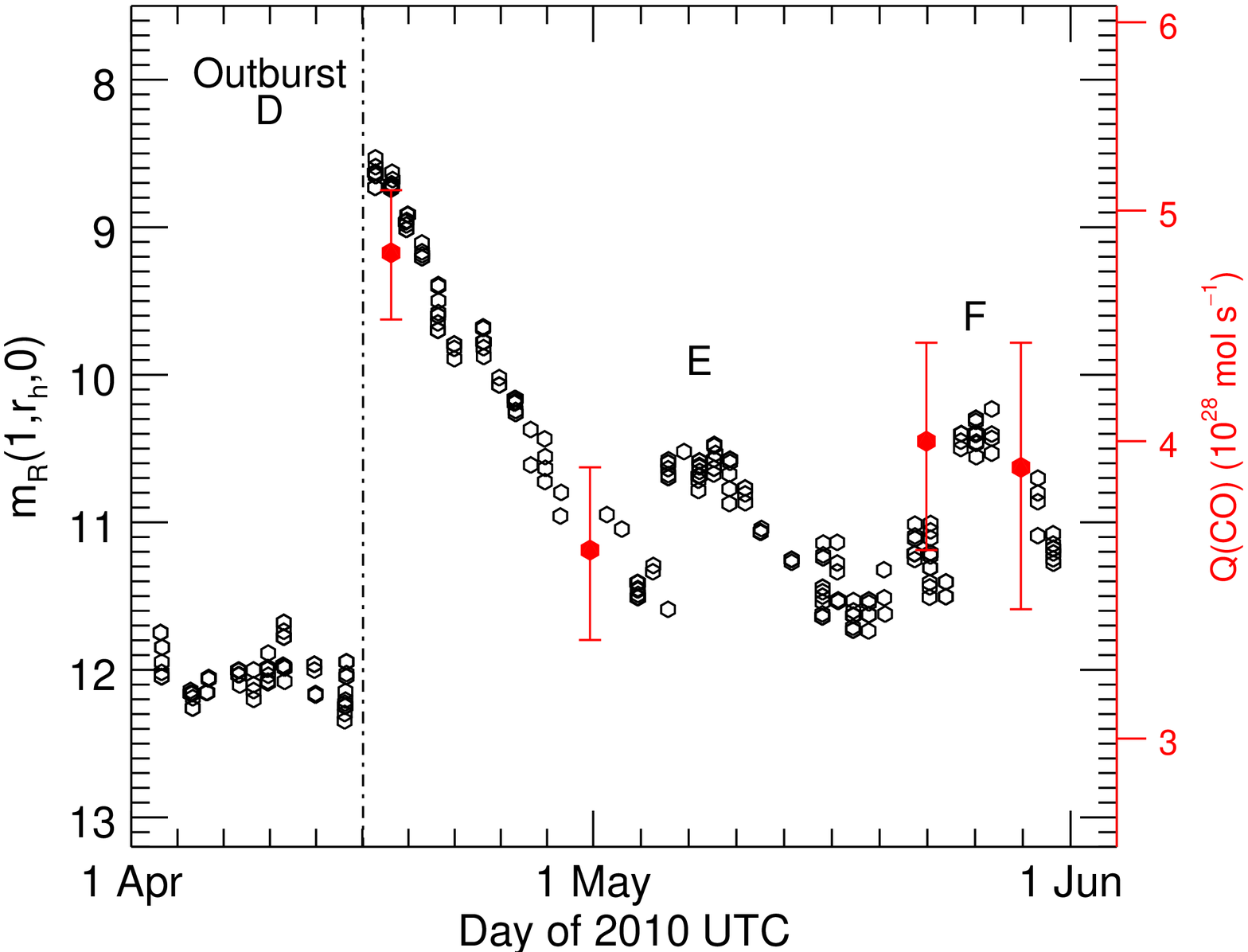}
\end{center}
\caption{CO production rates and reduced nuclear magnitudes in December 2007 and April-May 2010. The times of outbursts A and D are marked by dot-dashed lines. Outbursts E and F are also shown. The R nuclear magnitudes are measured inside an aperture with a diameter of 10\arcsec~(Spanish amateur data reported in Tables~\ref{tab:2007} and \ref{tab:2010}; homepage of M. Kidger).  The relation between the CO and magnitude scales is ${\rm log}_{\rm 10}(Q(\rm CO)) = 29.25 - 0.062 m_{\rm R}(1,r_{\rm h},0)$, consistent with Eq.~\ref{eq:mv-co}.} \label{fig:CO-mv-time}
\end{figure}

Figure~\ref{fig:CO-mv-time} shows the time evolution of the CO production rate and R nuclear magnitude $m_{\rm R}$ in December 2007 and April-May 2010. The CO production rate is higher for higher coma brightness. The decay of the coma brightness after outburst D coincides with a decrease in CO production. 

\begin{figure}[h]
\begin{center}
\includegraphics[width=9.0cm]{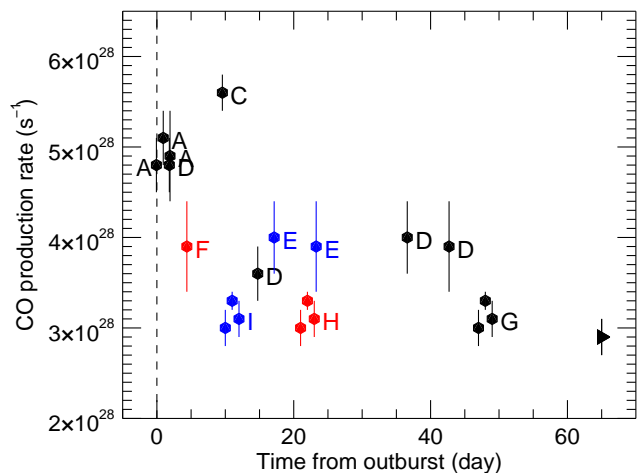}
\end{center}
 \caption{CO production rates  as a function of the elapsed time $\Delta T_{outburst}$ between outburst time and observation date.  The reference time for $\Delta T_{outburst}$ are outbursts A, C, D, and G (black dots), outbursts F and H (red dots), and  outbursts E and I (blue dots). The color code is such that when several outbursts are relevant to a CO measurement, black color is for the brightest, blue colour for the faintest, and red color is for the outburst with intermediate brightness. The black triangle (rightmost data point) refers to the January 2011 measurement obtained more than 93 d after an outburst.}
\label{fig:CO-timeoutburst}
\end{figure}

Figure~\ref{fig:CO-timeoutburst} plots the CO production rates as a function of the elapsed time $\Delta T_{\rm outburst}$ (Table~\ref{tab:PACS}) between outburst times and observing date, considering only IRAM data. The highest CO production rates are observed for $\Delta T_{\rm outburst}$ $\leq$ 10 days and are all about 5 $\times$ 10$^{28}$ mol s$^{-1}$. The figure might  suggest that in some instances, $Q$(CO) remains higher than the quiescent value up to 15--25 days (and even 40 days) after the most recent outbursts. However, the data points showing CO excess in this time range pertain to the observations of May 2010 with three consecutive outbursts (D, E, and F; bottom panel of Fig.~\ref{fig:CO-mv-time}). 

To quantify the significance of the correlation, we enlarged the sample, especially for measurements during quiescent activity, by considering the CO $J$(2--1) data acquired with the Arizona Radio Observatory 10 m Submillimeter Telescope (SMT) during the periods February-May 2016 and  November 2018 to  January 2019 \citep{2020AJ....159..136W}. For consistency, the CO production rates were recomputed using the published line areas, assuming a main-beam efficiency of 0.71, and using the same model and model parameters as were used to analyze the IRAM observations. The inferred CO production rates are very similar to those inferred by \citet{2020AJ....159..136W}.

\begin{figure}[h]
\begin{center}
\includegraphics[width=9.3cm]{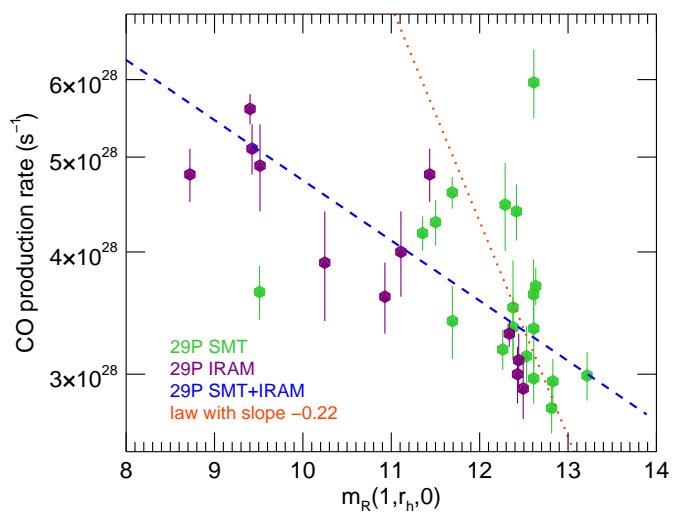}
\end{center}
 \caption{CO production rates as a function of $\Delta$-- and phase--corrected red nuclear magnitude $m_{\rm R}$(1,$r_{\rm h}$,0). Purple symbols show CO data from this work. Green symbols show CO data from \citet{2020AJ....159..136W}. The  dashed blue line shows the fit to all data (Eq.~\ref{eq:mv-co}). The dotted red line shows the curve ${\rm log}_{\rm 10}$($Q$(CO)) = $K$ -- 0.22 $m_{\rm R}$(1,$r_{\rm h}$,0), whose regression slope corresponds to that measured for comet Hale-Bopp ($K$ here is an arbitrary constant and not the constant appearing in the Hale-Bopp correlation equation Eq.~\ref{eq:mR-co-HB}).}
 \label{fig:CO-mv}
\end{figure}

Figure~\ref{fig:CO-mv} shows the CO production rate as a a function of the
the reduced magnitude $m_{\rm R}(1,r_{\rm h},0)$ defined in Sect.~\ref{sec:optical}. IRAM and SMT data are merged. A linear fit between ${\rm log}_{\rm 10}(Q(\rm CO))$ and $m_{\rm R}(1,r_{\rm h},0)$ gives
\begin{equation} 
{\rm log}_{\rm 10}(Q(\rm CO)) = (29.29 \pm 0.04) - (0.062 \pm 0.004) m_{\rm R}(1,r_{\rm h},0),
\label{eq:mv-co}
\end{equation}

\noindent
where the uncertainties do not consider magnitude errors. This fit is shown by a dashed line in Fig.~\ref{fig:CO-mv}. The Spearman rank correlation coefficient of $r_{\rm s}$ = --0.67 together with the small significance value of its deviation from zero ($p_{\rm r_s}$ = 0.002\%) and the number of standard deviations with respect to the null hypothesis ($z_{\rm D}$ = 3.8) are consistent with a moderate to strong correlation. The Spearman coefficient is $r_{\rm s}$ = --0.87 (with $p_{\rm r_s}$ = 0.03\%, $z_{\rm D}$=2.9) considering only IRAM data, and $r_{\rm s}$ = --0.54 (with $p_{\rm r_s}$ = 1.1\%, $z_{\rm D}$ = 2.4) for SMT data.

Several data points deviate significantly from the fit, and indeed \citet{2020AJ....159..136W} found that two dust outbursts coincided with a rise in CO, but two other outbursts occurred without any substantial increase in CO production. At quiescent magnitudes, $Q$(CO) is about 3 $\times$ 10$^{28}$ s$^{-1}$ (Fig.~\ref{fig:CO-mv}). We adopt in the following the central value of $Q_{\rm quiet}(\rm CO)$ = 2.9 $\times$ 10$^{28}$ s$^{-1}$ determined by \citet{2020AJ....159..136W} from 2016 CO data. The regression slope in the correlation equation (Eq.\ref{eq:mv-co}) is small (0.062), and it is three times smaller than the value established for comet Hale-Bopp \citep[0.22,][]{2021PSJ.....2...17W} (Appendix~\ref{appendix-HB}). This is illustrated in Fig.~\ref{fig:CO-mv} by the dot-dashed line.

Since at least two-thirds of the measured CO outgassing corresponds to permanent activity, we derived the correlation equation for the outburst material. The excess of CO production related to outbursts is given by
\begin{equation}
Q_{\rm out}(\rm CO) = Q(\rm CO) - Q_{\rm quiet}(\rm CO).
\end{equation}

\noindent
The nuclear magnitude of outburst dust ejecta is calculated according to
\begin{equation}
\begin{aligned}
m_{\rm R, out}(1,r_{\rm h},0)= & -2.5 {\rm log}_{\rm 10}\big(10^{-0.4m_{\rm R}(1,r_{\rm h},0)} \\
& -10^{-0.4m_{\rm R, quiet}(1,r_{\rm h},0)}-10^{-0.4m_{\rm R, nuc}(1,r_{\rm h},0)}\big),
\end{aligned}
\end{equation}

\noindent
where $m_{\rm R, quiet}(1,r_{\rm h},0)$ (= 13.4) is obtained from Eq.~\ref{eq:mv-co}. 
$m_{\rm R, nuc}(1,r_{\rm h},0)$ is the nucleus magnitude (equal to 14.04 at $r_{\rm h}$ = 6 au), derived from an expected $R$ absolute magnitude of 10.15, assuming a nucleus radius of 31 km and a $R$ geometric albedo of 0.044.

\begin{figure}[h]
\begin{center}
\includegraphics[width=9.3cm]{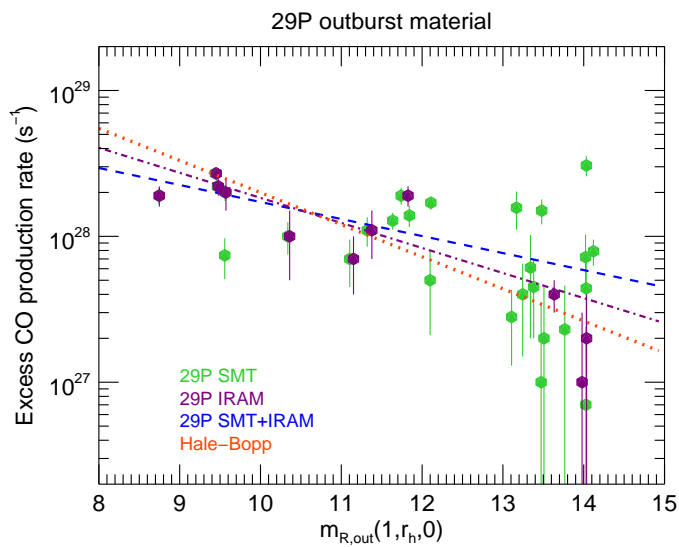}
\end{center}
 \caption{Same as Fig.~\ref{fig:CO-mv}, but considering the  contribution of outburst material. The  dashed blue line shows the fit to all 29P data (Eq.~\ref{eq:mv-co-outburst}). The  dot-dashed purple line shows the fit to IRAM 29P data (Eq.~\ref{eq:mv-co-outburst-IRAM}). The dotted red line shows the curve ${\rm log}_{\rm 10}$($Q$(CO)) = 30.5 -- 0.22 $m_{\rm R}$(1,$r_{\rm h}$,0) determined for comet Hale-Bopp (Appendix~\ref{appendix-HB}).}
 \label{fig:CO-mv-outburst}
\end{figure}

Using IRAM and SMT data, we obtain
  \begin{equation} 
{\rm log}_{\rm 10}(Q_{\rm out}(\rm CO)) = (29.40 \pm 0.04) - (0.127 \pm 0.003) m_{\rm R, out}(1,r_{\rm h},0),
\label{eq:mv-co-outburst}
\end{equation}
\noindent
and the Spearman rank correlation coefficient is $r_{\rm s}$ = --0.55 (with $p_{\rm r_s}$ = 0.2\%, $z_{\rm D}$ = 2.9). Using the IRAM data alone, we obtain
  \begin{equation} 
{\rm log}_{\rm 10}(Q_{\rm out}(\rm CO)) = (29.98 \pm 0.05) - (0.172 \pm 0.004) m_{\rm R, out}(1,r_{\rm h},0),
\label{eq:mv-co-outburst-IRAM}
\end{equation}
  \noindent
with $r_{\rm s}$ = --0.82, $p_{\rm r_s}$ = 0.2\%, $z_{\rm D}$ = 2.6. Figure~\ref{fig:CO-mv-outburst}  shows $Q_{\rm out}(\rm CO)$ as a function of $m_{\rm R, out}(1,r_{\rm h},0)$, and the linear fits given by the correlation equations Eq.~\ref{eq:mv-co-outburst} and \ref{eq:mv-co-outburst-IRAM}. The correlation law for 29P is very close to the $Q$(CO)/$m_{\rm R}(1,r_{\rm h},0)$ correlation established for comet Hale-Bopp at large heliocentric distances, where the activity was dominated by CO outgassing (Eq.~\ref{eq:mv-co-HB}, dotted red line).

\subsection{H$_2$O and HCN correlations with CO outgassing}

The two HIFI water detections were obtained 3 and 25 days after the major outburst D (Sect.~\ref{sec:optical}).
The signal decreased by a factor 1.45$\pm$ 0.42 between the two dates. The same decrease (by a factor 1.41$\pm$0.15) is observed for the CO line area; this is shown  by a comparison of the values at 1.8 and 15 d after outburst D. At the date of the H$_2$O nondetection (30 December 2010), 29P was quiescent (and the CO production rate measured 12 days later was at the quiescent value). These trends, together with the similarity between {\it Akari} and {\it Herschel} $Q$(H$_2$O)/$Q$(CO) measurements, suggest a correlation between water and CO production. On the other hand, there is no apparent correlation between the HCN and CO line areas, but the low signal-to-noise ratio of the HCN line area prevents any definitive conclusion.    

Taking into account that at least two-thirds of the measured CO outgassing is not related to recent outbursts, but corresponds to permanent activity, the constant H$_2$O/CO production rate ratio suggests that H$_2$O is present in the atmosphere of comet 29P even during quiescent phases. The H$_2$O/CO correlation (if confirmed) is surprising. As discussed in the next section, H$_2$O and HCN are released in the outer coma by long-lived icy grains, whereas CO molecules are outgassing from the near-nucleus region. This correlation could be explained if the dust-to-gas production rate ratio during outburst and quiescent phases were similar, but this contradicts the measurements (see the next section).

\subsection{Constraints on the origin of outbursts}

A well-documented outburst is the huge ($m_{\rm R}$ from 16.5 to 6.5) outburst of comet 17P/Holmes on 24 October 2007. A high CO production rate of 1.8 $\times$ 10$^{29}$ mol s$^{-1}$  was observed at the IRAM 30 m telescope two days after the onset of the outburst, followed by a steep decrease by a factor of 6.3 between $\Delta T_{\rm outburst}$ = 2 d and $\Delta T_{\rm outburst}$ = 7.5 d \citep{2008LPICo1405.8146B}. This is consistent with the rapid vaporization of icy debris and the short residence time of the CO molecules within the IRAM beam (typically 0.07 d for 17P at $\Delta \sim $1.62 au). In this time interval, $m_{\rm R}$ varied from 6.5 to 8.4. For 29P, the residence time of the CO molecules is 0.7 d, and the residence time is 1.7 d for the dust particles outflowing at 0.15 km s$^{-1}$ (Sect.~~\ref{sec:optical}). The constancy  of $Q$(CO) within 2--3 days after the December 2007 outburst (Fig.~\ref{fig:CO-mv-time}) suggests continuous CO production either from the outburst ejecta or from the nucleus surface areas from which the outburst was triggered. The amount of CO that was released during outbursts A and D can be roughly estimated by assuming that most of the production occurred within 5 days after outburst onset at a rate of 2 $\times$ 10$^{28}$ mol s$^{-1}$. The derived CO mass is $\sim$ 4$\times$10$^8$ kg, which corresponds to a 47 m radius sphere of pure CO ice. The few available estimates of the mass of dust in outburst ejectas give lower limits of 3--18 $\times$ 10$^8$ kg \citep{hosek2013,2017Icar..284..359S}. Assuming that CO is intimately mixed with nucleus material (with density $\rho_{\rm N}$ = 500 kg m$^{-3}$), the nucleus volume affected by CO vaporization is  0.64 10$^{-6}$ \% of the total volume of the nucleus.

The outbursts of 29P are observed with some periodicity (7.3 per year), which caused \citet{Trigo2010} to conclude that the triggering mechanism involves a periodic insolation of a particular region associated with the nucleus rotation with a presumed period $\sim$57 d. \citet{Miles2016} refined the analysis and suggested at least 6 discrete outburst sources that are grouped in longitude (within 15$^{\circ}$) on the surface of the nucleus.  The similarity of the CO line profiles during outburst and quiescent phases (Figs~\ref{fig:COspectra} and~\ref{fig:CO-blue-red}) confirms that outbursts occur in the subsolar region,  where CO outgassing predominantly and continuously operates. 

The established correlation laws between CO production rates and magnitudes, both in quiescent and outburst state, and the comparison with comet Hale-Bopp  provide insights into the properties of outbursting regions. We first mention that the size of comet Hale-Bopp \citep[37$\pm$3 km,][] {2012ApJ...761....8S} is similar to that of 29P, so that processes involving gravity, such as the dynamics of large particles, and their gravitational fallback, might be comparable.

The CO production rate of 29P during quiescent activity is very similar to that of comet Hale-Bopp at 6 au from the Sun \citep[$\sim$ 3 $\times$ 10$^{28}$ s$^{-1}$ inbound and $\sim$ 2 $\times$ 10$^{28}$ s$^{-1}$ outbound,][]{Biver2002}.
On the other hand,  with $m_{\rm R, quiet}(1,r_{\rm h},0)$ = 13.4 for 29P and $m_{\rm R}$(1,$r_{\rm h}$,0) $\sim$ 9 at $r_{\rm h}$ = 6 au for Hale-Bopp (Appendix~\ref{appendix-HB}), the quiescent dust activity of the two comets is different by more than one order of magnitude in brightness. This can be explained by two scenarios. The first scenario is differences in surface properties: A higher cohesion of the surface material of 29P could quench dust activity, or large particles on the surface (e.g. fallback particles) might reduce dust-gas coupling and thus dust lifting; see the discussion in \citet{2019A&A...630A..23T}. The second scenario is differences in size properties of the lifted dust particles. A deficiency in small particles in the  quiescent coma of 29P (i.e., a minimum particle size larger than in the coma of Hale-Bopp) would result in a lower coma brightness in the optical for the same dust production rate in kg/s; this would also imply different surface properties in terms of particle size distribution. The dust production rate of comet Hale-Bopp at large heliocentric distances is well constrained by mid-IR data \citep{2001A&A...377.1098G} and detailed modeling of optical data \citep{2003A&A...403..313W}. At 6 au outbound, the value determined by \citet{2003A&A...403..313W} is approximately 10$^3$ kg s$^{-1}$, about a factor of ten higher than the quiescent value for 29P (Sect.~\ref{sec:dustprod}). Therefore, this favors the first scenario, in which the dust activity of 29P (but not the gas activity) is quenched, possibly as a result of surface-subsurface processing induced by activity.      

In contrast, the outburst activity of 29P presents similarities with the continuous activity of Hale-Bopp.  The fact that the $Q$(CO) and visual magnitude correlations for the outburst material of 29P and for Hale-Bopp are very similar (Fig.~\ref{fig:CO-mv-outburst})  indicates a similar dust-to-gas flux ratio for the outburst ejecta of 29P and the continuous activity of Hale-Bopp (we refer here to dust particles that contribute to the scattering cross-section). Overall, this suggests  strong local heterogeneities on the surface of 29P, with quenched dust activity from most of the surface, but not in outbursting regions. 

Several triggering mechanisms for the 29P outbursts have been proposed, but the driving process remains unknown. The proposed scenarios include 1) the amorphous-to-crystalline phase transition of water,  and 2) the build-up of high-pressure pockets of hypervolatiles below the surface layers. On comet 67P, the spatial distribution of outburst locations on the nucleus correlates well with areas marked by steep scarps or cliffs \citep{2016MNRAS.462S.184V}, and 45\% of the 67P summer outbursts occurred near local noon. Some events were found to be initiated by the collapse of a cliff \citep{2017NatAs...1E..92P,2017MNRAS.469S.606A}, and thus to be simply related to erosion (scenario 3). As discussed by \citet{2016A&A...587A..14V}, activity from fractured cliffs leads to a weakening of the wall structure until it collapses. Cliffs should be more instable on larger bodies such as 29P. For these three scenarios, we expect an increase of CO outgassing correlated with dust release. The measured CO release shows that large areas on the 29P surface are affected during outbursts. We hypothesize that the slow (57 d) rotation of 29P plays a role for the driving mechanism, as it allows the heat wave to penetrate deeper into the subsurface layers. We propose a fourth scenario, namely that outbursts result from  fractures (or pits) on the 29P surface. From thermophysical modeling, \citet{2017A&A...608A.121H} showed that, through the effect of self-heating, fractures are an efficient heat trap  when the Sun  shines directly into the fracture, resulting in enhanced outgassing  with respect to a flat surface during illumination. This scenario could explain both the periodicity of the outbursts, and the higher dust-to-gas flux ratio observed during outbursts, if fracture floors are structurally less evolved than the remaining surface. For a 10-min-long outburst, the typical size of the illuminated fracture floor would be 25 m, but this would be 3 km for a one-day-long outburst.

\section{Water production and origin of H$_2$O and HCN}
\label{sec:gas-ori}

\subsection{Evidence for production by sublimating icy grains}

The observed water production rate might be explained by
outgassing from the nucleus surface. The thermal properties of
the nucleus of 29P have been constrained by 
multiwavelength {\it Spitzer} observations \citep{Stansberry2004,2015Icar..260...60S,2021PSJ.....2..126S}.  Using the Near Earth Asteroid Thermal Model  \citep[NEATM,][]{harris98}, \cite{2021PSJ.....2..126S} inferred an infrared beaming factor $\eta$=1.1$\pm$0.2, consistent with the mean value of 1.03$\pm$0.11 determined for an ensemble of 57 Jupiter-family comets \citep{2013Icar..226.1138F}. When we adopt $\eta$ = 1.03, a gray emissivity
of 0.95 and a Bond albedo of 0.012, the temperature of the subsolar point is equal to 158.59 K at $r_{\rm h}$ = 6.21 au. At this
temperature, a sublimating area of $\sim$2000 km$^2$ of crystalline ice is
needed to supply a rate of 4.1 $\times$ 10$^{27}$ s$^{-1}$ of water
molecules. However, with a rotation period of $\sim$57 days \citep[based on the periodicity of the outbursts,][]{Miles2016}, and an expected small thermal
inertia, as measured for other Centaurs and cometary nuclei
\citep{Groussin2013,Fornasier2013,Lellouch2013,2015Sci...347a0709G}, we expect
variations in the surface temperature with solar zenith angle, and
low temperatures on the night side. In order to compute the active fractional area of the nucleus surface that supplies the observed water production rate, we therefore applied the sublimation model of \cite{Cowan1979}, which computes the latitude dependence of the surface temperature and sublimation rate. We used the model outputs for a rotational pole pointed at the Sun, which is identical to both the nonrotating case and to the case of zero thermal inertia. It is therefore appropriate for investigating the activity of 29P. The derived active fractional area is 440\%, suggesting  that sublimating icy grains contribute mainly to water vapor release in the atmosphere of 29P. This active fractional area is in the upper range of values measured for hyperactive comets \citep{Lis2019}.  The vapor pressure of amorphous ice is one to two
orders of magnitude higher than for crystalline ice  \citep[see][and references therein]{Fray2009}, which means that the fractional
area of amorphous ice would be lower. However, we do not expect
water ice to be in amorphous form in the near-surface layers of the
nucleus of 29P \citep{Enzian1997,Kossacki2013}.

The low velocity offset observed for the H$_2$O line ($\Delta v$= --0.08$\pm$0.05 km s$^{-1}$,   Table~\ref{tab:2other}) also 
suggests that the nucleus contributes little to the water production.
Water sublimation is indeed expected to be most efficient near the subsolar point. Because
of the low phase angle ($\phi$ $<$ 10$^{\circ}$), such localized
outgassing would have resulted in a line shape that is blueshifted by a
fraction of kilometers per second, as observed for CO ($\Delta v$ between --0.3 and --0.2 km s$^{-1}$). 

HCN has a higher vapor pressure than water.  We calculated
that the observed production rate would correspond to an area of
sublimating HCN ice of 4 $\times$ 10$^{-3}$ km$^2$, assuming that this area is at the subsolar point. In this respect, the nucleus itself might therefore contribute to HCN production.  However, the HCN line also presents a small
velocity offset ($\Delta v$= --0.04$\pm$0.07 km s$^{-1}$; Table~\ref{tab:2other}), so that its production is likely associated with that of water.

It is thus very likely that both HCN and H$_2$O are the products
of icy-grain sublimation. Direct and indirect evidence for the
presence of icy grains in cometary atmospheres is now numerous
\citep[e.g.][]{Davies1997,Lellouch1998,Ahearn2011,Fougere2012,2014Icar..238..191P}. In
particular, the spectroscopic signature of water-ice grains has
been detected in comets at large $r_h$ as in C/1995 O1 (Hale-Bopp)
\citep[7 and 2.9 au,][]{Davies1997,Lellouch1998},
 C/2002 T7 (LINEAR) \citep[3.5 au,][]{Kawakita2004}, and C/2013 US$_{10}$ (Catalina) \citep[3.9 to 5.8 au,][]{2018ApJ...862L..16P}.

\begin{figure}[t!]
\begin{center}
\includegraphics[width=9.0cm]{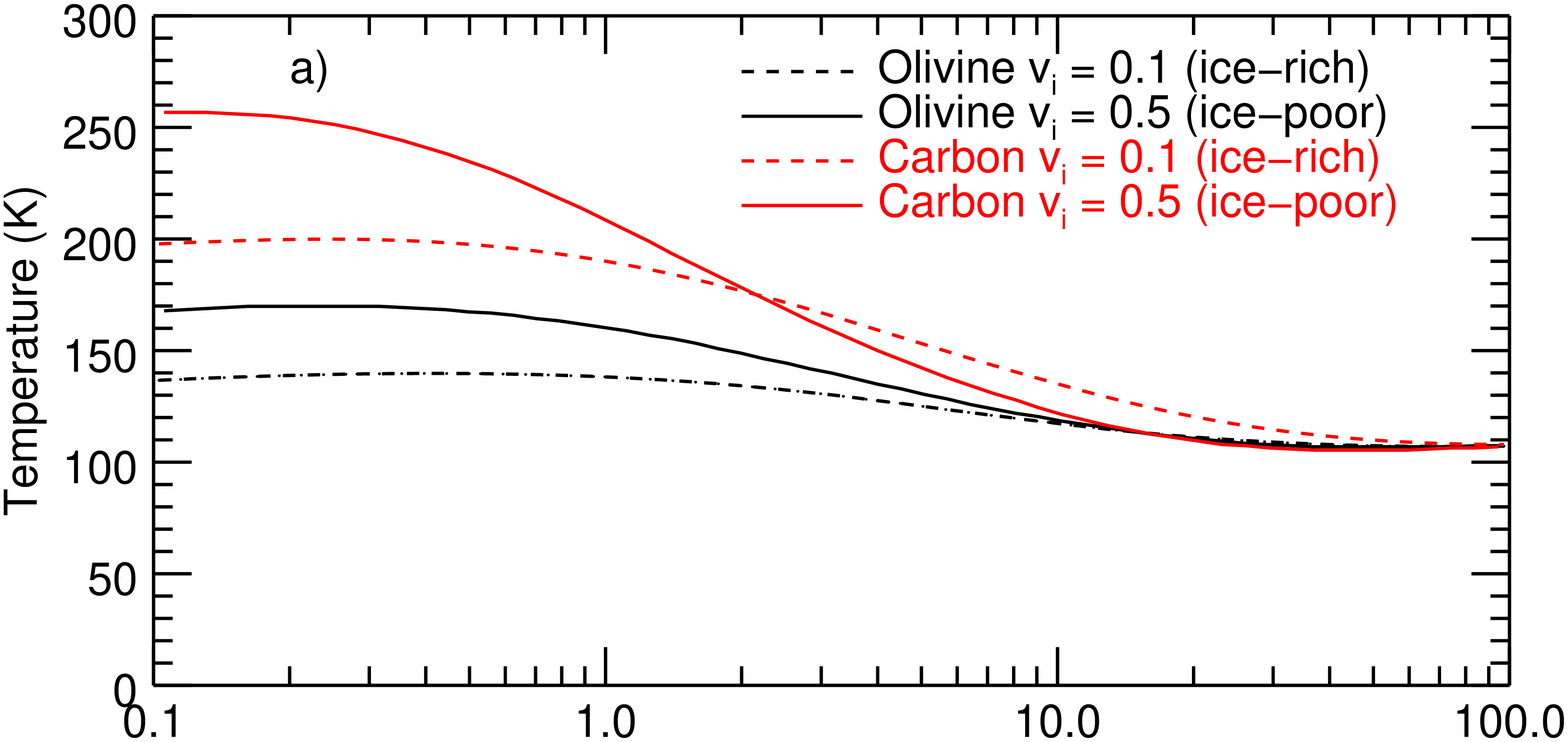}\vspace{-0.2cm}
\includegraphics[width=9.0cm]{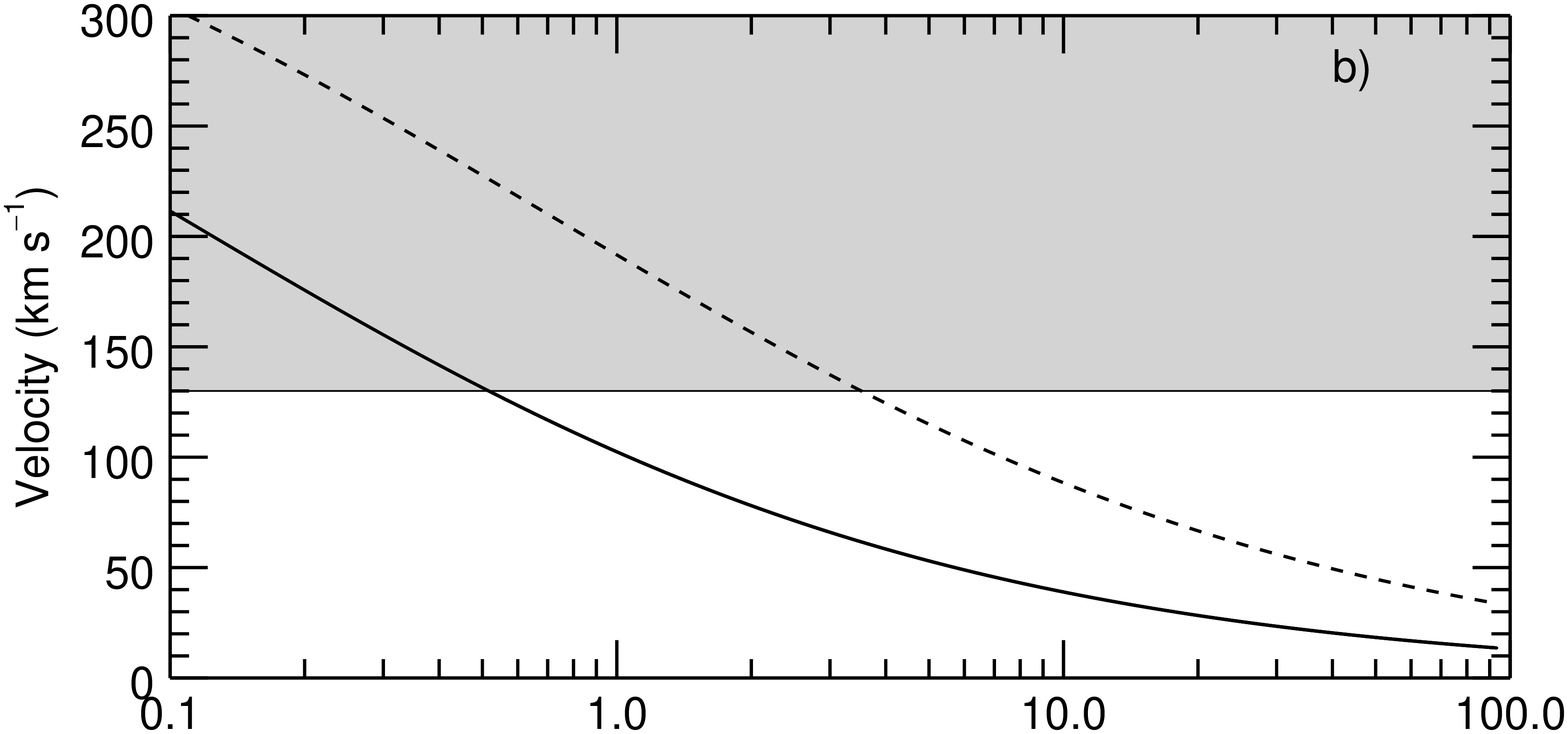}\vspace{-0.2cm}
\includegraphics[width=9.0cm]{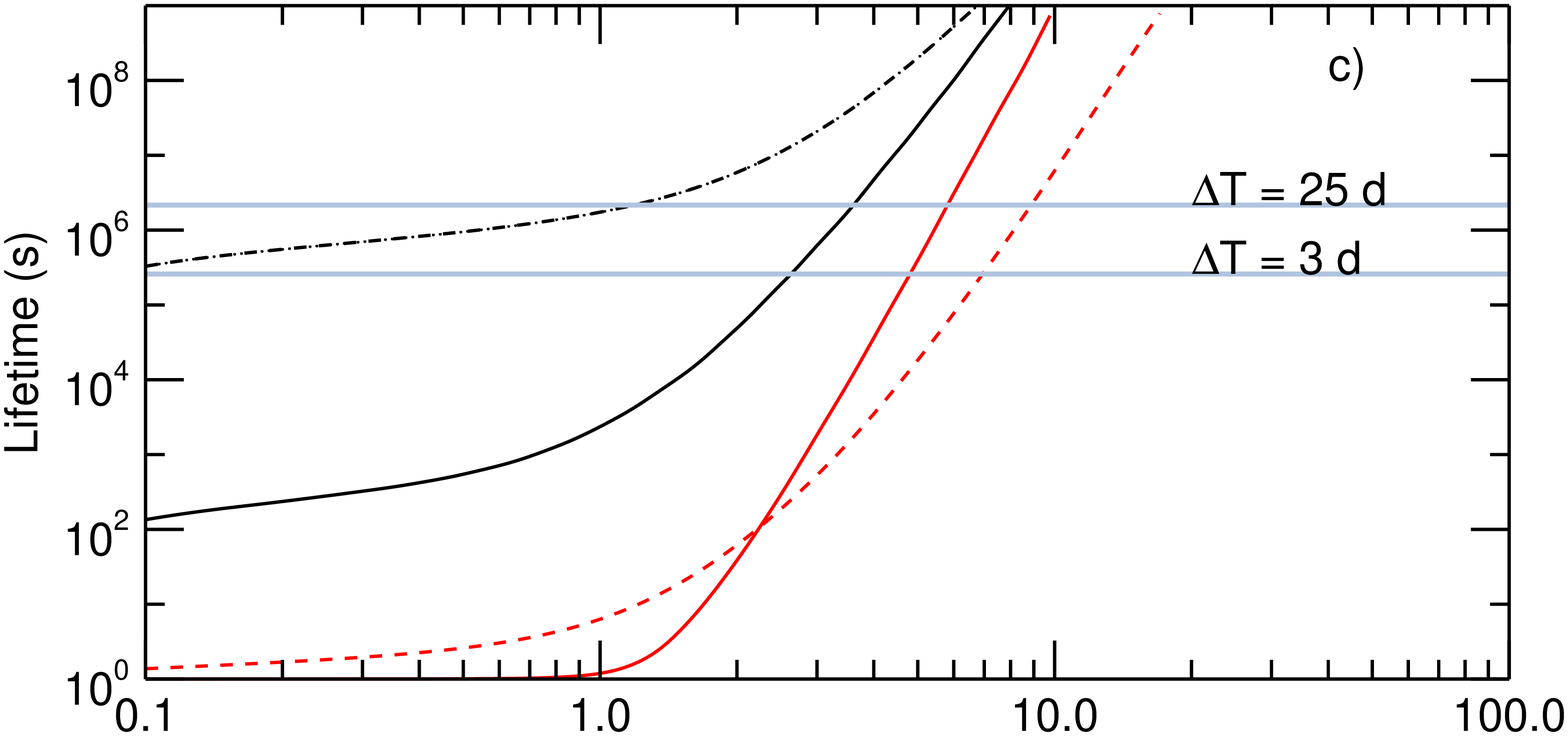}\vspace{-0.2cm}
\includegraphics[width=9.0cm]{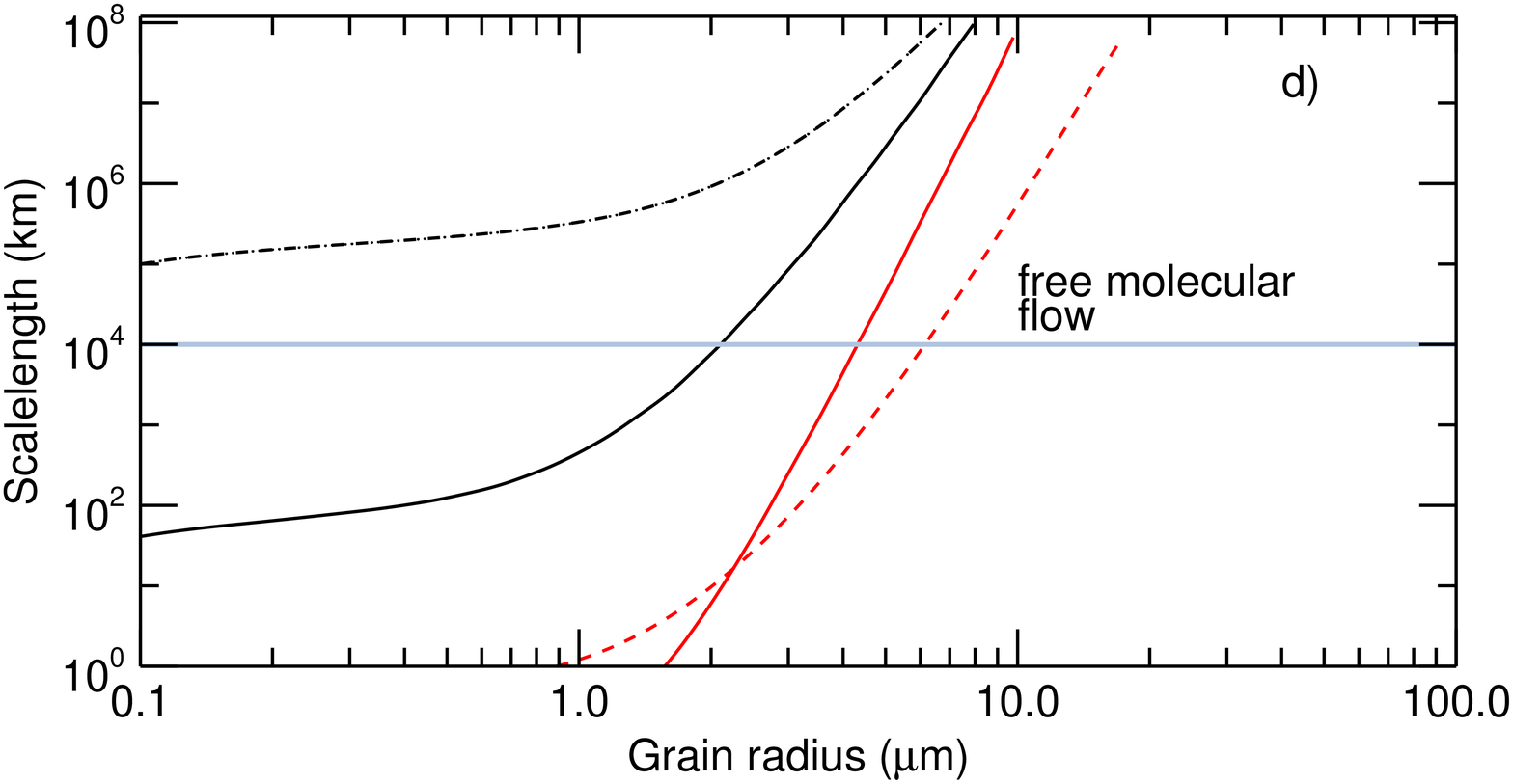}
\end{center}
 \caption{Properties of icy grains as a function of size. a) Grain temperature. b) Grain velocity computed
 assuming a CO production rate of 4 $\times$ 10$^{28}$ s$^{-1}$ emitted in
 a cone with a half-angle equal to $\psi$ = 45$^{\circ}$ (dashed curve) or $\psi$ = 180$^{\circ}$ (solid curve), and a dust and nucleus density
 equal to 500 kg m$^{-3}$. Velocities in the gray
 region are excluded at 1$\sigma$ from the velocity offset of the H$_2$O and HCN lines. c) Grain-sublimation lifetime; the two horizontal blue dotted lines
 correspond to the elapsed time between the 16.05 April 2010 outburst and the {\it Herschel} H$_2$O observations.
 d) Scale length defined as the product of the grain-sublimation lifetime times the grain velocity (calculated with $\psi$ =
 45$^{\circ}$); the horizontal line corresponds to the limit of CO free-molecular flow (see text). In panels a), c) and d)
the black and red curves correspond to crystalline ice mixed with
 olivine and carbon, respectively. Solid and dashed lines show ice-poor grains ($v_i$ = 0.5, ice mass fraction of 29\%) and ice-rich grains ($v_i$ = 0.1, ice mass fraction of 78\%), respectively.
}
\label{fig:grains}
\end{figure}

\subsection{Size constraints for sublimating icy grains}
\label{sec:4.2}

We computed (Appendix~\ref{appendix:dust}) the temperature, velocity, and
H$_2$O sublimation lifetime of icy grains as a function of size, for several grain
compositions (olivine or amorphous carbon, referred as dirt or impurities) and ice contents. The results are shown in Fig. \ref{fig:grains}, panels a--c. Velocities were computed for the initial mass (before water release) of the grains. The sublimation lifetime is defined as the time when ice is exhausted.
Calculations were made for volume fractions of dirt, $v_i$ ($i$ for
impurities), of 0.1 and 0.5, corresponding to ice mass fractions of
 78\% and 29\%, respectively (Appendix~\ref{appendix:dust-temp}). 
Figure~\ref{fig:grains}d plots the scale length of the sublimation of icy
grains, defined as the product of the grain-sublimation lifetime and the velocity
of the grains (i.e. it is assumed that the motion of the grains is radial). As shown by \citet{Gunnarsson2003}, grains exhaust
most of their ice content at a time similar to their lifetime.
Figure~\ref{fig:grains}a shows that, except for ice-rich olivine
mixtures ($v_i$ = 0.1), grains with sizes smaller than 1 $\mu$m
reach temperatures higher than 160~K, so that they lose their ice
content very quickly (in less than 1000 s). As expected, carbon grains reach higher
temperatures than olivine grains, and grains with a higher content
of dirt are generally warmer. The computed velocity of grains with a
radius of 10 $\mu$m is 85 m s$^{-1}$ considering CO anisotropic
outgassing, and 35 m s$^{-1}$ in the isotropic case
(Fig.~\ref{fig:grains}b). These values are similar to the few
measured values. For 29P in quiescent state, one estimate is 35 m s$^{-1}$ for
particles with $\beta$ = 400 (ratio of solar radiation pressure and solar gravity forces), corresponding to $a$ = 10 $\mu$m for
$\rho_d$ = 500 kg m$^{-3}$ \citep{Fulle1992}. Measurements after an outburst lead to
150 $\pm$ 50 m s$^{-1}$ \citep[][, see also Sect.~\ref{sec:optical}]{Feldman1996} to 250 $\pm$ 80 m
s$^{-1}$ \citep{Trigo2010} for typically 1 $\mu$m sized particles
which extrapolate to 25--110 m s$^{-1}$ for 10 $\mu$m grains.

The low velocity offset of the H$_2$O and HCN lines provides some constraints on the size of the icy particles. A significant contribution from small (radius $a$ $<$ 3 $\mu$m
according to Fig.~\ref{fig:grains}b) grains to the observed HCN
and H$_2$O molecules is excluded at the 1$\sigma$ level, because
their significant velocity would have resulted in a significant
negative velocity offset in the spectra. At the 3$\sigma$ level,
the limiting minimal size is $\sim$ 1 $\mu$m. We assumed here that the
grains originate from the sunlit hemisphere and are entrained by
the CO jet (i.e., we consider the dashed curve in Fig.~\ref{fig:grains}b).

The shapes of the HCN and H$_2$O lines are symmetric within the noise, unlike
the strongly asymmetric line of the main coma constituent CO. This also
indicates that these molecules are produced in a region in which 
collisions with CO molecules are rare. In the collisional region, extensive momentum exchange causes a coupling between its components, so that the distribution and kinetics of minor species follow those of the main constituent \citep[e.g.][]{2008ApJ...685..659T}. \citet{Crifo1999} showed
that the collisional region in comet 29P is much larger than 700
km. In their highly anisotropic case, the Knudsen number $K_n$ (the ratio of the molecular mean free path length to a representative physical length scale)  is
equal to a few 10$^{-2}$ at 500 km from the nucleus, which sets the
inner boundary for the almost free molecular flow  to
typically $\sim$ 10$^4$ km. Comparing this value to the grain-sublimation scale
length as a function of size (Fig.~\ref{fig:grains}d), we can
exclude a major water-outgassing contribution from short-lived olivine-rich grains
($v_i$ = 0.5) with $a$ $<$ 2 $\mu$m. For carbon-rich grains, excluded grains are those with  $a$ $<$ 6.1 $\mu$m ($v_i$ = 0.1, ice rich) and $a$ $<$ 4.3 $\mu$m ($v_i$ = 0.5, ice poor). However, the sub-$\mu$m
olivine-rich grains with a high ice content ($v_i$ = 0.1) sublimate
outside the collision zone. HCN is more volatile than H$_2$O and should be exhausted more rapidly than water if it is present as pure HCN ice in grains. The symmetric HCN line shape suggests that HCN production occurs in the collisionless region and is controlled by the sublimation of water ice. 

The H$_2$O and HCN line widths provide further constraints on the properties of the grains. Assuming isotropic ejection from the grains in a collisionless environment, the half-line width corresponds to the terminal velocity for free-molecular expansion, which is equal to the mean thermal speed:  $v_{\rm therm} = \sqrt{8 k_B T_{\rm d}/\pi m_{\rm H_2O}}$ for water, where $T_{\rm d}$ is the grain temperature and $m_{\rm H_2O}$ is the mass of one water molecule. The range of inferred $T_{\rm d}$ is  36--64 K using the measured H$_2$O line width and its 1$\sigma$ uncertainty (and $T_{\rm d}$ $>$ 62 K using the HCN line width). This is indicative of low-temperature grains. However, the inferred $T_{\rm d}$ is a factor of two lower than the equilibrium temperature expected for large ($>$ 10 $\mu$m) grains (Fig.~\ref{fig:grains}a). The low signal-to-noise ratio on the H$_2$O line is a possible explanation.

In conclusion, the characteristics of the HCN and H$_2$O line profiles suggest their production from long-lived icy grains with a size exceeding a few micrometers. We present in the next section results obtained from the strength of the water line.

\subsection{Sublimating icy grains: outburst contribution and production rate}
\label{sec:subliming}

We modeled the production of water molecules by icy grains in the coma during an outburst (Appendices~\ref{appendix:dust-sub}--\ref{appendix:dust-dyn}) with the aim to study the evolution of the H$_2$O signal in the HIFI beam from 19 April to 11 May 2010 after outburst D. 

\begin{figure}[!ht]
\begin{center}
\includegraphics[width=9.3cm]{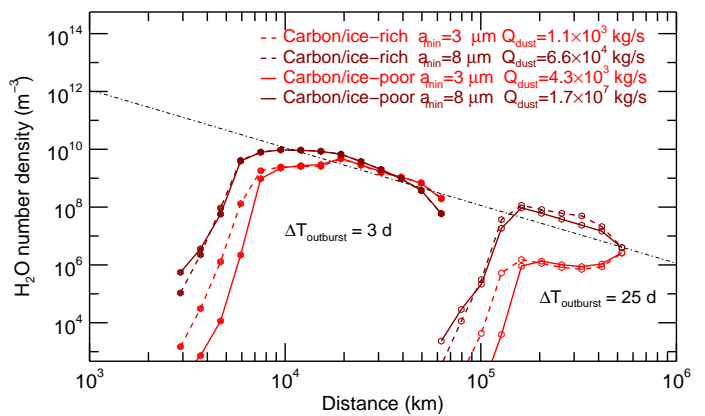}
\end{center}
 \caption{Radial H$_2$O number density in the coma of 29P after an outburst releasing icy grains  during two days at a rate $Q_{\rm dust}$.  The results are shown  for an elapsed time $\Delta T_{\rm outburst}$ = 3 d from the onset of the outburst (plain symbols, leftmost curves) and $\Delta T_{\rm outburst}$ = 25 d (open symbols, rightmost curves).  $Q_{\rm dust}$ (given in the legend) is set so that the number of molecules within the HIFI beam (whose projected radius is $\sim$ 8 $\times$ 10$^4$ km) is 10$^{33}$ molecules at time $\Delta T_{\rm outburst}$ = 3 d, corresponding to the 19.05 April 2010 measurement.  The results for $a_{\rm min}$  = 3 and 8 $\mu$m are shown in red and dark red, respectively. The maximum grain radius is $a_{\rm max}$  = 50 $\mu$m, and the size index is $\alpha$ = --3.5. Results are shown for both ice-poor ($v_i$ = 0.5) and ice-rich ($v_i$ = 0.1) carbon grains. The dot-dashed line is a Haser model with $Q$(H$_2$O) = 4.6 $\times$ 10$^{27}$ s$^{-1}$ and expansion velocity $v_{\rm H_2O}$ = 0.25 km s$^{-1}$.}  
 \label{fig:density}
\end{figure}

The outburst is described by a boxcar function defined by its duration and dust production rate $Q_{\rm dust}$. The number density of the H$_2$O molecules as a function of distance to nucleus was computed at a time interval with respect to outburst onset  $\Delta T_{\rm outburst}$ = 3 d and 25 d, for comparison with HIFI water observations (Tables~\ref{tab:1},  \ref{tab:2other}). Grain sublimation was modeled following Appendix~\ref{appendix:dust-sub}, considering the carbon/ice-rich and ice-poor mixtures presented in Sect.~\ref{sec:4.2}. Grains composed of olivine are too cold to produce significant amounts of water vapor (Fig.~\ref{fig:grains}a). The particle size distribution follows a power law  $n$($a$) $\propto$ $a^{\alpha}$, where $\alpha$ is the size index, and the particle radius takes values from $a_{\rm min}$ to $a_{\rm max}$. We ran the model with various sets  of parameters for the size distribution and the outburst duration. A small subset of the model results is given in Fig.~\ref{fig:density}, where the outburst duration is set to two days.

At $\Delta T_{\rm outburst}$ = 3 d from onset, all the molecules released by the outburst are at distances less than the radius of the FOV ($\sim$ 8.0 $\times$ 10$^4$ km). Therefore $Q_{\rm dust}$ can be readily estimated from the radial density profiles corresponding  to $\Delta T_{\rm outburst}$ = 3 d (leftmost curves in Fig.~\ref{fig:density}) to reproduce the number of molecules detected in the HIFI beam on 19.05 April 2010 (estimated as $\sim$ 10$^{33}$ molecules from the nucleus production model, Sect.~\ref{sec:H2O-prod}). In Fig.~\ref{fig:density} results are shown for $a_{\rm max}$ = 50 $\mu$m, $\alpha$ = --3.5,  carbon ice-poor and ice-rich mixtures, and two values of $a_{\rm min}$. The inferred $Q_{\rm dust}$ are given in the plot. For $a_{\rm min}$ = 3 $\mu$m,  the derived dust production rate is 1.1$\times$10$^3$ kg/s (ice rich) to 4.3 10$^3$ kg/s (ice poor), that is, $m_{\rm dust}$ of (2.0--7.4)$\times$10$^8$ kg released within two days.  It reaches 6.6$\times$10$^4$ kg/s (ice rich) to 1.7$\times$10$^7$  kg/s (ice poor) for $a_{\rm min}$ = 8 $\mu$m (i.e., $m_{\rm dust}$ of 1.1$\times$10$^{10}$ and 2.9$\times$10$^{12}$ kg, respectively, released within two days). The inferred $Q_{\rm dust}$ increases with increasing $a_{\rm min}$ since the grain temperature decreases with increasing size. $Q_{\rm dust}$ also increases for shallower size distributions: For example, for  $\alpha$ = --3.0, $a_{\rm min}$ = 3 $\mu$m, $a_{\rm max}$ = 50 $\mu$m, $Q_{\rm dust}$ is enhanced by a factor of two with respect to the case $\alpha$ = --3.5. The assumed maximum size  $a_{\rm max}$ also affects the results: For $a_{\rm min}$ = 3 $\mu$m, and $\alpha$ = --3.5,  $Q_{\rm dust}$ increases by a factor of 2.6  when  $a_{\rm max}$ is changed from 50 to 250 $\mu$m. This is an expected result as the largest particles contribute only weakly to water production. The value $a_{\rm max}$ = 250 $\mu$m corresponds to the maximum size that can be lifted from the nucleus of 29P (Sect.~\ref{sec:dustprod}).  In Sect.~\ref{sec:4.2} we show that the shape of the H$_2$O line profile suggests $a_{\rm min}$ $>$ $\sim$ 4 $\mu$m and  $a_{\rm min}$ $>$ $\sim$ 6 $\mu$m when we assume ice-poor and ice-rich particles, respectively.  Using these size constraints, we then derive a confident lower limit to the loss rate of icy particles during outburst D of  $\sim$1.0$\times$10$^4$ kg/s (ice poor) and $\sim$1.5$\times$10$^3$ kg/s (ice rich).

The density profiles at $\Delta T_{\rm outburst}$ = 3 d follow a Haser-type distribution for distances $>$ 2$\times$10$^4$ km (Fig.~\ref{fig:density}), but show a deficit in H$_2$O molecules at smaller distances. Molecules produced at small distances (essentially by small warm enough grains) moved to larger distances in the elapsed time since their production. The inner cutoff in the density profile is a function of the outburst duration and is no longer  observed when the outburst duration is set to a value equal to 3 d (i.e. equal to $\Delta T_{\rm outburst}$).
The calculated $Q_{\rm dust}$ (and $m_{\rm dust}$) does not vary much with the outburst duration when set to a value $\leq$ 3 d.

\begin{figure}[!ht]
\begin{center}
\includegraphics[width=9.0cm]{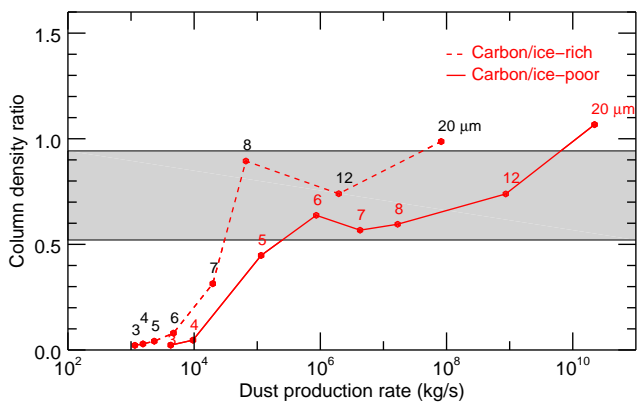}
\end{center}
 \caption{Model outputs for the ratio of the H$_2$O column densities within the HIFI beam at $\Delta T_{\rm outburst}$ = 25 d and  
 $\Delta T_{\rm outburst}$ = 3 d (referred as to $\zeta$ in the text). For comparison, the measured  intensity ratio of 0.73 $\pm$ 0.21 between 11 May 2010 and 19 April 2010 is indicated by the gray region. For all models, the number of molecules within the HIFI beam is 10$^{33}$ molecules at time $\Delta T_{\rm outburst}$ = 3 d, corresponding to the 19.05 April 2010 observation, and the derived dust production is given in the x-axis.  The model parameters are   the outburst duration of 2 d, $a_{\rm max}$  = 50 $\mu$m, and $\alpha$ = --3.5. $a_{\rm min}$ takes different values from 3 to 20 $\mu$m, which are indicated (in $\mu$m units) on the plot above the corresponding model results. Results are shown for both ice-poor ($v_i$ = 0.5) and ice-rich ($v_i$ = 0.1) carbon grains. } \label{fig:evolu}
\end{figure}

At $\Delta T_{\rm outburst}$ = 25 d, the water shell is far away from nucleus center ($>$ 10$^5$ km, rightmost curves in Fig.~\ref{fig:density}), and the total number of water molecules released by the icy grains increases. Only a fraction of them resides in the HIFI line of sight. Figure~\ref{fig:evolu} shows the ratio of the calculated H$_2$O column density within the HIFI beam at $\Delta T_{\rm outburst}$ = 25 d to the value at $\Delta T_{\rm outburst}$ = 3 d (this ratio is referred to as $\zeta$ in the following).  In this figure, $a_{\rm max}$  = 50 $\mu$m, $\alpha$ = --3.5, and $a_{\rm min}$ takes different values from 3 to 20 $\mu$m. The x-axis provides the $Q_{\rm dust}$ values reproducing the HIFI water measurement at $\Delta T_{\rm outburst}$ = 3 d. The measured H$_2$O intensity ratio of $\zeta$ = 0.73 $\pm$ 0.21 is shown with a gray box for comparison.  The calculated H$_2$O column density ratio $\zeta$ globally increases with increasing $a_{\rm min}$.  Figure~\ref{fig:evolu} shows that the model output and the measured evolution of the H$_2$O signal\footnote{Calculations considering the time evolution of H$_2$O excitation once released from grains (Sect.~\ref{sec:exci}), that is that all molecules are not in the same excitation state, lead to similar conclusions.} match well for values of $a_{\rm min}$ higher than typically 5--7 $\mu$m, depending on the ice content.  The $Q_{\rm dust}$ values consistent with the evolution of the H$_2$O signal are then $>$ 2$\times$10$^{4}$ kg/s (ice rich, $m_{\rm dust}$ $>$   3.5$\times$10$^{9}$ kg) and $>$ 1$\times$10$^{5}$ kg/s (ice poor, $m_{\rm dust}$ $>$  2$\times$10$^{10}$ kg).  The limiting $a_{\rm min}$ values consistent with the evolution of the H$_2$O signal are slightly higher than those obtained from the H$_2$O line shape ($a_{\rm min}$ $>$ 4--6 $\mu$m, Sect.~\ref{sec:4.2}). 
   
Outburst D was followed by minor outburst E on 5.5 May 2010. In addition, the activity of 29P remained above the quiescent value in the time interval between outburst D and the 11 May observation (Fig.~\ref{fig:CO-mv-time}). Both outburst E and this continuous activity possibly contributed to the water molecules detected on 11 May 2010.  Hence, the masses derived from the evolution of the H$_2$O signal might be overestimated.

In conclusion, the HIFI observations of water on 19 April 2010 suggest a $Q_{\rm dust}$ lower limit for outburst D ejecta of 1.5$\times$10$^3$ kg/s (2.6$\times$ 10$^8$ kg in two days). Compared with the excess of CO production (2 $\times$10$^{28}$ s$^{-1}$) related to the outburst, the inferred lower limit for the dust-to-CO production rate ratio (in mass) is about 1.6.  When we use the constraints obtained from the variation of the H$_2$O signal, we obtain $Q_{\rm dust}$/$Q$(CO) $>$ 22 (in mass).  Icy grains released during outbursts might contribute significantly to the water molecules present in 29P coma, even long after an outburst. This might explain the high production rate measured by {\it Akari} nine days after an outburst of moderate amplitude (Sect.~\ref{sec:H2O-prod}).

\section{Analysis of PACS data}
\label{sec:PACS-analysis}

We performed aperture photometry on the PACS 70 and 160 $\mu$m continuum images (Fig.~\ref{fig:PACS-IMAGE}) to provide estimates of the thermal flux detected from the nucleus and the dust coma. For the aperture photometry measurements, we applied two types of aperture corrections, depending on whether the flux within the aperture originated from the nucleus or the coma. The nucleus point-source contribution included aperture corrections based on the encircled energy fraction values presented in Table 7.4 of PACS handbook (version 4.0.1). For the coma, the aperture corrections were determined by comparing aperture photometry measurements of a synthetic 1/$\rho$ coma profile (where $\rho$ is the sky-plane projected cometocentric distance) versus a 1/$\rho$ profile convolved with the PACS point spread function (PSF; \cite{2016yCat..35910117B}). No color corrections were  applied to the measurements. After inspection of Table 7.5 of the PACS handbook, we determined these corrections to be at the $\sim$ 1\% level, well below the dominant uncertainty produced by the coma modeling and removal procedure.

\subsection{Modeling and removing the coma}
To obtain nucleus photometry measurements from the PACS images, the flux from the coma was modeled and removed. We used a well-established modeling technique \citep{lamy-toth_1995, lisse_1999, yan_phd_1999} for this procedure, where the coma brightness distribution with azimuth and radial distance is measured in regions outside of significant contribution from the nucleus PSF in order to generate a synthetic coma model. The flux contribution of the modeled coma is then subtracted from the observations resulting in an approximately bare-nucleus residual image. The PSF models used in the analysis were from \cite{2016yCat..35910117B}. 

\begin{figure*}
\includegraphics[width=6cm,clip]{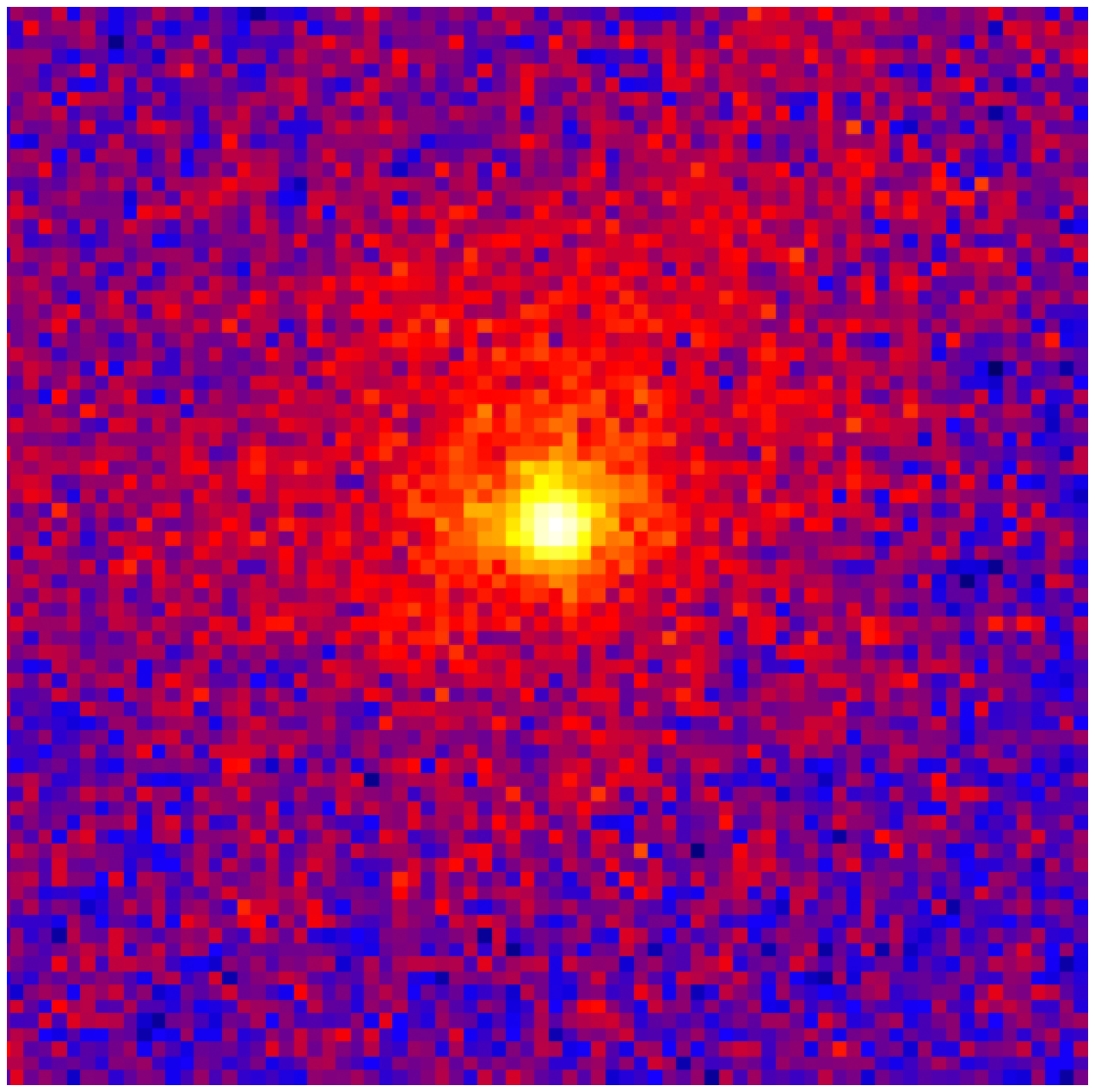}
\includegraphics[width=6cm,clip]{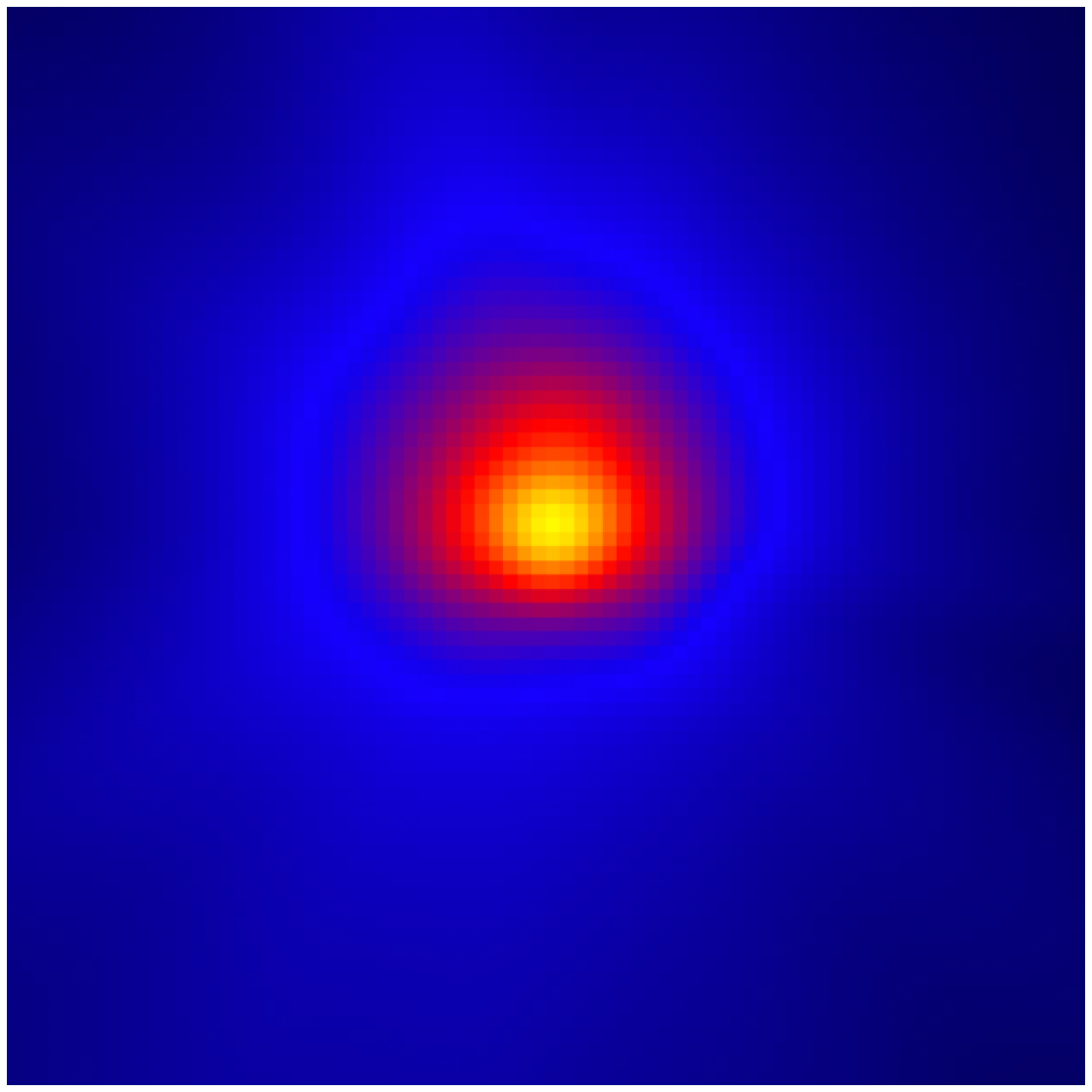}
\includegraphics[width=6cm,clip]{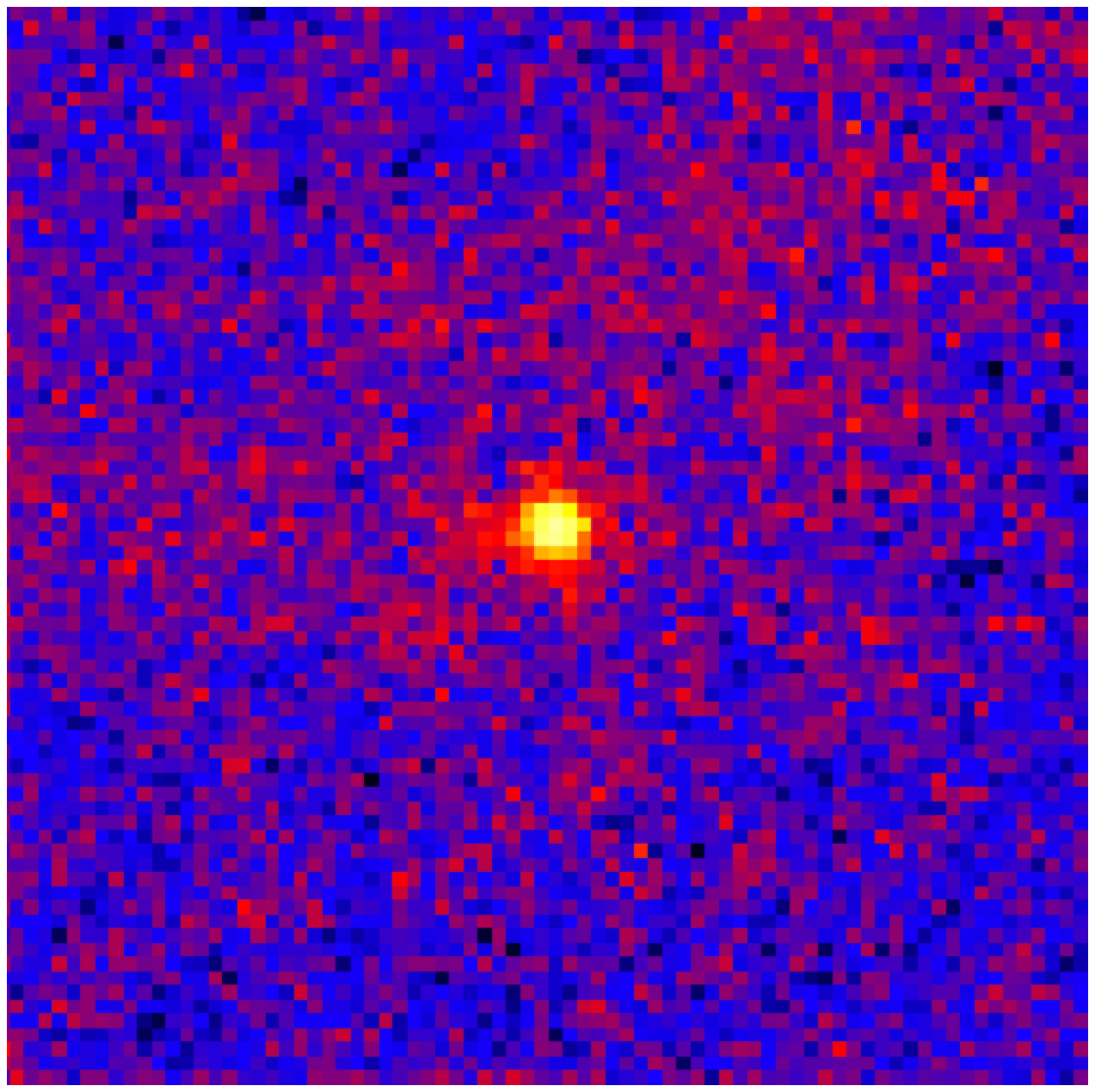}
\caption{PACS image in the 60-85 $\mu$m band from June 10, 2010. From left to right: Observed image ($2'$ $\times$ $2'$, 5.77$\times$10$^5$ km per side; equatorial north is up and east is to the left), the coma model, and the residuals after the coma model was subtracted.  \label{fig:coma_removal}}
\end{figure*}

The coma modeling and removal procedure was applied to each of the three 70 $\mu$m images from the three epochs of PACS data (Table~\ref{tab:1}), resulting in three independent measurements of the spectral flux density of the nucleus that are reported in Table \ref{tab:PACS}. Figure \ref{fig:coma_removal} provides an example of the results of the process.  The quality of the coma removal and nucleus flux extraction process can be seen by the consistent noise pattern present in the residual image (right panel of Fig.~\ref{fig:coma_removal}). The 160 $\mu$m data do not have sufficient detections of extended coma surface brightness for the application of this technique. For the 160 $\mu$m data, we therefore applied a different technique to disentangle the detected nucleus versus coma flux, which is described below.

\subsection{Nucleus thermal emission}
\label{sec:nucleus}

We applied the NEATM model  to each of the three epochs of 70 $\mu$m PACS images. Since extracted nucleus flux density measurements were only possible from the 70 $\mu$m data, our NEATM fits only included the effective radius of the nucleus as a free parameter. A value of $\eta$ = 1.03 was assumed for the beaming factor based on the results of the Survey of Ensemble Physical Properties of Cometary Nuclei \cite[SEPPCoN;][]{2013Icar..226.1138F}. Additionally, we used similar assumptions as SEPPCoN for the bolometric Bond albedo $A = 0.012$ (assuming a visible-wavelength geometrical albedo $p = 0.04$ and phase integral relation $q = 0.290 + 0.684 G$, \citet{2002aste.book..205H}), emissivity $\epsilon = 0.95$, and slope parameter $G = 0.05$. Using these assumptions, we derived three independent estimates of the effective radius of the nucleus that are reported in Table \ref{tab:PACS}. These estimates ($R_N$ between 30.3 and 31.9 km with 10\% uncertainty) are within the uncertainties of the recent values for 29P reported in \citet{2013ApJ...773...22B}  ($R_N$ = 23 $\pm$ 7.5 km)  and \citet{2021PSJ.....2..126S} ($R_N$ = 32.3 $\pm$ 3.1 km) that are based on {\it WISE} and {\it Spitzer} observations, respectively.

\subsection{Thermal emission of the dust coma}
\label{sec:dust-thermal}
The coma modeling of the 70 $\mu$m images yields measurements of the thermal flux emitted from the coma dust grains. We performed aperture photometry to each of the three datasets and provide the results in Table \ref{tab:PACS}. Our approach for separating the nucleus versus coma flux from the 160 $\mu$m data was to calculate the expected 160-$\mu$m NEATM nucleus flux density for each of the three epochs of PACS images using the nucleus radius derived from the 70 $\mu$m data (see Table~\ref{tab:PACS}), and to subtract it from the individual 160 $\mu$m images. The residual flux density after subtraction was attributed to the dust coma and the three values are presented in Table~\ref{tab:PACS}. The extracted coma flux densities are for aperture radii of 10\arcsec~(2010 and 2013 data) and 6\arcsec~(2011 data). A source close to the nucleus of 29P is indeed observed in the 2 January 2011 image (Fig.~\ref{fig:PACS-IMAGE}). 

\citet{2021PSJ.....2..126S} measured the coma flux density of 29P at 16, 24, and 70 $\mu$m using {\it Spitzer} observations undertaken on 23--24 November 2003 ($r_{\rm h}$ = 5.73 au, $\Delta$= 5.54 au). The uncertainty was large at 70 $\mu$m, but the measured value (102 $\pm$ 50 mJy in a 9\arcsec~radius aperture) is consistent with the {\it Herschel} measurements.
 
 The 70 $\mu$m image obtained on 10 June 2010 is more extended that those obtained on 2 January 2011 and 17 February 2013 (Fig.~\ref{fig:PACS-IMAGE}). A possible explanation is the presence of residual ejecta from the May 2010 outbursts (E and F) and possibly from the more productive April 2010 outburst D (Tables~\ref{tab:1} and \ref{tab:outburst}). These outbursts occurred between 17 to 46 days before the acquisition of the 10 June 2010 image, whereas the two other PACS images were obtained more than 42 and 84 days after a significant outburst. The average size of the outermost isophote ($\sim$ 1.2 10$^5$ km, Fig.~\ref{fig:PACS-IMAGE}, top left) implies projected dust velocities between 30--80 m/s, depending on which outburst is considered. 

\begin{table*}[t!]
\caption{PACS photometry results. \label{tab:PACS}}
\begin{center}
\begin{tabular}{cccccc}
\hline\hline\noalign{\smallskip}
UT Date & 70-$\mu$m Nuc. Flux$^{a}$ & 160-$\mu$m Nuc. Flux$^{b}$ & NEATM Nuc. Radius & 70-$\mu$m Coma Flux$^{c}$ &
160-$\mu$m Coma Flux$^{c}$ \\
(yyyy/mm/dd) & (mJy) & (mJy)  & (km) & (mJy) & (mJy) \\
\hline\noalign{\smallskip}
2010/06/10 & 102 $\pm$ 10 & 31 & 30.8 $\pm$ 3 & 146 $\pm$ 10 (10\arcsec) & 45 $\pm$ 10 (10\arcsec)   \\
2011/01/02 & 128 $\pm$ 10 & 39  & 30.3 $\pm$ 3 & 93 $\pm$ 10 (6\arcsec~) & 36 $\pm$ 10 (6\arcsec)  \\
2013/02/17 & 140 $\pm$ 10 & 43 & 31.9 $\pm$ 3 & 71 $\pm$ 10 (10\arcsec~) & 14 $\pm$ 5 (10\arcsec) \\
\hline
\end{tabular}
\end{center}

$^{a}${ The aperture-corrected total nucleus flux density measured in PACS 70-$\mu$m images.}
$^{b}${ 160 $\mu$m NEATM derived nucleus flux estimate based on the best-fit nucleus radius value derived from the 70 $\mu$m image analysis.}
$^{c}${ The radius of the photometric aperture is specified in brackets. The 2011 data used a smaller aperture due to the presence of a source close to the nucleus.}
\end{table*}

\subsection{Dust production rate}
\label{sec:dustprod}

To determine the dust production rate $Q_{\rm dust}$, we followed the approach used by \citet{2021PSJ.....2..126S} to analyze  {\it Spitzer} observations of the dust coma of 29P (see their Sect. 3.1.3). We applied the same model parameters. In summary, the model computes the thermal emission of an ensemble of particles defined by its size distribution, which is described by a power law  $n$($a$) $\propto$ $a^{\alpha}$, where $\alpha$ is the size index and the particle radius takes values from $a_{\rm min}$ to $a_{\rm max}$.
The maximum size that can be lifted from the surface of the nucleus of 29P is estimated to be $a_{\rm max}$ = 250 $\mu$m, for a CO-driven activity  restricted to a cone with a half-angle of 45$^{\circ}$ and a total CO production rate of 4 $\times$ 10$^{28}$ s$^{-1}$ and a 30 km radius nucleus \citep{2018Icar..312..121Z,2021Icar..35414091Z}.  The wavelength-dependent absorption coefficient and temperature of the dust particles was computed as a function of grain size using the Mie theory combined with an effective medium theory in order to consider mixtures of different materials following \citet{2017MNRAS.469S.443B} (see also Appendix~\ref{appendix:dust-temp}). We considered the two icy mixtures studied by \citet{2021PSJ.....2..126S}: 1) a matrix of crystalline ice with inclusions of amorphous carbon, and 2) a matrix of amorphous carbon with inclusions of crystalline ice. For the two mixtures,  the ice fraction by mass is $\sim$ 45\%. The dust temperatures inferred for mixture 1 are very similar to those of the ice-poor (29\% by mass) mixture considered in Sect.~\ref{sec:4.2}, whereas the grain temperatures for mixture 2 are intermediate between the temperatures of ice-rich and ice-poor grains shown in Fig.~\ref{fig:grains}. The dust density is taken equal to 500 kg/m$^3$. The dust velocity as a function of particle size varies $\propto$ $a^{-0.5}$, with a value of 60 m/s for 10-$\mu$m particles. The model output is the coma flux density for a given circular aperture and  wavelength.

In Table~\ref{tab:dust} we present $Q_{\rm dust}$ values derived from the measured 70 $\mu$m flux densities. Only results for particles made of amorphous carbon with inclusions of crystalline ice are given, because very similar results are obtained for the other icy mixture.  The values in Col.1 provide results for a range of $a_{\rm min}$ and size index. In Col. 2, only the size distributions providing a flux density ratio $F_{\rm 70}$/$F_{\rm 160}$ consistent with the observations (taking into account the large uncertainties at 160 $\mu$m) are considered. Here, $F_{\rm 70}$ and $F_{\rm 160}$ refer to the measured coma flux density at 70 and 160 $\mu$m, respectively. In Col. 3, we use the size distributions consistent with the flux density ratio $F_{\rm 16}$/$F_{\rm 24}$ measured by {\it Spitzer} \citep{2021PSJ.....2..126S}. The range of inferred dust production rates is similar for these three cases because the dust size distribution is poorly constrained by the {\it Spitzer} and {\it Herschel} data. For shallow size distributions, the derived $Q_{\rm dust}$ values are strongly dependent on the assumed maximum particle size. For example, the upper range of $Q_{\rm dust}$ values in Table ~\ref{tab:dust} is increased by 50$\%$ when we assume $a_{\rm max}$ = 500 $\mu$m.

The dust production rates measured from the {\it Herschel} 2010 and 2011 data ($\sim$ 60--120 kg s$^{-1}$) are similar to the values derived from the {\it Spitzer} 2003 data \citep{2021PSJ.....2..126S}. However, the PACS data indicate that comet 29P was a factor of 2.5 less productive in dust at the time of the 2013 {\it Herschel} observation. This low dust activity is not observed in the optical data. 29P was in a quiescent state during the three {\it Herschel}/PACS measurements with very similar nuclear magnitudes (Table~\ref{tab:1}).

\begin{table}[t!]
\caption{Dust production rates. \label{tab:dust}}
\begin{center}
\begin{tabular}{lccc}
\hline\hline\noalign{\smallskip}
UT Date & \multicolumn{3}{c}{Dust Production rate}\\
(yyyy/mm/dd) & \multicolumn{3}{c}{(kg s$^{-1}$)}\\
 \cline{2-4} \\
 & (1) & (2) & (3) \\
\hline\noalign{\smallskip}
2010/06/10 & 67--116 & 75--115 & 72--100 \\
2011/01/02 & 58--108 & 66--107 & 67--93 \\
2013/02/17 & 27--49 & 27--45 & 30--42 \\
\hline
\end{tabular}
\end{center}

Note: Results for dust particles made of a matrix of amorphous carbon with inclusions of crystalline ice, with an ice content of 45\% in mass. Column (1): Range of dust production rates inferred from the flux density at 70 $\mu$m for $a_{\rm min}$ = 0.5 to 10 $\mu$m, size index $\alpha$ from --4.5 to --2.5, and $a_{\rm max}$ = 250 $\mu$m. Column (2): Same as (1), but for $a_{\rm min}$ and $\alpha$ values providing a 70 $\mu$m to 160 $\mu$m flux density ratio $F_{\rm 70}$/$F_{\rm 160}$ consistent with the observations. Column (3): Same as (1),  but for $a_{\rm min}$ and $\alpha$ values providing a 16 $\mu$m to 24 $\mu$m flux density ratio $F_{\rm 16}$/$F_{\rm 24}$ consistent with the {\it Spitzer} observations undertaken on 23--24 November 2003 \citep{2021PSJ.....2..126S}.
\end{table}

\section{SPIRE data}
\label{sec:SPIRE-analysis}

 The images obtained on 10 June 2010 with the SPIRE photometer show a marginal signal at the position of  comet 29P, against a background that is crowded by astronomic sources (Fig.~\ref{fig:SPIRE-IMAGE}).
Based on the 70 $\mu$m PACS analysis (Sect.~\ref{sec:PACS-analysis}), the estimated thermal fluxes from the 29P nucleus in the SPIRE photometer bandpasses during the observations are 14 mJy (250 $\mu$m), 8 mJy (350 $\mu$m), and 4 mJy (500 $\mu$m). The dust fluxes, estimated from the coma PACS 70$\mu$m fluxes measured on the same date are 16 (250 $\mu$m), 11 (350 $\mu$m), and 7 mJy/beam (500 $\mu$m). The expected nucleus+coma fluxes are accordingly 30, 19, and 11 mJy/beam at 250, 350, and 500 $\mu$m respectively. 
 The measured signals at the position of the comet are 27.3$\pm$9.0, 9.0$\pm$7.5 and 9.0$\pm$10.8 mJy/beam at 250, 350 and 500 $\mu$m, respectively (we did not apply any color corrections). Therefore, this is consistent with the predictions. We have also to take into consideration the confusion limit of 5.8, 6.3, and 6.8 mJy/beam, at 250, 350 and 500 $\mu$m, respectively.  


\section{Summary}
\label{sec:summary}

Comet 29P is a fascinating object for understanding distant cometary activity and evolutionary processes that affect the surface and interior of Centaurs. Its distant orbit makes investigations of the composition of its atmosphere quite challenging. 

We used the HIFI and PACS instruments of the {\it Herschel} space observatory to observe the
H$_2$O 1$_{10}$--1$_{01}$ (557 GHz) and the NH$_3$ 1$_0$--0$_0$ (573 GHz) lines, and to image the coma at 70 and 160 $\mu$m.  {\it Herschel}/SPIRE images at 250, 350 and 500 $\mu$m were also acquired. Observations with the IRAM 30 m telescope were performed to monitor the CO production rate $Q$(CO) and to search for HCN, including at the time of the H$_2$O observations. HIFI and IRAM observations were performed soon after outbursts or during quiescent states. The following main results were obtained:
 
 \begin{itemize}
 \item CO production rates in the range (2.9--5.6) $\times$ 10$^{28}$ s$^{-1}$ (1400--2600 kg s$^{-1}$) are measured.
 
 \item A correlation between the CO production rate and dust brightness is observed (i.e., a higher CO production rate when the coma brightness is higher, e.g., at time of outbursts), with a regression slope between log$_{10}$($Q$(CO)) and the reduced nuclear magnitude $m_{\rm R}(1,r_{\rm h},0)$ equal to --0.062. During the quiescent states, the CO production rate is $\sim$ 3.0 $\times$ 10$^{28}$ s$^{-1}$ (1400 kg s$^{-1}$). From the comparison with Hale-Bopp activity at 6 au from the Sun, we showed that the dust activity of 29P (but not the gas activity) is quenched in the regions responsible for the quiescent activity, likely as a result of surface evolutionary processes induced by activity.
 
 \item We found a correlation between the excess of CO production and the excess of dust brightness  with respect to quiescent values. The correlation equation  
(${\rm log}_{\rm 10}(Q_{\rm out}(\rm CO) = 29.98 - 0.17 m_{\rm R, out}(1,r_{\rm h},0)$, considering IRAM data) is close to that established for the continuous activity of comet Hale-Bopp.  This is consistent with a similar dust-to-gas flux ratio in the outburst ejecta of 29P and in the coma of Hale-Bopp (referring to dust particles that contribute to the scattering cross-section).

 \item The similarity of the CO line profiles during outburst and quiescent phases confirms that outbursts occur in the subsolar region, where CO outgassing predominantly and continuously operates.
 
 \item The water line was detected on 19 April and 11 May 2010. Assuming near-nucleus production, the derived production rates $Q$(H$_2$O) are (4.6 $\pm$ 0.8) $\times$ 10$^{27}$ s$^{-1}$ and (3.5 $\pm$ 0.9) $\times$ 10$^{27}$ s$^{-1}$, respectively (about 120 kg s$^{-1}$). The mean $Q$(H$_2$O)/$Q$(CO) ratio is 10.0 $\pm$ 1.5\% and is similar to the value derived from {\it Akari} infrared data. The water line was not detected on 30 December 2010, and the derived 3-$\sigma$ upper limit is 
$Q$(H$_2$O)/$Q$(CO) $<$ 8\%.

\item HCN is  identified for the first time in the atmosphere of 29P. The relative production rates for the April-May 2010 period are $Q$(HCN)/$Q$(CO) = (0.12 $\pm$ 0.03)\% and $Q$(HCN)/$Q$(H$_2$O) = (1.2 $\pm$ 0.3)\% on average. The HCN abundance relative to water is a factor of 10 higher than values found in comets at 1 au the Sun.

\item NH$_3$ was not detected.  The derived 3-$\sigma$ upper limit for the average of
April and May 2010 data is 4.5 $\times$ 10$^{27}$ s$^{-1}$, leading to  $Q$(NH$_3$)/$Q$(CO) $<$ 10\% and $Q$(NH$_3$)/$Q$(H$_2$O) $<$ 110\%.

\item The H$_2$O and HCN lines are narrow and symmetric in the comet rest velocity frame, and strongly differ in shape from the CO line. The small (at most) velocity offset observed for the H$_2$O and HCN lines indicates that the nucleus contributes little to the production of these molecules which are instead released from sublimating icy grains. The characteristics of the  H$_2$O and HCN line profiles suggest that they are produced from dust particles that exceed a few micrometers in size. 

\item  The H$_2$O observations of 19 April 2010 and 11 May 2010 were obtained a few days to a few weeks after the major outburst D of 16.8 April 2010. Assuming an outburst duration of two days, a size index $\alpha$ = --3.5, and a maximum particle size of 50 $\mu$m in the dust ejecta, we showed from modeling that the weak decrease in the H$_2$O signal that is observed between the two dates can be explained if water is sublimating from large ($>$ 5--7 $\mu$m) icy carbon-grains, in line with the size constraints obtained from the H$_2$O line profile ($>$ 4--6 $\mu$m).  The lower limit for the mass of the icy ejecta is 2 $\times$ 10$^4$ kg/s, which corresponds to a dust-to-CO production rate ratio (in mass) $>$ 22 for the outburst. The calculations do not consider the minor outburst E that occurred on 5 May 2010. A  conservative lower limit of the dust-to-CO production rate ratio of 1.6 was obtained by considering only the  19 April 2010 data. 
 
\item Despite different production mechanisms, H$_2$O and CO productions are correlated, as suggested by the constant ratio of H$_2$O/CO line areas.  

\item We analyzed  the PACS 70 and 160-$\mu$m images to provide estimates of the thermal flux detected from the nucleus and the dust coma. For the 70-$\mu$m images, the relative contributions of the two components were extracted. The NEATM model applied to the measured nucleus 70-$\mu$m flux density allowed us to derive three independent estimates of the nucleus radius (on the order of 31 $\pm$ 3 km), which agree with recently published values based on {\it WISE} and {\it Spitzer} data \citep{2013ApJ...773...22B,2021PSJ.....2..126S}. This might suggest that 29P is an approximately spherical body. 

\item  The SPIRE images show marginal detections of the 29P thermal continuum.

\item We obtained three measurements of the dust production rate during  the quiescent state. The dust mass-loss rate was estimated to be in the range 60--120 kg s$^{-1}$ on 10 June 2010 and 2 January 2011, but a factor of 2.5 lower on 17 February 2013. The dust-to-gas production rate ratio in mass is thus $<$ 0.1  during quiescent phases. 

\item  An important finding of our study is the presence of strong local heterogeneities on the surface of 29P, with quenched dust activity from most of the surface, but not in outbursting regions.

 \end{itemize}
  
In the near future, the James Webb space telescope will provide the opportunity to investigate  the activity and atmospheric composition of comet 29P in unprecedented detail. Not only CO and water, but other species that possibly contribute to its distant activity will hopefully be revealed.

\begin{acknowledgements}
HIFI has been designed and built by a consortium of institutes and
university departments from across Europe, Canada and the United
States (NASA) under the leadership of SRON, Netherlands Institute
for Space Research, Groningen, The Netherlands, and with major
contributions from Germany, France and the US. Consortium members
are: Canada: CSA, U.Waterloo; France: CESR, LAB, LERMA, IRAM;
Germany: KOSMA, MPIfR, MPS; Ireland, NUI Maynooth; Italy: ASI,
IFSI-INAF, Osservatorio Astrofisico di Arcetri-INAF; Netherlands:
SRON, TUD; Poland: CAMK, CBK; Spain: Observatorio Astron\'omico
Nacional (IGN), Centro de Astrobiolog\'ia (CSIC-INTA). Sweden:
Chalmers University of Technology - MC2, RSS \& GARD; Onsala Space
Observatory; Swedish National Space Board, Stockholm University -
Stockholm Observatory; Switzerland: ETH Zurich, FHNW; USA:
Caltech, JPL, NHSC. 

PACS has been developed by a consortium of institutes led byMPE (Germany) and including UVIE (Austria); KUL, CSL, IMEC (Belgium);CEA,  OAMP  (France);  MPIA  (Germany);  IFSI,  OAP/AOT,  OAA/CAISMI,LENS,  SISSA  (Italy);  IAC  (Spain).  This  development  has  been  supported  by the funding agencies BMVIT (Austria), ESA-PRODEX (Belgium), CEA/CNES(France),  DLR  (Germany),  ASI  (Italy),  and  CICYT/MCYT  (Spain). 

IRAM  is  supported  by  the  Institut  National  des Sciences  de  l'Univers  (INSU)  of  the  French  Centre  national  de  la  recherche scientifique  (CNRS),  the  Max-Planck-Gesellschaft  (MPG,  Germany)  and  the Spanish  IGN  (Instituto  Geogr\'afico  Nacional).  We  gratefully  acknowledge  the support from the IRAM staff for its support during the observations.

Part of this research was carried out at the Jet Propulsion Laboratory, California Institute of Technology, under a contract with the National Aeronautics and Space Administration. MdVB acknowledges partial support from grants NSF
AST-1108686 and NASA NNX12AH91H. S.S. was supported by polish
MNiSW funds (181/N-HSO/2008/0). V.Z. was supported by the Italian Space Agency (ASI) within the ASI-INAF agreements I/032/05/0 and I/024/12/0.

We thank the amateur astronomers who provided optical support to the radio observations of 29P and supplied magnitudes to the MPC, LESIA and spanish comet databases.

\end{acknowledgements}

{}

\begin{appendix}
\renewcommand{\appendix}{%
  \par
  \setcounter{section}{0}%
  \renewcommand{\thesection}{\thechapter.\Alph{section}}%
}

\onecolumn
\section{SPIRE images}

\FloatBarrier

\begin{figure*}[!htb]  
\centering
\begin{minipage}[t]{9.5cm}
\includegraphics[width=9.5cm]{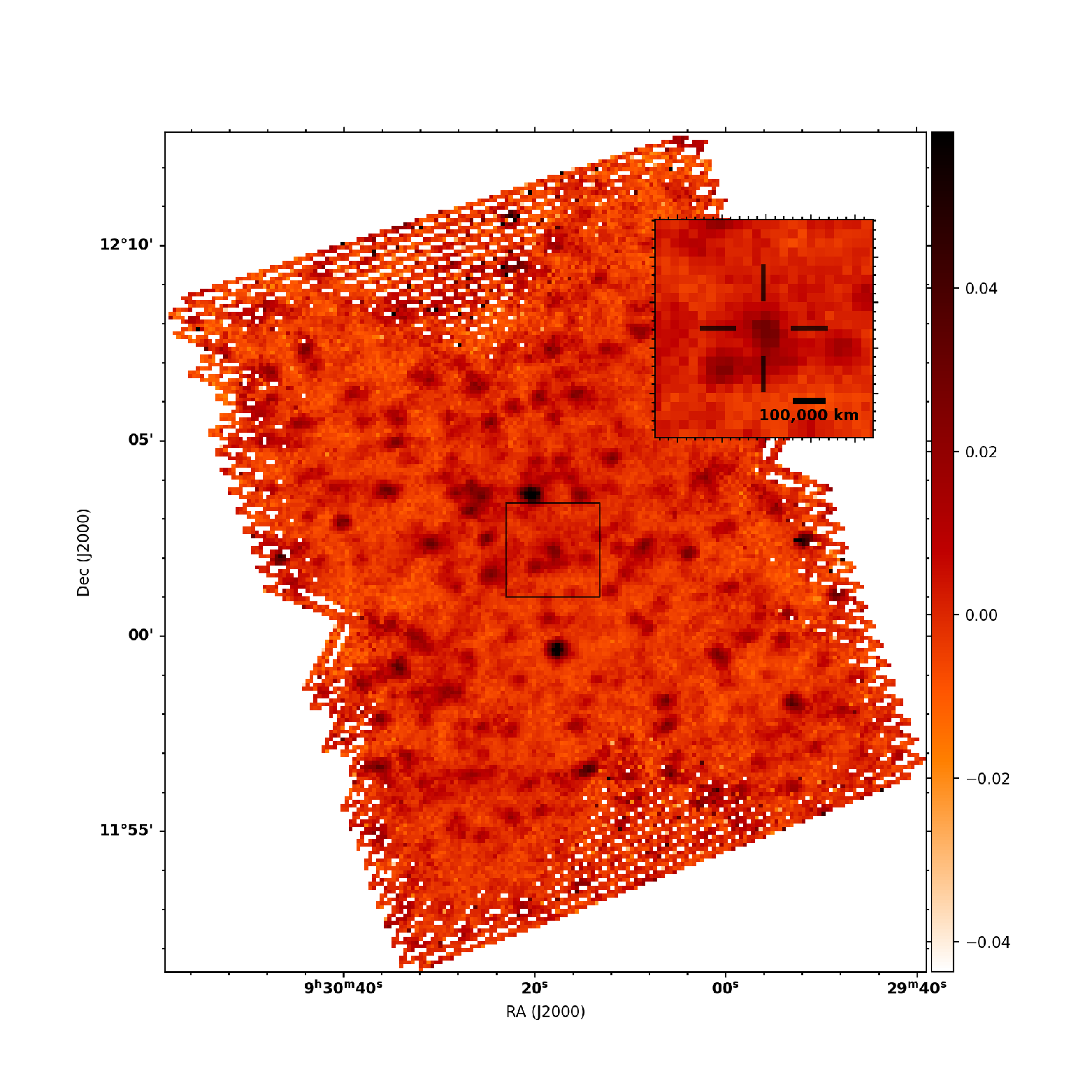}
\vspace*{-62ex}
\begin{center}
250 $\mu$m
\end{center}
\end{minipage} \hfill
\hspace{-0.8cm}
\begin{minipage}[t]{9.5cm}
\includegraphics[width=9.5cm]{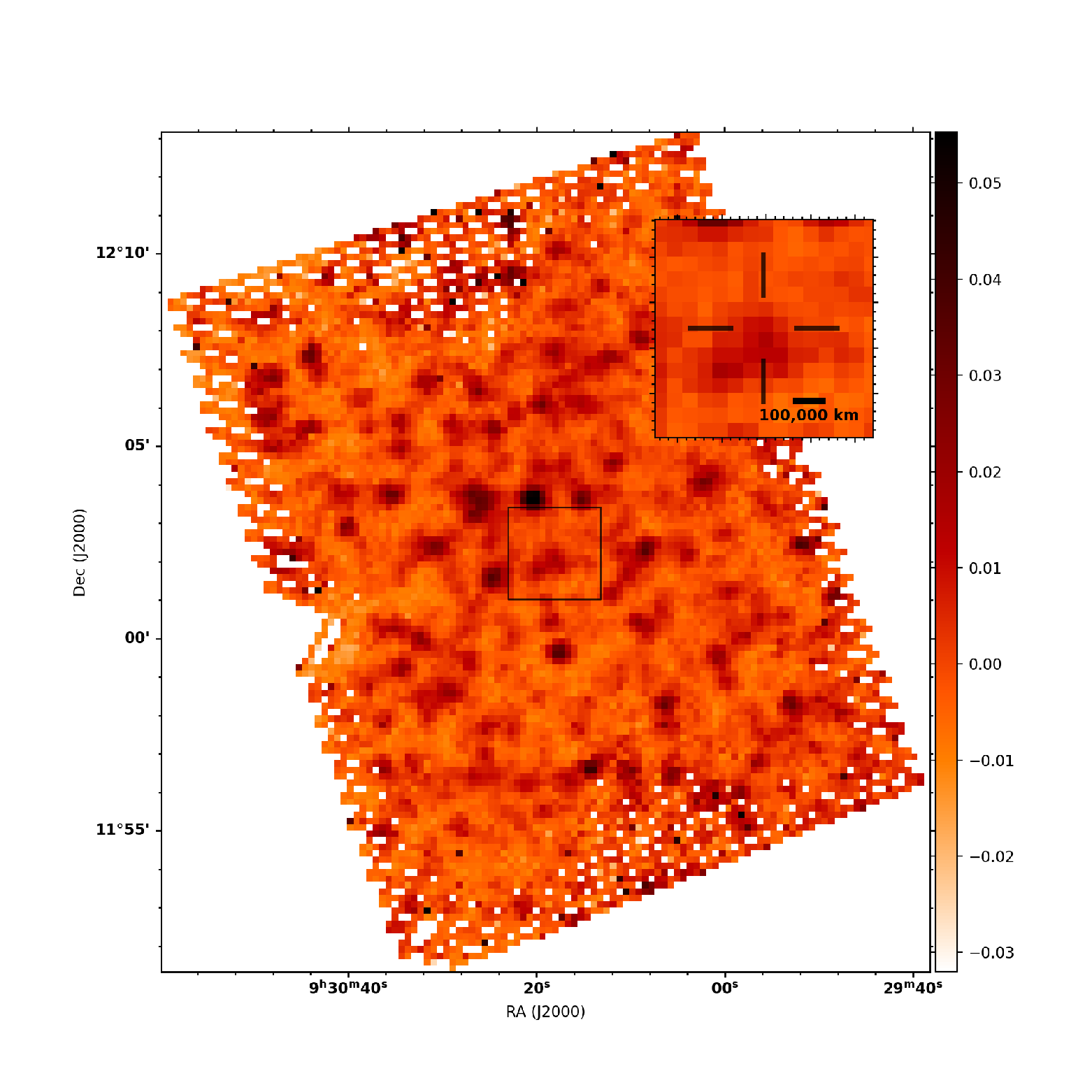}
\vspace*{-62ex}
\begin{center}
350 $\mu$m
\end{center}
\end{minipage} 
\begin{minipage}[t]{9.5cm}
\includegraphics[width=9.5cm]{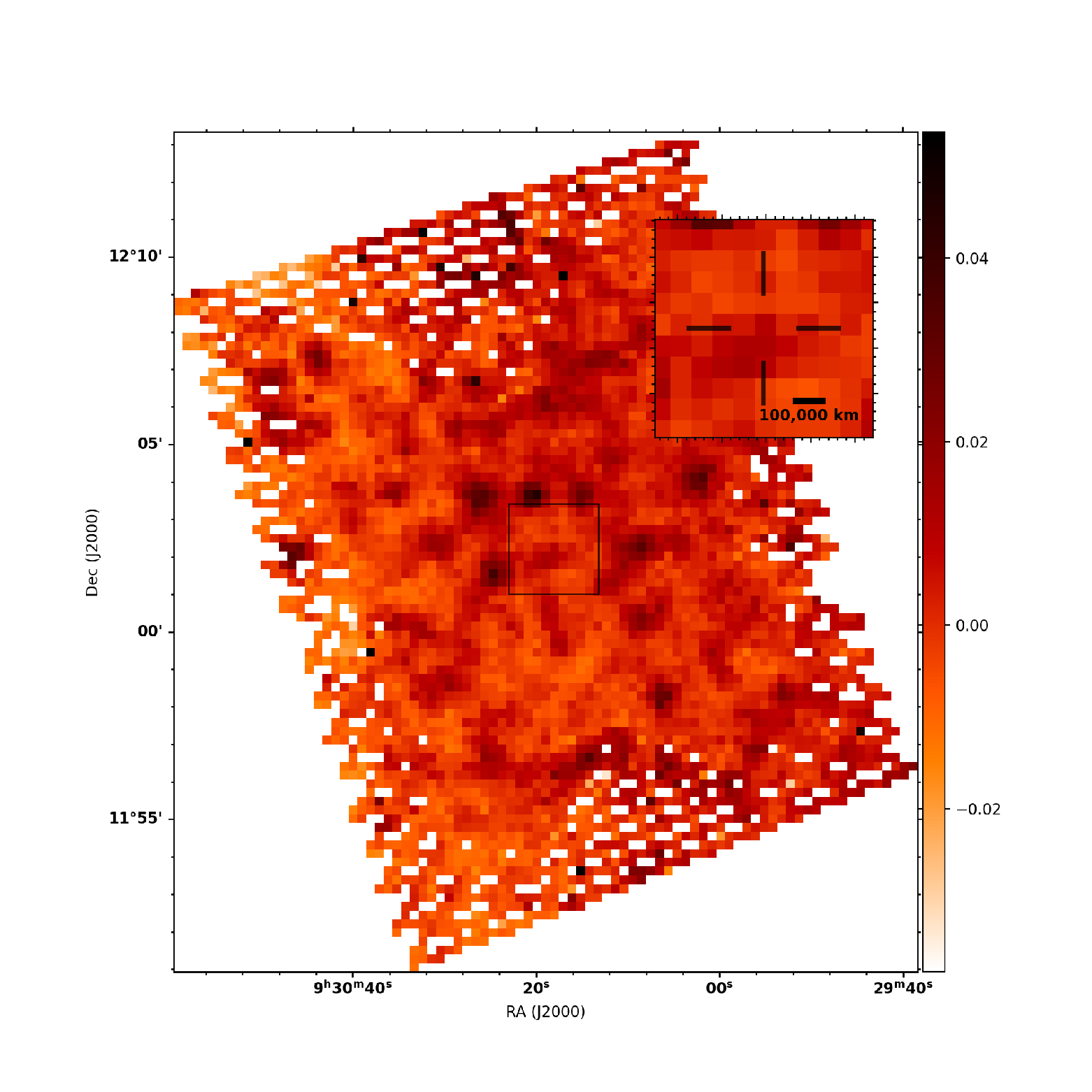}
\vspace*{-62ex}
\begin{center}
500 $\mu$m
\end{center}
\end{minipage}
\caption{SPIRE images of 29P in 250 (top left), 350 (top right) and 500 (bottom) $\mu$m filters obtained on 10 June 2010. Flux is given in Jy/beam (color bar). The insets show a zoom of the images centered on the comet position. Negative pixel values are the result of the local background subtraction.}
\label{fig:SPIRE-IMAGE}
\end{figure*}

\twocolumn
\clearpage

\section{C/1995 O1 (Hale-Bopp): Correlation between nuclear magnitude and CO production rate}
\label{appendix-HB}

\citet{2021PSJ.....2...17W} presented correlation equations between visual magnitudes and CO production rates for comet C/1995 O1 (Hale-Bopp),  
\begin{equation} 
\begin{aligned}
{\rm log}_{\rm 10}(Q(\rm CO)) & = 29.9 - 0.24 m_{\rm h} \\
& = 29.9 - 0.24(m_{\rm v} -5 {\rm log}_{\rm 10}(\Delta)),
\end{aligned}
\end{equation}
\noindent
where $m_{\rm h}$ is the heliocentric magnitude, and $m_{\rm v}$ is the total visual magnitude. A slightly different correlation is found when a phase correction is applied to the magnitudes (as done in Sect.~\ref{sec:optical} for 29P) and systematic differences between observers are corrected, \citep{2021PSJ.....2...17W}
\begin{equation} 
{\rm log}_{\rm 10}(Q(\rm CO))  = 29.71 - 0.22 m_{\rm shift}.
\label{eq:mv-co-HB}
\end{equation}

For 29P, the correlation equation for 29P involves the nuclear magnitude. An additional correction factor must therefore be applied to Eq.~\ref{eq:mv-co-HB} to compare the correlations of 29P and Hale-Bopp (nuclear magnitudes are not available for Hale-Bopp).

For a steady-state dust production, assuming a 1/$\rho$ decrease in coma brightness with projected distance $\rho$ from comet center, the integrated magnitude in an aperture of radius $\rho$ varies as
\begin{equation}
m(\rho) =  m(\rho_0) - 2.5{\rm log}_{\rm 10}(\frac{\rho}{\rho_0}).
\label{eq:mv-co-HB-2}
\end{equation}
At a distance of $\approx$ 6 au from the Sun, in October--November 1995 and
July--September 1998, the Hale-Bopp coma was 2--3\arcmin~in diameter, so that we estimate that the nuclear magnitudes in  a 10\arcsec~diameter aperture should have been  $+3\pm0.2$ magnitude fainter than the total magnitudes $m_{\rm v}$ used for the correlation with the CO production rate (on the other hand,
when the comet was twice as close to the Sun and Earth its coma was 10$\pm$5\arcmin~in diameter, and the nuclear magnitude in  a 20\arcsec~diameter aperture should have been
about 3.5$\pm$0.6 magnitudes fainter than $m_{\rm v}$).
As a consequence the correlation between $Q_{\rm CO}$
and total heliocentric magnitudes for Hale-Bopp (Eq.~\ref{eq:mv-co-HB-2} can  approximately be 
extrapolated to a correlation with nuclear 
heliocentric magnitudes (within an aperture diameter of $\sim$10\arcsec) by adding +0.8$\pm$0.2 to the constant parameter,
\begin{equation} 
{\rm log}_{\rm 10}(Q(\rm CO))  = 30.5(\pm 0.2) - 0.22 m_{\rm R}(1,r_{\rm h},0).
\label{eq:mR-co-HB}
\end{equation}

\noindent
At 6 au from the Sun, the total visual magnitude corrected from geocentric distance and phase of comet Hale-Bopp was $\sim$ 6 \citep{2020AJ....159..136W}, so that$m_{\rm R}(1,r_{\rm h},0)$ at 6 au is estimated to $\sim$ 9.

\section{Dust models}
\label{appendix:dust}

\subsection{Dust velocity} 
\label{appendix:dust-vel}
The dust velocity as a function of grain radius $a$ is computed following
\citet{1997Icar..127..319C},
\begin{equation}
v_{\rm a} = \frac{v_{\rm g}}{1.2 + 0.72 (a/a^*)^{0.5}}
\end{equation}
\noindent
with
\begin{equation}
a^* = \frac{m_{\rm CO}~Q_e(\rm CO)}{4\pi~r_{\rm N}~\rho_{\rm
 d}~v_{\rm g0}},
\end{equation}
\noindent where $v_g$ is the CO terminal velocity, taken equal to
0.5 km s$^{-1}$ (i.e. similar to the value measured in the jet component of the CO spectrum), and $v_{\rm g0}$ is the CO velocity at the
nucleus surface,
\begin{equation}
v_{\rm g0} = \sqrt{\frac{\gamma_{\rm CO} k_B T_N }{m_{\rm CO}}}.
\end{equation}

\noindent $r_N$ and $T_N$ are the nucleus radius and temperature
(assumed to be 30 km and 160 K for 29P). $\rho_d$
is the dust density, taken equal to 500 kg m$^{-3}$ to 
calculate the velocity of dust grains (Fig.\ref{fig:grains})
and dust production rate (Sect.~\ref{sec:dustprod}). $\gamma_{\rm
CO}$ is the heat capacity ratio of CO (= 1.4), $m_{CO}$ is the
mass of one CO molecule, and $Q_e$(CO) is the equivalent CO
production rate defined as $Q_e$(CO) = $Q$(CO) $\times$
4$\pi$/$\Omega$, where $\Omega$ is the solid angle of the jet and
$Q$(CO) is the total production rate.

\subsection{Grain composition and temperature}
\label{appendix:dust-temp}

Icy grains are modeled as constituted of a matrix of crystalline ice incorporating impurities (silicates or carbon). The relative fractions of each component are defined
by the fractional volume of impurities (also referred to as dirt)
$v_{\rm i}$ with respect to the total volume of the grain. The
total ice content by mass is computed assuming densities of 1000
kg m$^{-3}$ and 2500 kg m$^{-3}$ for ice and impurities,
respectively, and a grain porosity $p$ = 0.5. We consider in this
paper values of $v_{\rm i}$ of 0.1, and 0.5, dirt-to-ice mass ratios of 0.28, and 2.5, respectively. For 
$v_{\rm i}$ of 0.1 and 0.5, the ice mass fraction is thus 78\% and 29\%, respectively

The grain temperatures were computed assuming radiative
equilibrium. At 6 au from the Sun, grain temperatures are low, and
cooling by sublimation is negligible with respect to thermal
radiation \citep{Gunnarsson2003,2006Icar..180..473B}. The energy balance involves
grain absorption coefficients $Q_{\rm abs}$, which were computed
using the Mie theory. Refractive indices for mixtures were obtained
from the Maxwell-Garnett effective medium theory following
\citet{green90}. The Maxwell-Garnett effective medium theory was
also used to compute refractive indices of porous grains.
Refractive indices for silicates (namely olivine with 50\% Mg and
50\% Fe), carbon, and ice were taken from \citet{dors95}, \citet{edo83},
and \citet{war08}, respectively. More details on the model can be found in
\citet{2017MNRAS.469S.443B}. 

\subsection{Grain sublimation}
\label{appendix:dust-sub}
The sublimation of dirty icy grains results in an ice mass loss,
\begin{equation}
\frac{dm_{\rm ice}}{dt} = - (1 - v_{i}) 4 \pi a^2  Z_{\rm H_2O}  m_{\rm
H_2O}
\label{eq:A4}
\end{equation}
\noindent  where $m_{\rm H_2O}$ is the mass of one water molecule. We assumed that  
ice and impurities are intimately and homogeneously mixed. Therefore, the surfacic fraction of the ice is equal to the ice volumic fraction of $1-v_{i}$. $Z_{\rm H_2O}$ is the
water sublimation rate by unit of time and surface (m$^{-2}$ s$^{-1}$), derived from
the vapor pressure law of \citet{Fanale1984}, which depends on the grain temperature 
$T_{\rm d}$,
\begin{equation}
Z_{\rm H_2O} = A e^{-B/T_{\rm d}} \sqrt{ \frac{1}{2 \pi m_{\rm H_2O} k_B T_{\rm d}} },
\end{equation}
\noindent
with $A$ = 3.56 10$^{12}$ Pa, and $B$ = 6141.667 K. The temperature of the dust particles as a function of size was computed as described in Sect.~\ref{appendix:dust-temp}.

The variation of the grain radius with time follows
\begin{equation}
\frac{da}{dt}= - \frac{Z_{\rm H_2O}  m_{\rm H_2O}}{\rho_{\rm d}} 
\end{equation}
\noindent
where $\rho_{\rm d}$ is the grain density: 575 and 875 kg m$^{-3}$, for ice-rich ($v_{i}$ = 0.1) and ice-poor ($v_{i}$ = 0.5) grains, respectively, for a porosity of 0.5.
To compute the sublimation rate of the grains as a
function of time, and initial size, we took into account that
grains become hotter when their size diminishes due to sublimation.  The impurities embedded in the ice are released together with the gas as it sublimates. Sublimation stops when water ice is exhausted. This defines the grain-sublimation lifetime. Similar approaches were used by \citet{Gunnarsson2003} and \citet{2006Icar..180..473B}.

\subsection{Dynamics and number density of water molecules sublimated from grains}
\label{appendix:dust-dyn}

In order to interpret the number of water molecules detected
within the HIFI beam (Sect.~\ref{sec:subliming}), it is requisite to describe the dynamics of
the water molecules sublimated from grains.

Our model assumes that, when they are released from grains, H$_2$O
molecules expand radially outward in the coma at a velocity equal
to $v_{\rm H_2O}$ = 0.25 km s$^{-1}$. This corresponds to the H$_2$O
velocity derived from the width of the H$_2$O 557 GHz line profile
(Sect.~\ref{obs:hifi}).

The distance traveled by the molecules at time $t$ is given by
\begin{equation}
l = v_{\rm a}(t_{\rm a}-t_{\rm 0}) + v_{\rm H_2O}(t-t_{\rm
a})
\end{equation}
\noindent where $t_{\rm 0}$ is the time at which the grain is
released from the nucleus, and $t_{\rm a}$ is the time at which the molecule sublimates. $v_{\rm a}$ is the dust velocity defined in Sect.~\ref{appendix:dust-vel}. The number of molecules sublimating at $t_{\rm a}$ from grains of radius $a$ is derived from the ice mass-loss rate (Eq.~\ref{eq:A4}). 

The algorithm was adapted to simulate an outburst described by a boxcar function with the outburst duration as a free parameter. The injected dust particles follow a size distribution. For comparison with the observations, the algorithm computes the water distribution at an elapsed time with respect to outburst onset (at $t$ = 0 s). This was done by computing the number of molecules within spheres of increasing radius (nominally 40 spheres with radii from 10$^3$ to 10$^7$ km, with a logarithmic step). The number density (m$^{-3}$) throughout the coma can then easily be deduced. The expanding dust and H$_2$O
clouds were assumed to be isotropic. The number of molecules within the HIFI beam was computed by volume integration.

\newpage
\onecolumn
\section{Magnitude data}
\label{appendix:mv}

{\small
\begin{longtable}{lcccccc|lcccccc}
\caption{Magnitudes in an aperture with a diameter of 10$\arcsec$ in 2007--2008.}\label{tab:2007}\\
\hline\hline\noalign{\smallskip}
Date & Time & Obs$^a$ & $r_{\rm h}$ & $\Delta$ & $m_{\rm R}$ & $m_{\rm R}(1,r_{\rm h},0)$ & Date & Time & Obs & $r_{\rm h}$ & $\Delta$ & $m_{\rm R}$ & $m_{\rm R}(1,r_{\rm h},0)$ \\
 jj/mm/yyyy & hr:mn & & (au) & (au) & & & jj/mm/yyyy & hr:mn & & (au) & (au)  & & \\
\hline\noalign{\vskip 2mm}
\endfirsthead
\hline\hline\noalign{\smallskip}
Date & Time & Obs & $r_{\rm h}$ & $\Delta$ & $m_{\rm R}$ & $m_{\rm R}(1,r_{\rm h},0)$ & Date & Time & Obs & $r_{\rm h}$ & $\Delta$ & $m_{\rm R}$ & $m_{\rm R}(1,r_{\rm h},0)$ \\
 jj/mm/yyyy & hr:mn &  & (au) & (au)  & & jj/mm/yyyy & hr:mn & & (au) & (au)  & & \\
\hline\noalign{\vskip 2mm}
\endhead
\hline \multicolumn{14}{r}{\textit{Continued on next page}} \\
\endfoot
\hline
\endlastfoot
24/12/2007 & 23:05 &  442 & 5.979 & 5.003 & 16.12 & 12.57 & 30/12/2007 & 00:13 &
  J46 & 5.981 & 5.010 & 12.95 &  9.38 \\
25/12/2007 & 21:10 &  442 & 5.980 & 5.003 & 16.00 & 12.45 & 30/12/2007 & 00:14 &
  J46 & 5.981 & 5.010 & 12.96 &  9.39 \\
25/12/2007 & 21:18 &  213 & 5.980 & 5.003 & 15.90 & 12.35 & 30/12/2007 & 00:15 &
  J46 & 5.981 & 5.010 & 12.95 &  9.38 \\
25/12/2007 & 21:24 &  213 & 5.980 & 5.003 & 15.88 & 12.33 & 30/12/2007 & 00:16 &
  J46 & 5.981 & 5.010 & 12.96 &  9.39 \\
25/12/2007 & 21:25 &  213 & 5.980 & 5.003 & 15.90 & 12.35 & 30/12/2007 & 00:17 &
  J46 & 5.981 & 5.010 & 12.96 &  9.39 \\
25/12/2007 & 21:29 &  213 & 5.980 & 5.003 & 15.90 & 12.35 & 30/12/2007 & 00:18 &
  J46 & 5.981 & 5.010 & 12.95 &  9.38 \\
26/12/2007 & 23:04 &  442 & 5.980 & 5.004 & 16.06 & 12.51 & 30/12/2007 & 00:19 &
  J46 & 5.981 & 5.010 & 12.95 &  9.38 \\
27/12/2007 & 19:28 &  J36 & 5.980 & 5.006 & 15.80 & 12.24 & 30/12/2007 & 00:20 &
  J46 & 5.981 & 5.010 & 12.95 &  9.38 \\
27/12/2007 & 19:36 &  J36 & 5.980 & 5.006 & 15.83 & 12.27 & 30/12/2007 & 00:21 &
  J46 & 5.981 & 5.010 & 12.92 &  9.35 \\
27/12/2007 & 19:37 &  J36 & 5.980 & 5.006 & 15.80 & 12.24 & 30/12/2007 & 00:22 &
  J46 & 5.981 & 5.010 & 12.96 &  9.39 \\
27/12/2007 & 19:47 &  J36 & 5.980 & 5.006 & 15.80 & 12.24 & 30/12/2007 & 00:23 &
  J46 & 5.981 & 5.010 & 12.95 &  9.38 \\
27/12/2007 & 19:53 &  J76 & 5.980 & 5.006 & 16.00 & 12.44 & 30/12/2007 & 00:24 &
  J46 & 5.981 & 5.010 & 12.95 &  9.38 \\
27/12/2007 & 19:58 &  J76 & 5.980 & 5.006 & 16.00 & 12.44 & 30/12/2007 & 00:25 &
  J46 & 5.981 & 5.010 & 12.94 &  9.37 \\
27/12/2007 & 19:59 &  J76 & 5.980 & 5.006 & 15.97 & 12.41 & 30/12/2007 & 00:26 &
  J46 & 5.981 & 5.010 & 12.95 &  9.38 \\
27/12/2007 & 20:04 &  J76 & 5.980 & 5.006 & 16.00 & 12.44 & 30/12/2007 & 00:27 &
  J46 & 5.981 & 5.010 & 12.93 &  9.36 \\
27/12/2007 & 20:16 &  213 & 5.980 & 5.006 & 15.90 & 12.34 & 30/12/2007 & 00:28 &
  J46 & 5.981 & 5.010 & 12.95 &  9.38 \\
27/12/2007 & 20:23 &  213 & 5.980 & 5.006 & 15.91 & 12.35 & 30/12/2007 & 00:29 &
  J46 & 5.981 & 5.010 & 12.94 &  9.37 \\
27/12/2007 & 20:23 &  213 & 5.980 & 5.006 & 15.90 & 12.34 & 30/12/2007 & 00:30 &
  J46 & 5.981 & 5.010 & 12.95 &  9.38 \\
27/12/2007 & 20:31 &  213 & 5.980 & 5.006 & 15.90 & 12.34 & 30/12/2007 & 00:31 &
  J46 & 5.981 & 5.010 & 12.92 &  9.35 \\
28/12/2007 & 20:02 &  J46 & 5.980 & 5.007 & 15.80 & 12.24 & 30/12/2007 & 00:32 &
  J46 & 5.981 & 5.010 & 12.90 &  9.33 \\
28/12/2007 & 20:11 &  J46 & 5.980 & 5.007 & 15.80 & 12.24 & 30/12/2007 & 17:51 &
  B20 & 5.981 & 5.012 & 13.00 &  9.43 \\
28/12/2007 & 20:58 &  J47 & 5.980 & 5.007 & 15.90 & 12.34 & 30/12/2007 & 17:54 &
  B20 & 5.981 & 5.010 & 13.05 &  9.48 \\
28/12/2007 & 21:11 &  J47 & 5.980 & 5.007 & 15.90 & 12.34 & 30/12/2007 & 17:54 &
  B20 & 5.981 & 5.012 & 13.10 &  9.53 \\
28/12/2007 & 21:11 &  J47 & 5.980 & 5.007 & 15.90 & 12.34 & 30/12/2007 & 17:58 &
  B20 & 5.981 & 5.012 & 13.10 &  9.52 \\
28/12/2007 & 21:24 &  J47 & 5.980 & 5.007 & 15.90 & 12.34 & 30/12/2007 & 18:03 &
  213 & 5.981 & 5.012 & 12.90 &  9.32 \\
28/12/2007 & 22:37 &  J38 & 5.980 & 5.007 & 15.80 & 12.24 & 30/12/2007 & 18:05 &
  213 & 5.981 & 5.012 & 13.00 &  9.42 \\
28/12/2007 & 22:40 &  J38 & 5.980 & 5.007 & 15.80 & 12.24 & 30/12/2007 & 18:06 &
  213 & 5.981 & 5.010 & 12.97 &  9.40 \\
28/12/2007 & 22:43 &  J38 & 5.980 & 5.007 & 15.70 & 12.14 & 30/12/2007 & 18:10 &
  213 & 5.981 & 5.012 & 13.00 &  9.42 \\
28/12/2007 & 22:48 &  J38 & 5.980 & 5.007 & 15.80 & 12.24 & 30/12/2007 & 19:55 &
  J46 & 5.981 & 5.010 & 12.95 &  9.38 \\
28/12/2007 & 23:17 &  A06 & 5.980 & 5.007 & 15.90 & 12.34 & 30/12/2007 & 20:03 &
  170 & 5.981 & 5.012 & 13.00 &  9.42 \\
28/12/2007 & 23:21 &  A06 & 5.980 & 5.007 & 15.89 & 12.33 & 30/12/2007 & 20:21 &
  442 & 5.981 & 5.010 & 13.06 &  9.49 \\
28/12/2007 & 23:28 &  A06 & 5.980 & 5.007 & 15.90 & 12.34 & 30/12/2007 & 21:08 &
  X10 & 5.981 & 5.012 & 13.10 &  9.52 \\
28/12/2007 & 23:44 &  A06 & 5.980 & 5.007 & 15.90 & 12.34 & 30/12/2007 & 21:12 &
  X10 & 5.981 & 5.012 & 13.10 &  9.52 \\
29/12/2007 & 21:51 &  213 & 5.981 & 5.010 & 14.00 & 10.43 & 30/12/2007 & 21:13 &
  X10 & 5.981 & 5.010 & 13.09 &  9.52 \\
29/12/2007 & 21:56 &  213 & 5.981 & 5.010 & 13.80 & 10.23 & 30/12/2007 & 21:17 &
  X10 & 5.981 & 5.012 & 13.10 &  9.52 \\
29/12/2007 & 22:06 &  213 & 5.981 & 5.010 & 13.84 & 10.27 & 30/12/2007 & 21:25 &
  X10 & 5.981 & 5.012 & 13.10 &  9.52 \\
29/12/2007 & 22:13 &  213 & 5.981 & 5.010 & 13.78 & 10.21 & 30/12/2007 & 21:50 &
  J36 & 5.981 & 5.012 & 12.90 &  9.32 \\
29/12/2007 & 22:38 &  J46 & 5.981 & 5.010 & 13.41 &  9.84 & 30/12/2007 & 22:00 &
  J36 & 5.981 & 5.010 & 12.96 &  9.38 \\
29/12/2007 & 22:39 &  J46 & 5.981 & 5.010 & 13.38 &  9.81 & 30/12/2007 & 22:10 &
  J36 & 5.981 & 5.012 & 13.00 &  9.42 \\
29/12/2007 & 22:40 &  J46 & 5.981 & 5.010 & 13.39 &  9.82 & 30/12/2007 & 22:11 &
  J36 & 5.981 & 5.012 & 13.00 &  9.42 \\
29/12/2007 & 22:41 &  J46 & 5.981 & 5.010 & 13.38 &  9.81 & 30/12/2007 & 23:19 &
  J38 & 5.981 & 5.012 & 13.10 &  9.52 \\
29/12/2007 & 22:42 &  J46 & 5.981 & 5.010 & 13.38 &  9.81 & 30/12/2007 & 23:21 &
  A06 & 5.981 & 5.012 & 13.00 &  9.42 \\
29/12/2007 & 22:43 &  J46 & 5.981 & 5.010 & 13.33 &  9.76 & 30/12/2007 & 23:22 &
  J38 & 5.981 & 5.012 & 13.10 &  9.52 \\
29/12/2007 & 22:44 &  J46 & 5.981 & 5.010 & 13.35 &  9.78 & 30/12/2007 & 23:25 &
  J38 & 5.981 & 5.012 & 13.10 &  9.52 \\
29/12/2007 & 22:45 &  J46 & 5.981 & 5.010 & 13.33 &  9.76 & 30/12/2007 & 23:27 &
  J38 & 5.981 & 5.010 & 13.06 &  9.48 \\
29/12/2007 & 22:46 &  J46 & 5.981 & 5.010 & 13.35 &  9.78 & 30/12/2007 & 23:27 &
  A06 & 5.981 & 5.010 & 12.97 &  9.39 \\
29/12/2007 & 22:48 &  J46 & 5.981 & 5.010 & 13.30 &  9.73 & 30/12/2007 & 23:40 &
  A06 & 5.981 & 5.012 & 13.00 &  9.42 \\
29/12/2007 & 22:49 &  J46 & 5.981 & 5.010 & 13.28 &  9.71 & 30/12/2007 & 23:58 &
  A06 & 5.981 & 5.012 & 13.00 &  9.42 \\
29/12/2007 & 22:50 &  J46 & 5.981 & 5.010 & 13.26 &  9.69 & 31/12/2007 & 01:03 &
  945 & 5.981 & 5.012 & 13.00 &  9.42 \\
29/12/2007 & 22:51 &  J46 & 5.981 & 5.010 & 13.26 &  9.69 & 31/12/2007 & 01:06 &
  945 & 5.981 & 5.012 & 13.00 &  9.42 \\
29/12/2007 & 22:52 &  J46 & 5.981 & 5.010 & 13.25 &  9.68 & 31/12/2007 & 01:09 &
  945 & 5.981 & 5.012 & 13.10 &  9.52 \\
29/12/2007 & 22:53 &  J46 & 5.981 & 5.010 & 13.23 &  9.66 & 31/12/2007 & 01:09 &
  945 & 5.981 & 5.010 & 13.02 &  9.44 \\
29/12/2007 & 22:55 &  J46 & 5.981 & 5.010 & 13.22 &  9.65 & 31/12/2007 & 01:12 &
  945 & 5.981 & 5.012 & 13.10 &  9.52 \\
29/12/2007 & 22:56 &  J46 & 5.981 & 5.010 & 13.23 &  9.66 & 31/12/2007 & 01:15 &
  945 & 5.981 & 5.012 & 13.00 &  9.42 \\
29/12/2007 & 22:57 &  J46 & 5.981 & 5.010 & 13.21 &  9.64 & 31/12/2007 & 21:46 &
  J40 & 5.981 & 5.014 & 13.40 &  9.82 \\
29/12/2007 & 22:58 &  J46 & 5.981 & 5.010 & 13.20 &  9.63 & 31/12/2007 & 21:47 &
  J40 & 5.981 & 5.014 & 13.20 &  9.62 \\
29/12/2007 & 22:59 &  J46 & 5.981 & 5.010 & 13.19 &  9.62 & 31/12/2007 & 21:49 &
  J40 & 5.981 & 5.012 & 13.30 &  9.72 \\
29/12/2007 & 23:00 &  J46 & 5.981 & 5.010 & 13.20 &  9.63 & 31/12/2007 & 21:54 &
  J40 & 5.981 & 5.014 & 13.30 &  9.72 \\
29/12/2007 & 23:01 &  J46 & 5.981 & 5.010 & 13.16 &  9.59 & 01/01/2008 & 03:15 &
  213 & 5.982 & 5.017 & 13.20 &  9.61 \\
29/12/2007 & 23:02 &  J46 & 5.981 & 5.010 & 13.16 &  9.59 & 01/01/2008 & 03:17 &
  213 & 5.981 & 5.014 & 13.24 &  9.66 \\
29/12/2007 & 23:03 &  J46 & 5.981 & 5.010 & 13.17 &  9.60 & 01/01/2008 & 03:20 &
  213 & 5.982 & 5.017 & 13.30 &  9.71 \\
29/12/2007 & 23:04 &  J46 & 5.981 & 5.010 & 13.14 &  9.57 & 01/01/2008 & 21:12 &
  213 & 5.982 & 5.017 & 13.40 &  9.81 \\
29/12/2007 & 23:05 &  J46 & 5.981 & 5.010 & 13.15 &  9.58 & 01/01/2008 & 21:23 &
  213 & 5.982 & 5.017 & 13.45 &  9.86 \\
29/12/2007 & 23:06 &  J46 & 5.981 & 5.010 & 13.14 &  9.57 & 01/01/2008 & 21:33 &
  213 & 5.982 & 5.017 & 13.40 &  9.81 \\
29/12/2007 & 23:07 &  J46 & 5.981 & 5.010 & 13.11 &  9.54 & 01/01/2008 & 21:45 &
  J51 & 5.982 & 5.017 & 13.00 &  9.41 \\
29/12/2007 & 23:08 &  J46 & 5.981 & 5.010 & 13.10 &  9.53 & 01/01/2008 & 21:59 &
  J51 & 5.982 & 5.017 & 13.00 &  9.41 \\
29/12/2007 & 23:09 &  J46 & 5.981 & 5.010 & 13.11 &  9.54 & 01/01/2008 & 22:14 &
  J51 & 5.982 & 5.017 & 12.90 &  9.31 \\
29/12/2007 & 23:10 &  J46 & 5.981 & 5.010 & 13.10 &  9.53 & 02/01/2008 & 19:22 &
  J46 & 5.982 & 5.021 & 13.68 & 10.08 \\
29/12/2007 & 23:11 &  J46 & 5.981 & 5.010 & 13.10 &  9.53 & 02/01/2008 & 19:23 &
  J46 & 5.982 & 5.021 & 13.68 & 10.08 \\
29/12/2007 & 23:12 &  J46 & 5.981 & 5.010 & 13.11 &  9.54 & 02/01/2008 & 21:36 &
  J51 & 5.982 & 5.021 & 13.80 & 10.20 \\
29/12/2007 & 23:13 &  J46 & 5.981 & 5.010 & 13.08 &  9.51 & 02/01/2008 & 21:48 &
  J51 & 5.982 & 5.021 & 13.80 & 10.20 \\
29/12/2007 & 23:14 &  J46 & 5.981 & 5.010 & 13.11 &  9.54 & 02/01/2008 & 21:54 &
  J51 & 5.982 & 5.021 & 13.77 & 10.17 \\
29/12/2007 & 23:15 &  J46 & 5.981 & 5.010 & 13.09 &  9.52 & 02/01/2008 & 22:00 &
  J51 & 5.982 & 5.021 & 13.80 & 10.20 \\
29/12/2007 & 23:16 &  J46 & 5.981 & 5.010 & 13.07 &  9.50 & 02/01/2008 & 22:13 &
  J51 & 5.982 & 5.021 & 13.80 & 10.20 \\
29/12/2007 & 23:17 &  J46 & 5.981 & 5.010 & 13.07 &  9.50 & 03/01/2008 & 19:10 &
  J38 & 5.982 & 5.024 & 14.00 & 10.39 \\
29/12/2007 & 23:18 &  J46 & 5.981 & 5.010 & 13.07 &  9.50 & 03/01/2008 & 19:16 &
  J38 & 5.982 & 5.024 & 14.00 & 10.39 \\
29/12/2007 & 23:19 &  J46 & 5.981 & 5.010 & 13.06 &  9.49 & 03/01/2008 & 19:17 &
  J38 & 5.982 & 5.024 & 13.99 & 10.38 \\
29/12/2007 & 23:22 &  J46 & 5.981 & 5.010 & 13.04 &  9.47 & 03/01/2008 & 19:22 &
  J38 & 5.982 & 5.024 & 14.00 & 10.39 \\
29/12/2007 & 23:23 &  J46 & 5.981 & 5.010 & 13.04 &  9.47 & 03/01/2008 & 21:06 &
  213 & 5.982 & 5.024 & 13.70 & 10.09 \\
29/12/2007 & 23:24 &  J46 & 5.981 & 5.010 & 13.04 &  9.47 & 03/01/2008 & 21:15 &
  213 & 5.982 & 5.024 & 13.84 & 10.23 \\
29/12/2007 & 23:25 &  J46 & 5.981 & 5.010 & 13.03 &  9.46 & 03/01/2008 & 21:25 &
  213 & 5.982 & 5.024 & 13.80 & 10.19 \\
29/12/2007 & 23:26 &  J46 & 5.981 & 5.010 & 13.04 &  9.47 & 03/01/2008 & 21:28 &
  213 & 5.982 & 5.024 & 13.90 & 10.29 \\
29/12/2007 & 23:27 &  J46 & 5.981 & 5.010 & 13.03 &  9.46 & 03/01/2008 & 23:37 &
  J46 & 5.982 & 5.024 & 13.89 & 10.28 \\
29/12/2007 & 23:28 &  J46 & 5.981 & 5.010 & 13.02 &  9.45 & 06/01/2008 & 20:31 &
  213 & 5.983 & 5.037 & 13.10 &  9.47 \\
29/12/2007 & 23:30 &  J46 & 5.981 & 5.010 & 13.03 &  9.46 & 06/01/2008 & 20:35 &
  213 & 5.983 & 5.037 & 13.11 &  9.48 \\
29/12/2007 & 23:31 &  J46 & 5.981 & 5.010 & 13.00 &  9.43 & 06/01/2008 & 20:35 &
  213 & 5.983 & 5.037 & 13.10 &  9.47 \\
29/12/2007 & 23:32 &  J46 & 5.981 & 5.010 & 13.00 &  9.43 & 06/01/2008 & 20:39 &
  213 & 5.983 & 5.037 & 13.10 &  9.47 \\
29/12/2007 & 23:33 &  J46 & 5.981 & 5.010 & 13.00 &  9.43 & 06/01/2008 & 20:54 &
  B20 & 5.983 & 5.037 & 13.18 &  9.55 \\
29/12/2007 & 23:34 &  J46 & 5.981 & 5.010 & 13.01 &  9.44 & 06/01/2008 & 21:02 &
  J51 & 5.983 & 5.037 & 13.20 &  9.57 \\
29/12/2007 & 23:35 &  J46 & 5.981 & 5.010 & 13.02 &  9.45 & 06/01/2008 & 21:17 &
  J51 & 5.983 & 5.037 & 13.16 &  9.53 \\
29/12/2007 & 23:36 &  J46 & 5.981 & 5.010 & 13.00 &  9.43 & 06/01/2008 & 21:17 &
  J51 & 5.983 & 5.037 & 13.20 &  9.57 \\
29/12/2007 & 23:37 &  J46 & 5.981 & 5.010 & 13.00 &  9.43 & 06/01/2008 & 21:32 &
  J51 & 5.983 & 5.037 & 13.20 &  9.57 \\
29/12/2007 & 23:38 &  J46 & 5.981 & 5.010 & 13.02 &  9.45 & 06/01/2008 & 22:04 &
  J47 & 5.983 & 5.037 & 13.10 &  9.47 \\
29/12/2007 & 23:39 &  J46 & 5.981 & 5.010 & 13.00 &  9.43 & 06/01/2008 & 22:09 &
  J47 & 5.983 & 5.037 & 13.15 &  9.52 \\
29/12/2007 & 23:40 &  J46 & 5.981 & 5.010 & 12.99 &  9.42 & 06/01/2008 & 22:15 &
  J47 & 5.983 & 5.037 & 13.20 &  9.57 \\
29/12/2007 & 23:41 &  J46 & 5.981 & 5.010 & 12.98 &  9.41 & 07/01/2008 & 01:24 &
  J40 & 5.983 & 5.037 & 13.20 &  9.57 \\
29/12/2007 & 23:42 &  J46 & 5.981 & 5.010 & 12.99 &  9.42 & 07/01/2008 & 01:34 &
  J40 & 5.983 & 5.037 & 13.20 &  9.57 \\
29/12/2007 & 23:43 &  J46 & 5.981 & 5.010 & 12.97 &  9.40 & 07/01/2008 & 01:35 &
  J40 & 5.983 & 5.037 & 13.21 &  9.58 \\
29/12/2007 & 23:44 &  J46 & 5.981 & 5.010 & 13.00 &  9.43 & 07/01/2008 & 01:43 &
  J40 & 5.983 & 5.037 & 13.30 &  9.67 \\
29/12/2007 & 23:45 &  J46 & 5.981 & 5.010 & 12.99 &  9.42 & 07/01/2008 & 20:00 &
  J46 & 5.983 & 5.037 & 13.29 &  9.65 \\
29/12/2007 & 23:46 &  J46 & 5.981 & 5.010 & 13.00 &  9.43 & 07/01/2008 & 22:04 &
  J47 & 5.984 & 5.042 & 13.30 &  9.66 \\
29/12/2007 & 23:47 &  J46 & 5.981 & 5.010 & 12.98 &  9.41 & 07/01/2008 & 22:11 &
  J47 & 5.984 & 5.042 & 13.40 &  9.76 \\
29/12/2007 & 23:49 &  J46 & 5.981 & 5.010 & 12.98 &  9.41 & 07/01/2008 & 22:18 &
  J47 & 5.984 & 5.042 & 13.40 &  9.76 \\
29/12/2007 & 23:50 &  J46 & 5.981 & 5.010 & 12.99 &  9.42 & 07/01/2008 & 23:19 &
  945 & 5.984 & 5.042 & 13.30 &  9.66 \\
29/12/2007 & 23:51 &  J46 & 5.981 & 5.010 & 12.97 &  9.40 & 07/01/2008 & 23:22 &
  945 & 5.984 & 5.042 & 13.30 &  9.66 \\
29/12/2007 & 23:52 &  J46 & 5.981 & 5.010 & 12.99 &  9.42 & 07/01/2008 & 23:24 &
  945 & 5.984 & 5.042 & 13.32 &  9.68 \\
29/12/2007 & 23:53 &  J46 & 5.981 & 5.010 & 12.97 &  9.40 & 07/01/2008 & 23:25 &
  945 & 5.984 & 5.042 & 13.30 &  9.66 \\
29/12/2007 & 23:54 &  J46 & 5.981 & 5.010 & 12.97 &  9.40 & 07/01/2008 & 23:27 &
  945 & 5.984 & 5.042 & 13.30 &  9.66 \\
29/12/2007 & 23:55 &  J46 & 5.981 & 5.010 & 12.98 &  9.41 & 07/01/2008 & 23:29 &
  945 & 5.984 & 5.042 & 13.40 &  9.76 \\
29/12/2007 & 23:56 &  J46 & 5.981 & 5.010 & 12.98 &  9.41 & 08/01/2008 & 20:58 &
  213 & 5.984 & 5.047 & 13.70 & 10.05 \\
29/12/2007 & 23:57 &  J46 & 5.981 & 5.010 & 12.97 &  9.40 & 08/01/2008 & 23:05 &
  J46 & 5.984 & 5.047 & 13.56 &  9.91 \\
29/12/2007 & 23:58 &  J46 & 5.981 & 5.010 & 12.97 &  9.40 & 09/01/2008 & 20:05 &
  J46 & 5.984 & 5.052 & 13.83 & 10.17 \\
29/12/2007 & 23:59 &  J46 & 5.981 & 5.010 & 12.96 &  9.39 & 09/01/2008 & 22:37 &
  J47 & 5.984 & 5.052 & 13.90 & 10.24 \\
30/12/2007 & 00:00 &  J46 & 5.981 & 5.010 & 12.97 &  9.40 & 09/01/2008 & 22:41 &
  J47 & 5.984 & 5.052 & 13.80 & 10.14 \\
30/12/2007 & 00:01 &  J46 & 5.981 & 5.010 & 12.95 &  9.38 & 09/01/2008 & 22:44 &
  J47 & 5.984 & 5.052 & 13.89 & 10.23 \\
30/12/2007 & 00:02 &  J46 & 5.981 & 5.010 & 12.97 &  9.40 & 09/01/2008 & 22:55 &
  J47 & 5.984 & 5.052 & 13.90 & 10.24 \\
30/12/2007 & 00:03 &  J46 & 5.981 & 5.010 & 12.95 &  9.38 & 09/01/2008 & 23:55 &
  945 & 5.984 & 5.052 & 13.90 & 10.24 \\
30/12/2007 & 00:04 &  J46 & 5.981 & 5.010 & 12.96 &  9.39 & 09/01/2008 & 23:59 &
  945 & 5.984 & 5.052 & 14.00 & 10.34 \\
30/12/2007 & 00:05 &  J46 & 5.981 & 5.010 & 12.97 &  9.40 & 10/01/2008 & 00:01 &
  945 & 5.984 & 5.052 & 13.99 & 10.33 \\
30/12/2007 & 00:07 &  J46 & 5.981 & 5.010 & 12.98 &  9.41 & 10/01/2008 & 00:02 &
  945 & 5.984 & 5.052 & 14.00 & 10.34 \\
30/12/2007 & 00:08 &  J46 & 5.981 & 5.010 & 12.93 &  9.36 & 10/01/2008 & 00:10 &
  945 & 5.984 & 5.052 & 14.10 & 10.44 \\
30/12/2007 & 00:09 &  J46 & 5.981 & 5.010 & 12.96 &  9.39 & 10/01/2008 & 19:48 &
  213 & 5.985 & 5.058 & 14.30 & 10.63 \\
30/12/2007 & 00:10 &  J46 & 5.981 & 5.010 & 12.95 &  9.38 & 10/01/2008 & 19:53 &
  213 & 5.985 & 5.058 & 14.30 & 10.63 \\
30/12/2007 & 00:11 &  J46 & 5.981 & 5.010 & 12.98 &  9.41 & 10/01/2008 & 19:56 &
  213 & 5.985 & 5.058 & 14.31 & 10.64 \\
30/12/2007 & 00:12 &  J46 & 5.981 & 5.010 & 12.95 &  9.38 & 10/01/2008 & 20:04 &
  213 & 5.985 & 5.058 & 14.30 & 10.63 \\
\hline
\end{longtable}

\begin{minipage}{0.93\linewidth}
$^{a}$ Observer MPC code.
\end{minipage}
}

\onecolumn
{\small
\begin{longtable}{lcccccc|lcccccc}
\caption{Magnitudes in an aperture with a diameter of 10$\arcsec$ in 2010.}\label{tab:2010}\\
\hline\hline\noalign{\smallskip}
Date & Time & Obs$^a$ & $r_{\rm h}$ & $\Delta$ & $m_{\rm R}$ & $m_{\rm R}(1,r_{\rm h},0)$ & Date & Time & Obs & $r_{\rm h}$ & $\Delta$ & $m_{\rm R}$ & $m_{\rm R}(1,r_{\rm h},0)$ \\
 jj/mm/yyyy & hr:mn & & (au) & (au) & & & jj/mm/yyyy & hr:mn & & (au) & (au)  & & \\
\hline\noalign{\vskip 2mm}
\endfirsthead
\hline\hline\noalign{\smallskip}
Date & Time & Obs & $r_{\rm h}$ & $\Delta$ & $m_{\rm R}$ & $m_{\rm R}(1,r_{\rm h},0)$ & Date & Time & Obs & $r_{\rm h}$ & $\Delta$ & $m_{\rm R}$ & $m_{\rm R}(1,r_{\rm h},0)$ \\
 jj/mm/yyyy & hr:mn &  & (au) & (au)  & & jj/mm/yyyy & hr:mn & & (au) & (au)  & & \\
\hline\noalign{\vskip 2mm}
\endhead
\hline \multicolumn{14}{r}{\textit{Continued on next page}} \\
\endfoot
\hline
\endlastfoot
02/04/2010 &  21:19 &  J97 & 6.203 & 5.582 & 15.80 & 11.75 & 25/04/2010 &  22:28
 &  213 & 6.207 & 5.921 & 14.40 & 10.16 \\
02/04/2010 &  21:25 &  J97 & 6.203 & 5.582 & 15.80 & 11.75 & 25/04/2010 &  22:51
 &  945 & 6.207 & 5.921 & 14.40 & 10.16 \\
02/04/2010 &  21:26 &  J97 & 6.203 & 5.582 & 15.80 & 11.75 & 25/04/2010 &  22:56
 &  945 & 6.207 & 5.921 & 14.50 & 10.26 \\
02/04/2010 &  21:31 &  J97 & 6.203 & 5.582 & 15.80 & 11.75 & 25/04/2010 &  23:00
 &  945 & 6.207 & 5.921 & 14.40 & 10.16 \\
02/04/2010 &  23:03 &  J53 & 6.203 & 5.582 & 16.10 & 12.05 & 25/04/2010 &  23:02
 &  945 & 6.207 & 5.921 & 14.48 & 10.24 \\
02/04/2010 &  23:09 &  J53 & 6.203 & 5.582 & 16.07 & 12.02 & 25/04/2010 &  23:04
 &  945 & 6.207 & 5.921 & 14.50 & 10.26 \\
02/04/2010 &  23:12 &  J53 & 6.203 & 5.582 & 16.10 & 12.05 & 25/04/2010 &  23:09
 &  945 & 6.207 & 5.921 & 14.40 & 10.16 \\
02/04/2010 &  23:14 &  J53 & 6.203 & 5.582 & 16.10 & 12.05 & 26/04/2010 &  22:24
 &  945 & 6.207 & 5.921 & 14.61 & 10.37 \\
02/04/2010 &  23:17 &  J47 & 6.203 & 5.582 & 16.00 & 11.95 & 26/04/2010 &  22:36
 &  J38 & 6.207 & 5.921 & 14.85 & 10.61 \\
02/04/2010 &  23:40 &  J47 & 6.203 & 5.582 & 15.90 & 11.85 & 27/04/2010 &  20:13
 &  B20 & 6.208 & 5.937 & 14.68 & 10.43 \\
03/04/2010 &  00:01 &  J47 & 6.203 & 5.582 & 15.90 & 11.85 & 27/04/2010 &  20:37
 &  213 & 6.208 & 5.937 & 14.97 & 10.72 \\
04/04/2010 &  19:51 &  213 & 6.198 & 5.283 & 16.10 & 12.16 & 27/04/2010 &  21:02
 &  J47 & 6.208 & 5.937 & 14.80 & 10.55 \\
04/04/2010 &  20:14 &  213 & 6.198 & 5.283 & 16.08 & 12.14 & 27/04/2010 &  21:22
 &  J38 & 6.208 & 5.937 & 14.88 & 10.63 \\
04/04/2010 &  20:19 &  213 & 6.198 & 5.283 & 16.10 & 12.16 & 28/04/2010 &  20:45
 &  C12 & 6.208 & 5.953 & 15.21 & 10.96 \\
04/04/2010 &  20:33 &  213 & 6.198 & 5.283 & 16.10 & 12.16 & 28/04/2010 &  21:59
 &  213 & 6.208 & 5.953 & 15.05 & 10.80 \\
04/04/2010 &  22:37 &  945 & 6.198 & 5.283 & 16.10 & 12.16 & 01/05/2010 &  21:13
 &  J36 & 6.208 & 6.000 & 15.22 & 10.95 \\
04/04/2010 &  22:41 &  945 & 6.198 & 5.283 & 16.10 & 12.16 & 04/05/2010 &  21:35
 &  J40 & 6.209 & 6.065 & 15.59 & 11.29 \\
04/04/2010 &  22:48 &  945 & 6.198 & 5.283 & 16.10 & 12.16 & 05/05/2010 &  20:35
 &  J47 & 6.209 & 6.081 & 14.90 & 10.60 \\
04/04/2010 &  22:50 &  945 & 6.198 & 5.283 & 16.10 & 12.16 & 05/05/2010 &  20:47
 &  J47 & 6.209 & 6.081 & 14.94 & 10.64 \\
04/04/2010 &  22:54 &  945 & 6.198 & 5.283 & 16.10 & 12.16 & 05/05/2010 &  20:57
 &  J38 & 6.209 & 6.081 & 14.98 & 10.68 \\
04/04/2010 &  23:03 &  945 & 6.198 & 5.283 & 16.10 & 12.16 & 05/05/2010 &  20:59
 &  J38 & 6.209 & 6.081 & 15.00 & 10.70 \\
04/04/2010 &  23:17 &  J53 & 6.198 & 5.283 & 16.20 & 12.26 & 05/05/2010 &  21:02
 &  J38 & 6.209 & 6.081 & 15.00 & 10.70 \\
04/04/2010 &  23:40 &  J53 & 6.198 & 5.283 & 16.13 & 12.19 & 05/05/2010 &  21:03
 &  J47 & 6.209 & 6.081 & 14.90 & 10.60 \\
04/04/2010 &  23:43 &  J53 & 6.198 & 5.283 & 16.20 & 12.26 & 05/05/2010 &  21:05
 &  J38 & 6.209 & 6.081 & 15.00 & 10.70 \\
04/04/2010 &  23:49 &  J53 & 6.198 & 5.283 & 16.10 & 12.16 & 05/05/2010 &  21:17
 &  J40 & 6.209 & 6.081 & 14.88 & 10.58 \\
05/04/2010 &  21:07 &  J47 & 6.198 & 5.290 & 16.10 & 12.15 & 06/05/2010 &  21:30
 &  B74 & 6.209 & 6.097 & 14.83 & 10.52 \\
05/04/2010 &  21:13 &  J47 & 6.198 & 5.290 & 16.10 & 12.15 & 07/05/2010 &  19:44
 &  213 & 6.210 & 6.113 & 15.00 & 10.69 \\
05/04/2010 &  21:22 &  J47 & 6.198 & 5.290 & 16.10 & 12.15 & 07/05/2010 &  19:53
 &  213 & 6.210 & 6.113 & 15.03 & 10.72 \\
05/04/2010 &  21:51 &  J51 & 6.198 & 5.290 & 16.10 & 12.15 & 07/05/2010 &  20:02
 &  213 & 6.210 & 6.113 & 15.10 & 10.79 \\
05/04/2010 &  22:10 &  J51 & 6.198 & 5.290 & 16.10 & 12.15 & 07/05/2010 &  21:09
 &  J47 & 6.210 & 6.113 & 14.90 & 10.59 \\
05/04/2010 &  22:30 &  J51 & 6.198 & 5.290 & 16.10 & 12.15 & 07/05/2010 &  21:25
 &  J40 & 6.210 & 6.113 & 14.93 & 10.62 \\
06/04/2010 &  00:20 &  945 & 6.198 & 5.290 & 16.00 & 12.05 & 07/05/2010 &  21:31
 &  J47 & 6.210 & 6.113 & 14.94 & 10.63 \\
06/04/2010 &  00:24 &  945 & 6.198 & 5.290 & 16.00 & 12.05 & 07/05/2010 &  21:32
 &  212 & 6.210 & 6.113 & 15.00 & 10.69 \\
06/04/2010 &  00:26 &  945 & 6.198 & 5.290 & 16.01 & 12.06 & 07/05/2010 &  21:33
 &  J47 & 6.210 & 6.113 & 14.90 & 10.59 \\
06/04/2010 &  00:28 &  945 & 6.198 & 5.290 & 16.00 & 12.05 & 07/05/2010 &  21:40
 &  212 & 6.210 & 6.113 & 14.97 & 10.66 \\
06/04/2010 &  00:32 &  945 & 6.198 & 5.290 & 16.00 & 12.05 & 07/05/2010 &  21:42
 &  212 & 6.210 & 6.113 & 14.90 & 10.59 \\
06/04/2010 &  00:36 &  945 & 6.198 & 5.290 & 16.00 & 12.05 & 07/05/2010 &  21:47
 &  212 & 6.210 & 6.113 & 15.00 & 10.69 \\
07/04/2010 &  22:16 &  J53 & 6.204 & 5.650 & 16.10 & 12.00 & 07/05/2010 &  21:51
 &  J47 & 6.210 & 6.113 & 15.00 & 10.69 \\
07/04/2010 &  22:29 &  J53 & 6.204 & 5.650 & 16.13 & 12.03 & 08/05/2010 &  19:45
 &  213 & 6.210 & 6.145 & 15.00 & 10.67 \\
07/04/2010 &  22:30 &  J53 & 6.204 & 5.650 & 16.10 & 12.00 & 08/05/2010 &  20:00
 &  213 & 6.210 & 6.145 & 14.90 & 10.57 \\
07/04/2010 &  22:44 &  J53 & 6.204 & 5.650 & 16.10 & 12.00 & 08/05/2010 &  20:00
 &  213 & 6.210 & 6.145 & 14.96 & 10.63 \\
07/04/2010 &  23:51 &  I96 & 6.204 & 5.650 & 16.10 & 12.00 & 08/05/2010 &  20:15
 &  213 & 6.210 & 6.145 & 14.90 & 10.57 \\
08/04/2010 &  00:22 &  I96 & 6.204 & 5.650 & 16.10 & 12.00 & 08/05/2010 &  20:23
 &  213 & 6.210 & 6.145 & 14.96 & 10.63 \\
08/04/2010 &  00:26 &  I96 & 6.204 & 5.650 & 16.13 & 12.03 & 08/05/2010 &  20:31
 &  B74 & 6.210 & 6.145 & 14.80 & 10.47 \\
08/04/2010 &  01:05 &  I96 & 6.204 & 5.650 & 16.20 & 12.10 & 08/05/2010 &  20:35
 &  B74 & 6.210 & 6.145 & 14.80 & 10.47 \\
08/04/2010 &  22:26 &  J53 & 6.204 & 5.650 & 16.10 & 12.00 & 08/05/2010 &  20:37
 &  B74 & 6.210 & 6.145 & 14.81 & 10.48 \\
08/04/2010 &  22:32 &  J53 & 6.204 & 5.650 & 16.20 & 12.10 & 08/05/2010 &  20:42
 &  B74 & 6.210 & 6.145 & 14.80 & 10.47 \\
08/04/2010 &  22:32 &  J53 & 6.204 & 5.650 & 16.24 & 12.14 & 08/05/2010 &  22:43
 &  I32 & 6.210 & 6.145 & 14.86 & 10.53 \\
08/04/2010 &  22:38 &  J53 & 6.204 & 5.650 & 16.30 & 12.20 & 09/05/2010 &  20:12
 &  J98 & 6.210 & 6.145 & 15.20 & 10.87 \\
09/04/2010 &  19:35 &  A06 & 6.205 & 5.679 & 16.10 & 11.99 & 09/05/2010 &  20:13
 &  J98 & 6.210 & 6.145 & 15.10 & 10.77 \\
09/04/2010 &  20:00 &  A06 & 6.205 & 5.679 & 16.11 & 12.00 & 09/05/2010 &  20:14
 &  J98 & 6.210 & 6.145 & 15.00 & 10.67 \\
09/04/2010 &  20:12 &  A06 & 6.205 & 5.679 & 16.10 & 11.99 & 09/05/2010 &  21:04
 &  945 & 6.210 & 6.145 & 14.90 & 10.57 \\
09/04/2010 &  20:39 &  213 & 6.205 & 5.679 & 16.20 & 12.09 & 09/05/2010 &  21:08
 &  945 & 6.210 & 6.145 & 14.90 & 10.57 \\
09/04/2010 &  20:49 &  213 & 6.205 & 5.679 & 16.19 & 12.08 & 09/05/2010 &  21:12
 &  945 & 6.210 & 6.145 & 14.90 & 10.57 \\
09/04/2010 &  20:49 &  213 & 6.205 & 5.679 & 16.20 & 12.09 & 09/05/2010 &  21:14
 &  945 & 6.210 & 6.145 & 14.92 & 10.59 \\
09/04/2010 &  20:59 &  213 & 6.205 & 5.679 & 16.20 & 12.09 & 09/05/2010 &  21:16
 &  945 & 6.210 & 6.145 & 14.90 & 10.57 \\
09/04/2010 &  21:06 &  J97 & 6.205 & 5.679 & 16.10 & 11.99 & 09/05/2010 &  21:24
 &  945 & 6.210 & 6.145 & 14.90 & 10.57 \\
09/04/2010 &  21:14 &  J97 & 6.205 & 5.679 & 16.00 & 11.89 & 10/05/2010 &  20:42
 &  J47 & 6.210 & 6.161 & 15.10 & 10.77 \\
09/04/2010 &  21:15 &  J97 & 6.205 & 5.679 & 16.10 & 11.99 & 10/05/2010 &  20:49
 &  J47 & 6.210 & 6.161 & 15.10 & 10.77 \\
09/04/2010 &  21:20 &  J97 & 6.205 & 5.679 & 16.10 & 11.99 & 10/05/2010 &  20:49
 &  J47 & 6.210 & 6.161 & 15.14 & 10.81 \\
09/04/2010 &  21:20 &  B20 & 6.205 & 5.679 & 16.15 & 12.04 & 10/05/2010 &  20:57
 &  J47 & 6.210 & 6.161 & 15.20 & 10.87 \\
10/04/2010 &  20:27 &  B20 & 6.205 & 5.693 & 16.09 & 11.97 & 11/05/2010 &  20:40
 &  J47 & 6.210 & 6.161 & 15.40 & 11.07 \\
10/04/2010 &  21:21 &  945 & 6.205 & 5.693 & 15.90 & 11.78 & 11/05/2010 &  20:54
 &  J47 & 6.210 & 6.161 & 15.40 & 11.07 \\
10/04/2010 &  21:34 &  945 & 6.205 & 5.693 & 15.80 & 11.68 & 11/05/2010 &  20:57
 &  J47 & 6.210 & 6.177 & 15.40 & 11.06 \\
10/04/2010 &  21:37 &  945 & 6.205 & 5.693 & 15.86 & 11.74 & 11/05/2010 &  21:18
 &  J47 & 6.210 & 6.161 & 15.40 & 11.07 \\
10/04/2010 &  21:41 &  945 & 6.205 & 5.693 & 15.90 & 11.78 & 11/05/2010 &  22:12
 &  J40 & 6.210 & 6.177 & 15.38 & 11.04 \\
10/04/2010 &  21:46 &  945 & 6.205 & 5.693 & 15.80 & 11.68 & 13/05/2010 &  21:11
 &  J38 & 6.211 & 6.210 & 15.60 & 11.25 \\
10/04/2010 &  21:51 &  945 & 6.205 & 5.693 & 15.90 & 11.78 & 13/05/2010 &  21:14
 &  J38 & 6.211 & 6.210 & 15.60 & 11.25 \\
10/04/2010 &  22:43 &  J53 & 6.205 & 5.693 & 16.10 & 11.98 & 13/05/2010 &  21:18
 &  J38 & 6.211 & 6.210 & 15.60 & 11.25 \\
10/04/2010 &  22:49 &  J53 & 6.205 & 5.693 & 16.11 & 11.99 & 13/05/2010 &  21:19
 &  J38 & 6.211 & 6.210 & 15.62 & 11.27 \\
10/04/2010 &  22:49 &  J53 & 6.205 & 5.693 & 16.00 & 11.88 & 15/05/2010 &  20:58
 &  213 & 6.211 & 6.242 & 16.00 & 11.64 \\
10/04/2010 &  22:56 &  J53 & 6.205 & 5.693 & 16.20 & 12.08 & 15/05/2010 &  21:10
 &  213 & 6.211 & 6.242 & 15.98 & 11.62 \\
12/04/2010 &  21:06 &  J38 & 6.205 & 5.722 & 16.10 & 11.96 & 15/05/2010 &  21:10
 &  213 & 6.211 & 6.242 & 16.00 & 11.64 \\
12/04/2010 &  21:11 &  J38 & 6.205 & 5.722 & 16.10 & 11.96 & 15/05/2010 &  21:15
 &  213 & 6.211 & 6.242 & 16.00 & 11.64 \\
12/04/2010 &  21:16 &  J38 & 6.205 & 5.722 & 16.14 & 12.00 & 15/05/2010 &  21:25
 &  J53 & 6.211 & 6.242 & 15.90 & 11.54 \\
12/04/2010 &  21:24 &  J38 & 6.205 & 5.722 & 16.10 & 11.96 & 15/05/2010 &  21:27
 &  J53 & 6.211 & 6.242 & 15.80 & 11.44 \\
12/04/2010 &  22:44 &  J47 & 6.205 & 5.722 & 16.30 & 12.16 & 15/05/2010 &  21:27
 &  213 & 6.211 & 6.242 & 15.86 & 11.50 \\
12/04/2010 &  23:03 &  J47 & 6.205 & 5.722 & 16.30 & 12.16 & 15/05/2010 &  21:30
 &  J53 & 6.211 & 6.242 & 15.83 & 11.47 \\
12/04/2010 &  23:04 &  J47 & 6.205 & 5.722 & 16.31 & 12.17 & 15/05/2010 &  21:31
 &  J53 & 6.211 & 6.242 & 15.80 & 11.44 \\
12/04/2010 &  23:28 &  J47 & 6.205 & 5.722 & 16.30 & 12.16 & 15/05/2010 &  22:15
 &  A01 & 6.211 & 6.242 & 15.60 & 11.24 \\
14/04/2010 &  20:27 &  213 & 6.205 & 5.752 & 16.50 & 12.35 & 15/05/2010 &  22:18
 &  A01 & 6.211 & 6.242 & 15.60 & 11.24 \\
14/04/2010 &  20:32 &  213 & 6.205 & 5.752 & 16.45 & 12.30 & 15/05/2010 &  22:21
 &  A01 & 6.211 & 6.242 & 15.58 & 11.22 \\
14/04/2010 &  20:37 &  213 & 6.205 & 5.752 & 16.40 & 12.25 & 15/05/2010 &  22:29
 &  A01 & 6.211 & 6.242 & 15.50 & 11.14 \\
14/04/2010 &  20:57 &  J38 & 6.205 & 5.752 & 16.40 & 12.25 & 16/05/2010 &  20:18
 &  J98 & 6.211 & 6.258 & 15.50 & 11.14 \\
14/04/2010 &  21:00 &  J38 & 6.205 & 5.752 & 16.40 & 12.25 & 16/05/2010 &  20:20
 &  J98 & 6.211 & 6.258 & 15.64 & 11.28 \\
14/04/2010 &  21:04 &  J38 & 6.205 & 5.752 & 16.38 & 12.23 & 16/05/2010 &  20:22
 &  J98 & 6.211 & 6.258 & 15.70 & 11.34 \\
14/04/2010 &  21:09 &  J38 & 6.205 & 5.752 & 16.40 & 12.25 & 16/05/2010 &  21:41
 &  J53 & 6.211 & 6.258 & 15.90 & 11.54 \\
14/04/2010 &  21:42 &  J47 & 6.205 & 5.752 & 16.40 & 12.25 & 16/05/2010 &  21:43
 &  J53 & 6.211 & 6.258 & 15.90 & 11.54 \\
14/04/2010 &  21:46 &  J47 & 6.205 & 5.752 & 16.30 & 12.15 & 16/05/2010 &  21:45
 &  J53 & 6.211 & 6.258 & 15.89 & 11.53 \\
14/04/2010 &  21:47 &  J47 & 6.205 & 5.752 & 16.36 & 12.21 & 16/05/2010 &  21:49
 &  J53 & 6.211 & 6.258 & 15.90 & 11.54 \\
14/04/2010 &  21:52 &  J47 & 6.205 & 5.752 & 16.40 & 12.25 & 16/05/2010 &  21:57
 &  J40 & 6.211 & 6.258 & 15.90 & 11.54 \\
14/04/2010 &  23:07 &  945 & 6.205 & 5.752 & 16.10 & 11.95 & 17/05/2010 &  20:49
 &  213 & 6.211 & 6.274 & 16.00 & 11.63 \\
14/04/2010 &  23:12 &  945 & 6.205 & 5.752 & 16.20 & 12.05 & 17/05/2010 &  20:55
 &  213 & 6.211 & 6.274 & 16.08 & 11.71 \\
14/04/2010 &  23:17 &  945 & 6.205 & 5.752 & 16.10 & 11.95 & 17/05/2010 &  21:00
 &  213 & 6.211 & 6.274 & 16.10 & 11.73 \\
14/04/2010 &  23:19 &  945 & 6.205 & 5.752 & 16.18 & 12.03 & 17/05/2010 &  21:05
 &  213 & 6.211 & 6.274 & 16.10 & 11.73 \\
14/04/2010 &  23:21 &  945 & 6.205 & 5.752 & 16.20 & 12.05 & 17/05/2010 &  21:32
 &  J38 & 6.211 & 6.274 & 15.90 & 11.53 \\
14/04/2010 &  23:25 &  945 & 6.205 & 5.752 & 16.10 & 11.95 & 17/05/2010 &  21:36
 &  J38 & 6.211 & 6.274 & 15.97 & 11.60 \\
16/04/2010 &  19:22 &  A06 & 6.206 & 5.782 & 12.80 &  8.63 & 17/05/2010 &  21:38
 &  J38 & 6.211 & 6.274 & 16.00 & 11.63 \\
16/04/2010 &  19:45 &  A06 & 6.206 & 5.782 & 12.80 &  8.63 & 17/05/2010 &  21:44
 &  J38 & 6.211 & 6.274 & 16.00 & 11.63 \\
16/04/2010 &  19:51 &  A06 & 6.206 & 5.782 & 12.90 &  8.73 & 18/05/2010 &  21:04
 &  J40 & 6.212 & 6.289 & 16.11 & 11.74 \\
16/04/2010 &  20:08 &  A06 & 6.206 & 5.782 & 12.90 &  8.73 & 18/05/2010 &  21:31
 &  945 & 6.212 & 6.289 & 16.00 & 11.63 \\
16/04/2010 &  20:28 &  213 & 6.206 & 5.782 & 12.70 &  8.53 & 18/05/2010 &  21:35
 &  945 & 6.212 & 6.289 & 15.90 & 11.53 \\
16/04/2010 &  20:35 &  A06 & 6.206 & 5.782 & 12.80 &  8.63 & 18/05/2010 &  21:39
 &  945 & 6.212 & 6.289 & 16.00 & 11.63 \\
16/04/2010 &  20:39 &  C12 & 6.206 & 5.782 & 12.80 &  8.63 & 18/05/2010 &  21:41
 &  945 & 6.212 & 6.289 & 15.92 & 11.55 \\
16/04/2010 &  20:42 &  213 & 6.206 & 5.782 & 12.80 &  8.63 & 18/05/2010 &  21:47
 &  945 & 6.212 & 6.289 & 15.90 & 11.53 \\
16/04/2010 &  20:42 &  213 & 6.206 & 5.782 & 12.76 &  8.59 & 18/05/2010 &  21:54
 &  945 & 6.212 & 6.289 & 16.00 & 11.63 \\
16/04/2010 &  20:43 &  C12 & 6.206 & 5.782 & 12.80 &  8.63 & 19/05/2010 &  21:06
 &  B74 & 6.212 & 6.305 & 15.70 & 11.32 \\
16/04/2010 &  20:45 &  C12 & 6.206 & 5.782 & 12.82 &  8.65 & 19/05/2010 &  21:39
 &  J38 & 6.212 & 6.305 & 15.89 & 11.51 \\
16/04/2010 &  20:47 &  C12 & 6.206 & 5.782 & 12.80 &  8.63 & 19/05/2010 &  22:57
 &  J40 & 6.212 & 6.305 & 16.00 & 11.62 \\
16/04/2010 &  20:49 &  213 & 6.206 & 5.782 & 12.80 &  8.63 & 21/05/2010 &  20:37
 &  B20 & 6.212 & 6.337 & 15.60 & 11.21 \\
16/04/2010 &  20:51 &  C12 & 6.206 & 5.782 & 12.80 &  8.63 & 21/05/2010 &  20:42
 &  A06 & 6.212 & 6.337 & 15.60 & 11.21 \\
16/04/2010 &  20:56 &  J38 & 6.206 & 5.782 & 12.80 &  8.63 & 21/05/2010 &  20:58
 &  A06 & 6.212 & 6.337 & 15.60 & 11.21 \\
16/04/2010 &  21:00 &  J38 & 6.206 & 5.782 & 12.82 &  8.65 & 21/05/2010 &  21:01
 &  A06 & 6.212 & 6.337 & 15.64 & 11.25 \\
16/04/2010 &  21:01 &  J38 & 6.206 & 5.782 & 12.80 &  8.63 & 21/05/2010 &  21:14
 &  A06 & 6.212 & 6.337 & 15.60 & 11.21 \\
16/04/2010 &  21:04 &  J38 & 6.206 & 5.782 & 12.80 &  8.63 & 21/05/2010 &  21:23
 &  J38 & 6.212 & 6.337 & 15.40 & 11.01 \\
17/04/2010 &  20:32 &  C12 & 6.206 & 5.797 & 12.90 &  8.72 & 21/05/2010 &  21:26
 &  J38 & 6.212 & 6.337 & 15.50 & 11.11 \\
17/04/2010 &  20:37 &  C12 & 6.206 & 5.797 & 12.90 &  8.72 & 21/05/2010 &  21:30
 &  J38 & 6.212 & 6.337 & 15.48 & 11.09 \\
17/04/2010 &  20:38 &  C12 & 6.206 & 5.797 & 12.92 &  8.74 & 21/05/2010 &  21:32
 &  J38 & 6.212 & 6.337 & 15.50 & 11.11 \\
17/04/2010 &  20:41 &  C12 & 6.206 & 5.797 & 12.90 &  8.72 & 22/05/2010 &  20:15
 &  213 & 6.212 & 6.353 & 15.90 & 11.51 \\
17/04/2010 &  20:47 &  B74 & 6.206 & 5.782 & 12.90 &  8.73 & 22/05/2010 &  20:25
 &  213 & 6.212 & 6.353 & 15.80 & 11.41 \\
17/04/2010 &  20:49 &  B74 & 6.206 & 5.782 & 12.88 &  8.71 & 22/05/2010 &  20:37
 &  213 & 6.212 & 6.353 & 15.83 & 11.44 \\
17/04/2010 &  20:49 &  B74 & 6.206 & 5.782 & 12.90 &  8.73 & 22/05/2010 &  20:57
 &  213 & 6.212 & 6.353 & 15.70 & 11.31 \\
17/04/2010 &  20:50 &  B74 & 6.206 & 5.782 & 12.90 &  8.73 & 22/05/2010 &  21:27
 &  945 & 6.212 & 6.353 & 15.60 & 11.21 \\
17/04/2010 &  20:56 &  213 & 6.206 & 5.797 & 12.90 &  8.72 & 22/05/2010 &  21:34
 &  945 & 6.212 & 6.353 & 15.70 & 11.31 \\
17/04/2010 &  21:01 &  213 & 6.206 & 5.797 & 12.90 &  8.72 & 22/05/2010 &  21:38
 &  945 & 6.212 & 6.353 & 15.60 & 11.21 \\
17/04/2010 &  21:01 &  213 & 6.206 & 5.797 & 12.90 &  8.72 & 22/05/2010 &  21:40
 &  945 & 6.212 & 6.353 & 15.62 & 11.23 \\
17/04/2010 &  21:06 &  213 & 6.206 & 5.797 & 12.90 &  8.72 & 22/05/2010 &  21:43
 &  945 & 6.212 & 6.353 & 15.60 & 11.21 \\
17/04/2010 &  22:03 &  J47 & 6.206 & 5.797 & 12.90 &  8.72 & 22/05/2010 &  21:50
 &  J38 & 6.212 & 6.353 & 15.45 & 11.06 \\
17/04/2010 &  22:07 &  J47 & 6.206 & 5.797 & 12.90 &  8.72 & 22/05/2010 &  21:52
 &  945 & 6.212 & 6.353 & 15.60 & 11.21 \\
17/04/2010 &  22:14 &  J47 & 6.206 & 5.782 & 12.89 &  8.72 & 22/05/2010 &  21:55
 &  J38 & 6.212 & 6.353 & 15.50 & 11.11 \\
17/04/2010 &  22:14 &  J47 & 6.206 & 5.797 & 12.90 &  8.72 & 22/05/2010 &  21:58
 &  J38 & 6.212 & 6.353 & 15.40 & 11.01 \\
17/04/2010 &  22:17 &  J38 & 6.206 & 5.782 & 12.90 &  8.73 & 23/05/2010 &  21:26
 &  945 & 6.212 & 6.368 & 15.90 & 11.50 \\
17/04/2010 &  22:20 &  J47 & 6.206 & 5.797 & 12.90 &  8.72 & 23/05/2010 &  21:42
 &  945 & 6.212 & 6.368 & 15.80 & 11.40 \\
17/04/2010 &  22:23 &  J38 & 6.206 & 5.782 & 12.80 &  8.63 & 23/05/2010 &  21:51
 &  945 & 6.212 & 6.368 & 15.90 & 11.50 \\
17/04/2010 &  22:24 &  J38 & 6.206 & 5.782 & 12.88 &  8.71 & 23/05/2010 &  21:57
 &  945 & 6.212 & 6.368 & 15.80 & 11.40 \\
17/04/2010 &  22:24 &  J47 & 6.206 & 5.797 & 12.90 &  8.72 & 24/05/2010 &  21:02
 &  213 & 6.213 & 6.384 & 14.90 & 10.50 \\
17/04/2010 &  22:28 &  J38 & 6.206 & 5.782 & 12.90 &  8.73 & 24/05/2010 &  21:06
 &  213 & 6.213 & 6.384 & 14.85 & 10.45 \\
17/04/2010 &  23:15 &  J24 & 6.206 & 5.782 & 12.85 &  8.68 & 24/05/2010 &  21:10
 &  213 & 6.213 & 6.384 & 14.80 & 10.40 \\
18/04/2010 &  19:29 &  A06 & 6.206 & 5.812 & 13.15 &  8.96 & 24/05/2010 &  21:47
 &  213 & 6.213 & 6.384 & 14.80 & 10.40 \\
18/04/2010 &  20:31 &  213 & 6.206 & 5.812 & 13.20 &  9.01 & 24/05/2010 &  21:50
 &  213 & 6.213 & 6.384 & 14.80 & 10.40 \\
18/04/2010 &  20:42 &  213 & 6.206 & 5.812 & 13.17 &  8.98 & 24/05/2010 &  21:52
 &  213 & 6.213 & 6.384 & 14.80 & 10.40 \\
18/04/2010 &  20:45 &  213 & 6.206 & 5.812 & 13.20 &  9.01 & 25/05/2010 &  20:08
 &  213 & 6.213 & 6.400 & 14.80 & 10.40 \\
18/04/2010 &  20:54 &  213 & 6.206 & 5.812 & 13.20 &  9.01 & 25/05/2010 &  20:14
 &  213 & 6.213 & 6.400 & 14.80 & 10.40 \\
18/04/2010 &  21:15 &  213 & 6.206 & 5.812 & 13.14 &  8.95 & 25/05/2010 &  20:14
 &  213 & 6.213 & 6.400 & 14.81 & 10.41 \\
18/04/2010 &  22:18 &  A02 & 6.206 & 5.812 & 13.10 &  8.91 & 25/05/2010 &  20:15
 &  B74 & 6.213 & 6.400 & 14.80 & 10.40 \\
18/04/2010 &  22:19 &  A02 & 6.206 & 5.812 & 13.10 &  8.91 & 25/05/2010 &  20:27
 &  B74 & 6.213 & 6.400 & 14.72 & 10.32 \\
18/04/2010 &  22:23 &  A02 & 6.206 & 5.812 & 13.10 &  8.91 & 25/05/2010 &  20:29
 &  B74 & 6.213 & 6.400 & 14.70 & 10.30 \\
18/04/2010 &  23:08 &  A02 & 6.206 & 5.812 & 13.10 &  8.91 & 25/05/2010 &  20:30
 &  213 & 6.213 & 6.400 & 14.70 & 10.30 \\
18/04/2010 &  23:11 &  A02 & 6.206 & 5.812 & 13.10 &  8.91 & 25/05/2010 &  20:37
 &  213 & 6.213 & 6.400 & 14.80 & 10.40 \\
18/04/2010 &  23:14 &  A02 & 6.206 & 5.812 & 13.10 &  8.91 & 25/05/2010 &  20:37
 &  B74 & 6.213 & 6.400 & 14.70 & 10.30 \\
18/04/2010 &  23:17 &  A02 & 6.206 & 5.812 & 13.10 &  8.91 & 25/05/2010 &  20:51
 &  213 & 6.213 & 6.400 & 14.80 & 10.40 \\
19/04/2010 &  20:54 &  213 & 6.206 & 5.827 & 13.30 &  9.11 & 25/05/2010 &  21:06
 &  J30 & 6.213 & 6.400 & 14.96 & 10.56 \\
19/04/2010 &  21:02 &  213 & 6.206 & 5.827 & 13.36 &  9.17 & 25/05/2010 &  21:10
 &  J30 & 6.213 & 6.400 & 14.87 & 10.47 \\
19/04/2010 &  21:05 &  213 & 6.206 & 5.827 & 13.40 &  9.21 & 25/05/2010 &  21:10
 &  J30 & 6.213 & 6.400 & 14.80 & 10.40 \\
19/04/2010 &  21:10 &  213 & 6.206 & 5.827 & 13.40 &  9.21 & 25/05/2010 &  21:14
 &  J30 & 6.213 & 6.400 & 14.86 & 10.46 \\
19/04/2010 &  21:19 &  213 & 6.206 & 5.827 & 13.38 &  9.19 & 26/05/2010 &  21:10
 &  J30 & 6.213 & 6.415 & 14.94 & 10.53 \\
20/04/2010 &  21:38 &  C12 & 6.207 & 5.843 & 13.90 &  9.70 & 26/05/2010 &  21:14
 &  J30 & 6.213 & 6.415 & 14.81 & 10.40 \\
20/04/2010 &  21:47 &  C12 & 6.207 & 5.843 & 13.80 &  9.60 & 26/05/2010 &  21:14
 &  J30 & 6.213 & 6.415 & 14.64 & 10.23 \\
20/04/2010 &  21:48 &  C12 & 6.207 & 5.843 & 13.85 &  9.65 & 26/05/2010 &  21:18
 &  J30 & 6.213 & 6.415 & 14.84 & 10.43 \\
20/04/2010 &  21:51 &  C12 & 6.207 & 5.843 & 13.90 &  9.70 & 29/05/2010 &  20:58
 &  J30 & 6.213 & 6.461 & 15.51 & 11.09 \\
20/04/2010 &  21:58 &  C12 & 6.207 & 5.843 & 13.90 &  9.70 & 29/05/2010 &  21:03
 &  J30 & 6.213 & 6.461 & 15.22 & 10.80 \\
20/04/2010 &  22:14 &  B20 & 6.207 & 5.843 & 13.80 &  9.60 & 29/05/2010 &  21:03
 &  J30 & 6.213 & 6.461 & 15.28 & 10.86 \\
20/04/2010 &  22:18 &  945 & 6.207 & 5.843 & 13.60 &  9.40 & 29/05/2010 &  21:08
 &  J30 & 6.213 & 6.461 & 15.12 & 10.70 \\
20/04/2010 &  22:23 &  945 & 6.207 & 5.843 & 13.60 &  9.40 & 30/05/2010 &  20:52
 &  B74 & 6.214 & 6.476 & 15.50 & 11.08 \\
20/04/2010 &  22:25 &  945 & 6.207 & 5.843 & 13.59 &  9.39 & 30/05/2010 &  20:55
 &  B74 & 6.214 & 6.476 & 15.60 & 11.18 \\
20/04/2010 &  22:28 &  B20 & 6.207 & 5.843 & 13.78 &  9.58 & 30/05/2010 &  20:55
 &  B74 & 6.214 & 6.476 & 15.57 & 11.15 \\
20/04/2010 &  22:28 &  945 & 6.207 & 5.843 & 13.60 &  9.40 & 30/05/2010 &  21:02
 &  J40 & 6.214 & 6.476 & 15.67 & 11.25 \\
20/04/2010 &  22:31 &  945 & 6.207 & 5.843 & 13.60 &  9.40 & 30/05/2010 &  21:04
 &  B74 & 6.214 & 6.476 & 15.60 & 11.18 \\
20/04/2010 &  22:40 &  945 & 6.207 & 5.843 & 13.60 &  9.40 & 30/05/2010 &  21:15
 &  213 & 6.214 & 6.476 & 15.60 & 11.18 \\
20/04/2010 &  22:41 &  B20 & 6.207 & 5.843 & 13.70 &  9.50 & 30/05/2010 &  21:19
 &  213 & 6.214 & 6.476 & 15.63 & 11.21 \\
21/04/2010 &  22:57 &  J47 & 6.207 & 5.858 & 14.00 &  9.79 & 30/05/2010 &  21:19
 &  213 & 6.214 & 6.476 & 15.70 & 11.28 \\
21/04/2010 &  23:20 &  J47 & 6.207 & 5.858 & 14.03 &  9.82 & 30/05/2010 &  21:24
 &  213 & 6.214 & 6.476 & 15.60 & 11.18 \\
21/04/2010 &  23:29 &  J47 & 6.207 & 5.858 & 14.10 &  9.89 & 02/06/2010 &  20:36
 &  B74 & 6.214 & 6.522 & 15.50 & 11.05 \\
21/04/2010 &  23:36 &  J47 & 6.207 & 5.858 & 14.00 &  9.79 & 03/06/2010 &  20:34
 &  B74 & 6.209 & 6.049 & 15.70 & 11.41 \\
23/04/2010 &  20:10 &  I99 & 6.207 & 5.890 & 13.90 &  9.68 & 03/06/2010 &  20:41
 &  B74 & 6.209 & 6.049 & 15.80 & 11.51 \\
23/04/2010 &  20:14 &  I99 & 6.207 & 5.890 & 13.90 &  9.68 & 03/06/2010 &  20:46
 &  B74 & 6.209 & 6.049 & 15.70 & 11.41 \\
23/04/2010 &  20:21 &  I99 & 6.207 & 5.890 & 13.91 &  9.69 & 03/06/2010 &  20:54
 &  B74 & 6.209 & 6.049 & 15.70 & 11.41 \\
23/04/2010 &  20:21 &  J47 & 6.207 & 5.890 & 14.00 &  9.78 & 03/06/2010 &  21:08
 &  J53 & 6.209 & 6.049 & 15.80 & 11.51 \\
23/04/2010 &  20:23 &  I99 & 6.207 & 5.890 & 13.90 &  9.68 & 03/06/2010 &  21:13
 &  J53 & 6.209 & 6.049 & 15.80 & 11.51 \\
23/04/2010 &  20:32 &  I99 & 6.207 & 5.890 & 13.90 &  9.68 & 03/06/2010 &  21:24
 &  J53 & 6.209 & 6.049 & 15.75 & 11.46 \\
23/04/2010 &  20:39 &  J47 & 6.207 & 5.890 & 14.00 &  9.78 & 03/06/2010 &  21:24
 &  945 & 6.209 & 6.049 & 15.70 & 11.41 \\
23/04/2010 &  20:59 &  J47 & 6.207 & 5.890 & 14.04 &  9.82 & 03/06/2010 &  21:30
 &  945 & 6.209 & 6.049 & 15.70 & 11.41 \\
23/04/2010 &  21:18 &  J47 & 6.207 & 5.890 & 14.10 &  9.88 & 03/06/2010 &  21:34
 &  945 & 6.209 & 6.049 & 15.74 & 11.45 \\
23/04/2010 &  21:38 &  J47 & 6.207 & 5.890 & 14.00 &  9.78 & 03/06/2010 &  21:36
 &  945 & 6.209 & 6.049 & 15.80 & 11.51 \\
24/04/2010 &  21:12 &  J51 & 6.207 & 5.905 & 14.30 & 10.07 & 03/06/2010 &  21:38
 &  J53 & 6.209 & 6.049 & 15.70 & 11.41 \\
24/04/2010 &  21:23 &  J51 & 6.207 & 5.905 & 14.25 & 10.02 & 03/06/2010 &  21:44
 &  J38 & 6.209 & 6.049 & 15.80 & 11.51 \\
24/04/2010 &  21:36 &  J51 & 6.207 & 5.905 & 14.30 & 10.07 & 03/06/2010 &  21:45
 &  945 & 6.209 & 6.049 & 15.80 & 11.51 \\
25/04/2010 &  21:00 &  J51 & 6.207 & 5.921 & 14.40 & 10.16 & 03/06/2010 &  21:45
 &  J38 & 6.209 & 6.049 & 15.78 & 11.49 \\
25/04/2010 &  21:15 &  J51 & 6.207 & 5.921 & 14.43 & 10.19 & 03/06/2010 &  21:47
 &  J38 & 6.209 & 6.049 & 15.80 & 11.51 \\
25/04/2010 &  21:30 &  J51 & 6.207 & 5.921 & 14.40 & 10.16 & 03/06/2010 &  21:51
 &  945 & 6.209 & 6.049 & 15.80 & 11.51 \\
25/04/2010 &  22:17 &  213 & 6.207 & 5.921 & 14.40 & 10.16 & 03/06/2010 &  21:53
 &  J38 & 6.209 & 6.049 & 15.80 & 11.51 \\
25/04/2010 &  22:23 &  213 & 6.207 & 5.921 & 14.42 & 10.18 & 04/06/2010 &  20:54
 &  B74 & 6.215 & 6.551 & 15.80 & 11.34 \\
25/04/2010 &  22:23 &  213 & 6.207 & 5.921 & 14.40 & 10.16 & 05/06/2010 &  20:36
 &  213 & 6.215 & 6.566 & 16.06 & 11.59 \\
\hline
\end{longtable}

\begin{minipage}{0.93\linewidth}
$^{a}$ Observer MPC code.
\end{minipage}
}
\twocolumn
\end{appendix}


\begin{thebibliography}{}

\bibitem[Agarwal et al.(2017)]{2017MNRAS.469S.606A} Agarwal, J., Della Corte, V., Feldman, P.~D., et al.\ 2017, \mnras, 469, s606. doi:10.1093/mnras/stx2386

\bibitem[A'Hearn et al.(2011)]{Ahearn2011} A'Hearn, M.~F., Belton,
M.~J.~S., Delamere, W.~A., et al.\ 2011, Science, 332, 1396

\bibitem[A'Hearn et al.(2012)]{Ahearn2012} A'Hearn, M.~F., Feaga,
L.~M., Keller, H.~U., et al.\ 2012, \apj, 758, 29

\bibitem[Bauer et al.(2013)]{2013ApJ...773...22B} Bauer, J.~M., Grav, T., Blauvelt, E., et al.\ 2013, \apj, 773, 22. doi:10.1088/0004-637X/773/1/22

\bibitem[Beer et al.(2006)]{2006Icar..180..473B} Beer, E.~H., Podolak, M., \& Prialnik, D.\ 2006, \icarus, 180, 473. doi:10.1016/j.icarus.2005.10.018


\bibitem[Biver(1997)]{Biver1997} Biver, N.\ 1997, Ph.D.~Thesis,
University Paris 7

\bibitem[Biver et al.(1999)]{Biver1999} Biver, N.,
Bockel{\'e}e-Morvan, D., Crovisier, J., et al.\ 1999, \aj, 118,
1850

\bibitem[Biver et
al.(2002)]{Biver2002} Biver, N., Bockel{\'e}e-Morvan, D., Colom, P., et al.\ 2002, Earth Moon and Planets, 90, 5

\bibitem[Biver et al.(2007)]{2007P&SS...55.1058B} Biver, N., Bockel{\'e}e-Morvan, D., Crovisier, J., et al.\ 2007, \planss, 55, 1058. doi:10.1016/j.pss.2006.11.010

\bibitem[Biver et al.(2008)]{2008LPICo1405.8146B} Biver, N., Bockel{\'e}e-Morvan, D., Wiesemeyer, H., et al.\ 2008, Asteroids, Comets, Meteors 2008, 1405, 8146


\bibitem[Biver et al.(2012)]{Biver2012} Biver, N., Crovisier, J.,
Bockel{\'e}e-Morvan, D., et al.\ 2012, \aap, 539, A68


\bibitem[Bocchio et al.(2016)]{2016yCat..35910117B} Bocchio, M., Bianchi, A., \& Abergel, S.\ 2016, VizieR Online Data Catalog, J/A+A/591/A117

\bibitem[Bockel{\'e}e-Morvan et al.(2004)]{dbm2004}
Bockel{\'e}e-Morvan, D., Crovisier, J., Mumma, M.~J.,
\& Weaver, H.~A.\ 2004, Comets II, 391

\bibitem[Bockel{\'e}e-Morvan et al.(2010a)]{2010DPS....42.0304B} Bockel\'ee-Morvan, D., Biver, N., Crovisier, J., et al. \ 2010, Bull.  Amer.  Astron.  Soc.  42, 946\ 

\bibitem[Bockel{\'e}e-Morvan et al.(2010b)]{2010A&A...518L.149B} Bockel{\'e}e-Morvan, D., Hartogh, P., Crovisier, J., et al.\ 2010, \aap, 518, L149. doi:10.1051/0004-6361/201014655

\bibitem[Bockel{\'e}e-Morvan et al.(2012)]{2012A&A...544L..15B} Bockel{\'e}e-Morvan, D., Biver, N., Swinyard, B., et al.\ 2012, \aap, 544, L15. doi:10.1051/0004-6361/201219744

\bibitem[Bockel{\'e}e-Morvan et al.(2014)]{2014A&A...562A...5B} Bockel{\'e}e-Morvan, D., Biver, N., Crovisier, J., et al.\ 2014, \aap, 562, A5. doi:10.1051/0004-6361/201322939

\bibitem[Bockel{\'e}e-Morvan et al.(2017)]{2017MNRAS.469S.443B} Bockel{\'e}e-Morvan, D., Rinaldi, G., Erard, S., et al.\ 2017, \mnras, 469, S443. doi:10.1093/mnras/stx1950

\bibitem[Cochran
\& Cochran(1991)]{Cochran1991} Cochran, A.~L., \& Cochran, W.~D.\
1991, \icarus, 90, 172

\bibitem[Cowan \& A'Hearn(1979)]{Cowan1979} Cowan, J.~J. \& A'Hearn, M.~F.\ 1979, Moon and Planets, 21, 155. doi:10.1007/BF00897085

\bibitem[Crifo \& Rodionov(1997)]{1997Icar..127..319C} Crifo, J.~F. \& Rodionov, A.~V.\ 1997, \icarus, 127, 319. doi:10.1006/icar.1997.5690

\bibitem[Crifo et al.(1999)]{Crifo1999} Crifo, J.~F., Rodionov,
A.~V., \& Bockel{\'e}e-Morvan, D.\ 1999, \icarus, 138, 85


\bibitem[Crovisier et al.(1995)]{Crovisier1995} Crovisier, J., Biver,
N., Bockel\'ee-Morvan, D., et al.\ 1995, \icarus, 115, 213

\bibitem[Davies et al.(1997)]{Davies1997} Davies, J.~K., Roush,
T.~L., Cruikshank, D.~P., et al.\ 1997, \icarus, 127, 238

\bibitem[de Graauw et al.(2010)]{2010HIFI}de Graauw, Th., Helmich, F.P., Phillips, T.G., et al. 2010, \aap, 518, L6

\bibitem[Dello Russo et al.(2016)]{Dello16} Dello Russo, N., Kawakita, H., Vervack, R.~J., et al.\ 2016, \icarus, 278, 301. doi:10.1016/j.icarus.2016.05.039

\bibitem[de Val-Borro et al.(2010)]{2010A&A...521L..50D} de Val-Borro, M., Hartogh, P., Crovisier, J., et al.\ 2010, \aap, 521, L50. doi:10.1051/0004-6361/201015161

\bibitem[de Val-Borro et al.(2014)]{2014A&A...564A.124D} de Val-Borro, M., Bockel{\'e}e-Morvan, D., Jehin, E., et al.\ 2014, \aap, 564, A124. doi:10.1051/0004-6361/201423427

\bibitem[DiSanti et al.(2017)]{Disanti2017} DiSanti, M.~A., Bonev, B.~P., Russo, N.~D., et al.\ 2017, \aj, 154, 246. doi:10.3847/1538-3881/aa8639


\bibitem[Dorschner et al.(1995)]{dors95} Dorschner, J., Begemann, B., Henning, T., Jaeger, C., \& Mutschke, H.\ 1995, \aap, 300, 503

\bibitem[Edoh(1983)]{edo83} Edoh, J. H. 1983, Ph.D. thesis, Univ. Arizona

\bibitem[Enzian et al.(1997)]{Enzian1997} Enzian, A., Cabot, H., \& Klinger, J.\
1997, \aap, 319, 995

\bibitem[Exter et al.(2018)]{Exter2018}Exter, K., Balog, Z., Calzoletti, L., et al. 2018, The Photodetector Array Camera and Spectrometer (PACS) Handbook, Herschel-HSC-DOC-2101

\bibitem[Fanale \& Salvail(1984)]{Fanale1984} Fanale, F.~P., \& Salvail, J.~R.\
1984, \icarus, 60, 476

\bibitem[Feldman et al.(1996)]{Feldman1996} Feldman, P.~D.,
McPhate, J.~B., Weaver, H.~A., Tozzi, G.-P., \& A'Hearn, M.~F.\
1996, Bulletin of the American Astronomical Society, 28, 1084

\bibitem[Fernandez(1999)]{yan_phd_1999} Fernandez, Y.~R.\ 1999, Ph.D. Thesis, 6150

\bibitem[Fern{\'a}ndez et al.(2013)]{2013Icar..226.1138F} Fern{\'a}ndez, Y.~R., Kelley, M.~S., Lamy, P.~L., et al.\ 2013, \icarus, 226, 1138. doi:10.1016/j.icarus.2013.07.021

\bibitem[Festou et al.(2001)]{Festou2001} Festou, M.~C.,
Gunnarsson, M., Rickman, H., Winnberg, A., \& Tancredi, G.\ 2001,
\icarus, 150, 140

\bibitem[Fornasier et
al.(2013)]{Fornasier2013} Fornasier, S., Lellouch, E., M{\"u}ller, T., et al.\ 2013, \aap, 555, A15

\bibitem[Fougere et al.(2012)]{Fougere2012} Fougere, N., Combi,
M.~R., Tenishev, V., et al.\ 2012, \icarus, 221, 174

\bibitem[Fray \& Schmitt(2009)]{Fray2009} Fray, N., \& Schmitt, B.\ 2009,
\planss, 57, 2053

\bibitem[Fulle(1992)]{Fulle1992} Fulle, M.\ 1992, \nat, 359, 42


\bibitem[\protect\citeauthoryear{Greenberg
\& Hage}{1990}]{green90} Greenberg, J.~M.,\& Hage, J.~I. 1990,
ApJ, 361, 260

\bibitem[Griffin et al.(2010)]{2010A&A...518L...3G} Griffin, M.~J., Abergel, A., Abreu, A., et al.\ 2010, \aap, 518, L3. doi:10.1051/0004-6361/201014519


\bibitem[Groussin et al.(2013)]{Groussin2013} Groussin, O.,
Sunshine, J.~M., Feaga, L.~M., et al.\ 2013, \icarus, 222, 580

\bibitem[Gr{\"u}n et al.(2001)]{2001A&A...377.1098G} Gr{\"u}n, E., Hanner, M.~S., Peschke, S.~B., et al.\ 2001, \aap, 377, 1098. doi:10.1051/0004-6361:20011139

\bibitem[Gulkis et al.(2015)]{2015Sci...347a0709G} Gulkis, S., Allen, M., von Allmen, P., et al.\ 2015, Science, 347, aaa0709


\bibitem[Gunnarsson et al.(2002)]{Gunnarsson2002} Gunnarsson, M., Rickman, H., Festou, M.~C., Winnberg, A., \& Tancredi, G.\ 2002,
\icarus, 157, 309

\bibitem[Gunnarsson(2003)]{Gunnarsson2003} Gunnarsson, M.\ 2003, \aap, 398, 353
\bibitem[Gunnarsson et al.(2008)]{Gunnarsson2008} Gunnarsson, M., Bockel{\'e}e-Morvan,
D., Biver, N., Crovisier, J., \& Rickman, H.\ 2008, \aap, 484, 537

\bibitem[Harris(1998)]{harris98} Harris, A. W. 1998, \icarus, 131, 291

\bibitem[Harris \& Lagerros(2002)]{2002aste.book..205H} Harris, A.~W. \& Lagerros, J.~S.~V.\ 2002, Asteroids III, 205

\bibitem[Hartogh et al.(2009)]{hart09} Hartogh, P., Lellouch, E., Crovisier, J. et
al.\ 2009, P\&SS, 57, 1596

\bibitem[H{\"o}fner et al.(2017)]{2017A&A...608A.121H} H{\"o}fner, S., Vincent, J.-B., Blum, J., et al.\ 2017, \aap, 608, DA121. doi:10.1051/0004-6361/201628726


\bibitem[Hosek et al., 2013]{hosek2013} Hosek, M.~W., Blaauw, R.~C., Cooke, W.~J., Suggs, R.~M. 2013, AJ, 145, 122


\bibitem[Ivanova et al.(2016)]{2016P&SS..121...10I} Ivanova, O.~V., Luk`yanyk, I.~V., Kiselev, N.~N., et al.\ 2016, \planss, 121, 10. doi:10.1016/j.pss.2015.12.001

\bibitem[Jehin et al., 2011]{jehin2011} Jehin, E., Gillon, M., Queloz, D., et al. 2011, The Messenger, 145, 2 

\bibitem[Kawakita et al.(2004)]{Kawakita2004} Kawakita, H.,
Watanabe, J.-i., Ootsubo, T., et al.\ 2004, \apjl, 601, L191


\bibitem[Korsun et al.(2008)]{Korsun2008} Korsun, P.~P., Ivanova, O.~V., \& Afanasiev, V.~L.\ 2008, \icarus, 198, 465

\bibitem[Kossacki \& Szutowicz(2013)]{Kossacki2013} Kossacki, K.~J., \& Szutowicz, S.\ 2013, \icarus, 225, 111

\bibitem[Lamy \& Toth(1995)]{lamy-toth_1995} Lamy, P.~L. \& Toth, I.\ 1995, \aap, 293, L43

\bibitem[Lellouch et al.(1998)]{Lellouch1998} Lellouch, E., Crovisier, J., Lim, T., et
al.\ 1998, \aap, 339, L9


\bibitem[Lellouch et
al.(2013)]{Lellouch2013} Lellouch, E., Santos-Sanz, P., Lacerda, P., et al.\ 2013, \aap, 557, A60

\bibitem[Lippi et al.(2021)]{Lippi21} Lippi, M., Villanueva, G.~L., Mumma, M.~J., et al.\ 2021, \aj, 162, 74. doi:10.3847/1538-3881/abfdb7

\bibitem[Lis et al.(2019)]{Lis2019} Lis, D.~C., Bockel{\'e}e-Morvan, D., G{\"u}sten, R., et al.\ 2019, \aap, 625, L5. doi:10.1051/0004-6361/201935554

\bibitem[Lisse et al.(1999)]{lisse_1999} Lisse, C.~M., Fern{\'a}ndez, Y.~R., Kundu, A., et al.\ 1999, \icarus, 140, 189. doi:10.1006/icar.1999.6131

\bibitem[Marshall et al.(2018)]{2018A&A...616A.122M} Marshall, D., Groussin, O., Vincent, J.-B., et al.\ 2018, \aap, 616, A122

\bibitem[Miles(2016)]{Miles2016} Miles, R.\ 2016, \icarus, 272, 387. doi:10.1016/j.icarus.2015.11.011

\bibitem[Naves \& Camp\`as(2007)]{Naves2007}
Naves, R., \& Camp\`as, M. 2007, MPEC 2007-Y59, http://www.observadores-cometas.com

\bibitem[Ootsubo et al.(2012)]{Ootsubo2012} Ootsubo, T., Kawakita,
H., Hamada, S., et al.\ 2012, \apj, 752, 15

\bibitem[Paganini et al.(2013)]{Paganini2013} Paganini, L., Mumma, M.~J., Boehnhardt, H., et al.\ 2013, \apj, 766, 100

\bibitem[Pajola et al.(2017)]{2017NatAs...1E..92P} Pajola, M., H{\"o}fner, S., Vincent, J.~B., et al.\ 2017, Nature Astronomy, 1, 0092. doi:10.1038/s41550-017-0092

\bibitem[Pilbratt et al.(2010)]{Pilbratt2010} Pilbratt, G.~L., Riedinger, J.~R., Passvogel, T., et al.\ 2010, \aap, 518, L1

\bibitem[Poglitsch et al.(2010)]{Pacs2010} Poglitsch, A., Waelkens, C., Geis, N., et al.\ 2010, \aap, 518, L2

\bibitem[Prialnik \& Bar-Nun(1987)]{1987ApJ...313..893P} Prialnik, D. \& Bar-Nun, A.\ 1987, \apj, 313, 893. doi:10.1086/165029

\bibitem[Prialnik \& Bar-Nun(1990)]{1990ApJ...363..274P} Prialnik, D. \& Bar-Nun, A.\ 1990, \apj, 363, 274. doi:10.1086/169339

\bibitem[Protopapa et al.(2014)]{2014Icar..238..191P} Protopapa, S., Sunshine, J.~M., Feaga, L.~M., et al.\ 2014, \icarus, 238, 191

\bibitem[Protopapa et al.(2018)]{2018ApJ...862L..16P} Protopapa, S., Kelley, M.~S.~P., Yang, B., et al.\ 2018, \apjl, 862, L16

\bibitem[Sarid et al.(2019)]{2019ApJ...883L..25S} Sarid, G., Volk, K., Steckloff, J.~K., et al.\ 2019, \apjl, 883, L25. doi:10.3847/2041-8213/ab3fb3

\bibitem[Schambeau et al.(2015)]{2015Icar..260...60S} Schambeau, C.~A., Fern{\'a}ndez, Y.~R., Lisse, C.~M., et al.\ 2015, \icarus, 260, 60

\bibitem[Schambeau et al.(2017)]{2017Icar..284..359S} Schambeau, C.~A., Fern{\'a}ndez, Y.~R., Samarasinha, N.~H., et al.\ 2017, \icarus, 284, 359. doi:10.1016/j.icarus.2016.11.026

\bibitem[Schambeau et al.(2021)]{2021PSJ.....2..126S} Schambeau, C.~A., Fern{\'a}ndez, Y.~R., Samarasinha, N.~H., et al.\ 2021, PSJ, 2, 126. doi:10.3847/PSJ/abfe6f

\bibitem[Schleicher(2009)]{2009AJ....138.1062S} Schleicher, D.~G.\ 2009, \aj, 138, 1062. doi:10.1088/0004-6256/138/4/1062

\bibitem[Schleicher \& Bair(2011)]{2011AJ....141..177S} Schleicher, D.~G. \& Bair, A.~N.\ 2011, \aj, 141, 177. doi:10.1088/0004-6256/141/6/177

\bibitem[Senay \& Jewitt(1994)]{Senay1994} Senay, M.~C., \& Jewitt, D.\ 1994,
\nat, 371, 229

\bibitem[Shipman et al.(2017)]{Shipman2017} Shipman, R.~F., Beaulieu, S.~F., Teyssier, D., et al.\ 2017, \aap, 608, A49

\bibitem[Stansberry et al.(2004)]{Stansberry2004} Stansberry, J.~A., Van Cleve, J., Reach, W.~T., et al.\ 2004, \apjs, 154, 463

\bibitem[Swinyard et al.(2010)]{2010A&A...518L...4S} Swinyard, B.~M., Ade, P., Baluteau, J.-P., et al.\ 2010, \aap, 518, L4. doi:10.1051/0004-6361/201014605

\bibitem[Szab{\'o} et al.(2012)]{2012ApJ...761....8S} Szab{\'o}, G.~M., Kiss, L.~L., P{\'a}l, A., et al.\ 2012, \apj, 761, 8. doi:10.1088/0004-637X/761/1/8

\bibitem[Tenishev et al.(2008)]{2008ApJ...685..659T} Tenishev, V., Combi, M., \& Davidsson, B.\ 2008, \apj, 685, 659. doi:10.1086/590376

\bibitem[Teyssier et al.(2017)]{Teyssier2017}
Teyssier, D., Avruch, I., Beaulieu, S., et al. 2017, HIFI Handbook, HERSCHEL-HSC-DOC-2097

\bibitem[Trigo-Rodr{\'{\i}}guez et al.(2008)]{Trigo2008} Trigo-Rodr{\'{\i}}guez, J.~M., Garc{\'{\i}}a-Melendo, E., Davidsson, B.~J.~R., et al.\ 2008, \aap, 485, 599

\bibitem[Trigo-Rodr{\'{\i}}guez et al.(2010)]{Trigo2010} Trigo-Rodr{\'{\i}}guez, J.~M., Garc{\'{\i}}a-Hern{\'a}ndez, D.~A., S{\'a}nchez, A., et al.\ 2010, \mnras, 409, 1682

\bibitem[Tubiana et al.(2019)]{2019A&A...630A..23T} Tubiana, C., Rinaldi, G., G{\"u}ttler, C., et al.\ 2019, \aap, 630, A23. doi:10.1051/0004-6361/201834869

\bibitem[Valchanov(2017)]{spire_handbook}
Valchanov, I. 2017, Herschel Explanatory Supplement vol. IV, HERSCHEL-HSC-DOC-0798

\bibitem[Vincent et al.(2016)]{2016MNRAS.462S.184V} Vincent, J.-B., A'Hearn, M.~F., Lin, Z.-Y., et al.\ 2016, \mnras, 462, S184. doi:10.1093/mnras/stw2409

\bibitem[Vincent et al.(2016)]{2016A&A...587A..14V} Vincent, J.-B., Oklay, N., Pajola, M., et al.\ 2016, \aap, 587, A14. doi:10.1051/0004-6361/201527159

\bibitem[\protect\citeauthoryear{Warren \& Brandt}{2008}]{war08}
Warren, S. G., \& Brandt, R. E. 2008, JGR 113, D14220

\bibitem[Weiler et al.(2003)]{2003A&A...403..313W} Weiler, M., Rauer, H., Knollenberg, J., et al.\ 2003, \aap, 403, 313. doi:10.1051/0004-6361:20030289

\bibitem[Wierzchos \& Womack(2020)]{2020AJ....159..136W} Wierzchos, K. \& Womack, M.\ 2020, \aj, 159, 136

\bibitem[Womack et al.(2021)]{2021PSJ.....2...17W} Womack, M., Curtis, O., Rabson, D.~A., et al.\ 2021, PSJ, 2, 17. doi:10.3847/PSJ/abd32c

\bibitem[Zakharov et al.(2007)]{zakharov2007} Zakharov, V., Bockel{\'e}e-Morvan, D.,
Biver, N., Crovisier, J., \& Lecacheux, A.\ 2007, \aap, 473, 303

\bibitem[Zakharov et al.(2018)]{2018Icar..312..121Z} Zakharov, V.~V., Ivanovski, S.~L., Crifo, J.-F., et al.\ 2018, \icarus, 312, 121. doi:10.1016/j.icarus.2018.04.030

\bibitem[Zakharov et al.(2021)]{2021Icar..35414091Z} Zakharov, V.~V., Rodionov, A.~V., Fulle, M., et al.\ 2021, \icarus, 354, 114091. doi:10.1016/j.icarus.2020.114091


\end{thebibliography}
\end{document}